\pdfoutput=1

\def\useplain{1}

\ifx \useplain\undefined
	\documentclass[stsy, nonblindrev]{informs-stsy}
	\usepackage{informs_set}
	\input{informs_title}
\else
	\documentclass{article}
	\usepackage[margin=1.5in]{geometry}
	\usepackage{plain_macro}
	\title{Anonymous Stochastic Routing}

\author{
Mine Su Erturk\\
Graduate School of Business\\
Stanford University\\
   \texttt{mserturk@stanford.edu} 
  \and
Kuang Xu\\
Graduate School of Business\\
Stanford University\\
   \texttt{kuangxu@stanford.edu} 
}

\date{}
\fi

\usepackage{kuang_macro}
\usepackage{minesu_macro}
\usepackage{algpseudocode}
\usepackage{algorithm}

\def\neig{\mathcal{N}}

\begin{document}

\ifx \useplain\undefined
	\input{informs_title}
\else
	\title{Anonymous Stochastic Routing}

\author{
Mine Su Erturk\\
Graduate School of Business\\
Stanford University\\
   \texttt{mserturk@stanford.edu} 
  \and
Kuang Xu\\
Graduate School of Business\\
Stanford University\\
   \texttt{kuangxu@stanford.edu} 
}

\date{}
%\date{\today}
\fi

\ifx \useplain\undefined
\else
\maketitle
\fi

\def\abst_txt{We propose and analyze a recipient-anonymous stochastic routing model to study a fundamental trade-off between anonymity and routing delay. An agent wants to quickly reach a goal vertex in a network through a sequence of routing actions, while an overseeing adversary observes the agent's entire trajectory and tries to identify her goal among those vertices traversed. We are interested in understanding the probability that the adversary can correctly identify the agent's goal (anonymity), as a function of the time it takes the agent to reach it (delay). A key feature of our model is the presence of intrinsic uncertainty in the environment, so that each of the agent's intended steps is subject to random perturbation and thus may not materialize as planned. Using large-network asymptotics, our main results provide near-optimal characterization of the anonymity-delay trade-off under a number of network topologies. Our main technical contributions are centered around a new class of ``noise-harnessing'' routing strategies that adaptively combine intrinsic uncertainty from the environment with additional artificial randomization to achieve provably efficient obfuscation.\footnote{August 2020; revised: November 2020. This is an updated version of an earlier manuscript titled “Dynamically Protecting Privacy, under Uncertainty”.}

\emph{Keywords}: {\kwd}}

\def\kwd{anonymity, stochastic routing, networks, prediction, random walk.}

\ifx \useplain\undefined
\ABSTRACT{\abst_txt}
\else
\begin{abstract}
\abst_txt
\end{abstract}
\fi

\ifx \useplain\undefined
\maketitle
\fi

\vspace{-2em}
\section{Introduction}\label{sec:intro}

The advancement in machine learning and data collection infrastructure has made it increasingly effortless for companies and governmental entities to collect and analyze the behaviors and actions of individuals or competitors (\cite{valentino2018, mayer2016evaluating, de2013unique}). Such analysis enables the entity to make powerful predictions on sensitive information that an individual under monitoring would like to keep private. These emerging trends have spurred in recent years a growing literature on designing privacy-aware decision-making policies, whereby the decision maker would deliberately employ randomization in order to obfuscate certain sensitive information from an adversary  (cf.~\cite{fanti2015spy,luo2016infection, tossou2016algorithms, tsitsiklis2018delay, tang2020privacy}). 

%Models in this area typically assume that an adversary can (partially) observe actions of a decision maker, and in response, the decision maker would inject carefully curated {artificial randomization} in her actions in order to make it difficult for the adversary to accurately infer some sensitive hidden feature or input. 

Motivated by privacy concerns in applications arising in networking and secure logistics, and inspired by an anonymous path-planning problem studied in \cite{tsitsiklis2018delay}, we propose and analyze in this paper a stochastic routing problem that protects the anonymity of the recipient.  The model concerns an agent who traverses an underlying network to deliver an item or package, and her goal is to obfuscate the identity of the true recipient of the item even when the route is observable by an adversary. 
Our model departs from the prevailing literature in a major aspect: the routing dynamics in our model are subject to \emph{intrinsic uncertainty}, in the sense that the agent's routing trajectory is perturbed by random shocks in the environment and may deviate from the agent's intended movements. The majority of existing work on anonymity-aware decision-making, on the other hand, assume the decision maker has perfect control, so that their action maps to changes in the system in a deterministic manner; examples include the state updates in \cite{tsitsiklis2018delay}, routing of payments in \cite{tang2020privacy}, and dissemination of messages in \cite{fanti2015spy}. To the best of our knowledge, our work is the first to provide formal performance and anonymity guarantees in a decision-making problem under  intrinsic uncertainty. 

There are at least two main reasons why it is important to consider models with intrinsic uncertainty. First, random perturbations and stochastic shocks are inherent in a wide array of real-world systems \citep{sezer2015towards,jaillet2016routing,  flajolet2017robust}. For example, in a distributed network, computer nodes may experience random outages, rendering a link occasionally non-available, and similarly, roads may become difficult to traverse due to accidents, traffic, or unforeseen maintenance. Models that assume perfect control thus fail to capture these situations. Second, going from a model of perfect control to one with intrinsic uncertainty requires a fundamental redesign of policies, rather than minor modifications of existing algorithms. Our analysis suggests that having perfect control in a routing problem significantly simplifies the design of anonymity-aware policies, largely because it allows the decision maker to precisely control the type of randomization she wishes to implement. In contrast, intrinsic uncertainty tends to severely limit the range of randomization at the decision maker's disposal. Indeed, a main finding of our work is that designing efficient decision policies in the presence of uncertainty requires new ideas, such as noise harnessing and sequential randomization, which are generally unnecessary, and thus absent, in models with perfect control.  While our work focuses on the problem domain of routing, it appears likely that these findings have analogues in other dynamic decision-making problems, as well. 

We begin with an informal description of the Anonymous Stochastic Routing model. The decision maker is an \emph{agent} who operates in discrete time and whose state in each time period corresponds to a vertex in an undirected graph $G$. By traversing along the edges of $G$, the agent's main objective is to reach a {goal} vertex, $D$, drawn uniformly at random from $\calV$. The agent's performance is measured by the {\bf delay}, defined as the expected number of steps before she reaches the goal vertex for the first time. In the context of routing, the  agent can be thought of as a vessel (e.g., a data package containing encrypted information or a physical vehicle) that traverses an underlying network $G$ (e.g., a computer network or physical transportation network) in order to deliver a package or item to a recipient, the goal vertex. Crucially, the agent does not have perfect control over her trajectory due to intrinsic uncertainty: in each time period, the agent may choose a vertex, $v$, among those connected to her current state as her intended next step, which we refer to as her action. In the following period, her state becomes $v$ with probability $1-\varepsilon$, and is set to a random neighboring vertex, otherwise, where $\varepsilon$ is the {\bf noise level} which captures the degree of intrinsic uncertainty in the environment.

% For the purpose of anonymity, to be explained shortly, we will allow the agent to continue traversing the graph even after her goal has been reached. This makes it possible that the end of her trajectory is not necessarily the goal vertex, and thereby making anonymity possible for someone who observes her trajectory. 

The above routing model intends to capture the stochastic obstructions the agent faces as she traverses the network, such as those induced by random link failures, non-cooperating nodes, or traffic congestion. This model of uncertainty is the same as the stochastic routing problem considered in \cite{croucher1978note}, which in turn is a special case of what has come to be known as the stochastic shortest-path problem \citep{eaton1962optimal, bertsekas1991analysis}, which considers more general random shocks. We will explore more in detail how such uncertainty manifests in some example applications in Section \ref{sec:motExamp}. 

We now introduce the notion of anonymity. We assume that an overseeing \emph{adversary} obtains the entire actual trajectory traversed by the agent (but not her actions, which may or may not have materialized), which may not necessarily terminate in the goal vertex $D$. The objective of the adversary is to leverage this knowledge to predict the identity of $D$ among those vertices visited by the agent. If such a prediction can be performed with a high-level of accuracy, then we say that the agent's goal is not anonymous, and conversely, if we can demonstrate that the adversary cannot come up with an accurate prediction, then we say that the agent's routing strategy affords a high level of anonymity. 

Being wary of the adversary, the agent seeks a routing strategy that will minimize the probability that the goal can be predicted by the adversary, which we refer to as the {\bf prediction risk}, subject to an upper bound on the delay. The agent thus faces a delay-anonymity trade-off: on the one hand, she needs to employ randomization in her actions to obfuscate the goal vertex, while on the other hand, excessive obfuscation could substantially increase delay. 

The main contribution of this paper is to provide near-optimal characterizations for the fundamental trade-off between delay and prediction risk. At a high level,  a main finding of our paper is that, in the presence of intrinsic uncertainty, such trade-off takes on the form
\begin{equation}
\mbox{prediction risk}  = \mcal{O} \left(\frac{1}{\mbox{delay} - \mbox{cost of uncertainty}} \right).
\label{eq:high_lel_trade-off}
\end{equation}
where ``cost of uncertainty'' is a term that increases as the level of uncertainty, $\varepsilon$, grows, but does not depend on the delay budget. We establish trade-offs of this type in complete graphs, linearly dense non-complete graphs that are generated by a random graph model, and a family of sparse graphs, and show that they are achievable using a novel family of routing strategies, the Water-Filling Strategies. Importantly, Eq.~\eqref{eq:high_lel_trade-off} shows that while higher levels of intrinsic uncertainty will force the agent to incur longer delays, the additional overhead is only {additive}. 

To be clear, many assumptions in our model, such as the structure of the random shocks or the prior distribution of the goal vertex, are stylized, and as such applying the strategies we propose in an application would require caution and necessary modifications. However, it appears that a rigorous study of the kind we carry out on how anonymous routing policies perform in a stochastic environment will, at the least, yield valuable insights into the nature of intrinsic uncertainty and help guide policy design in real-world systems. 

\subsection{Why does intrinsic uncertainty matter: stochastic versus deterministic routing}
\label{sec:why_uncertainty}

Before stating our main results, let us first provide some intuition as to 
\begin{enumerate}
\item why the presence of  uncertainty fundamentally changes the nature of anonymous routing, and 
\item why the additive overhead in Eq.~\eqref{eq:high_lel_trade-off} may not be trivial to obtain. 
\end{enumerate}

To this end, it is instructive to contrast our setup with one without any intrinsic uncertainty. Our anonymous stochastic routing model is inspired by, and generalizes, the Goal Prediction Game proposed in \cite{tsitsiklis2018delay}. They consider the delay-anonymity trade-off in a \emph{deterministic} routing problem, where the agent has perfect control over her trajectory, and prove that the optimal prediction risk scales inversely proportionally with respect to the delay budget. The model in \cite{tsitsiklis2018delay} is therefore similar\footnote{To be more precise, \cite{tsitsiklis2018delay} consider a different, online version of the problem, where the adversary aims to predict the agent's goal in real-time \emph{before} it is reached, whereas in our application of routing, such predictions are made offline after the agent's routing actions have been completed. However, the distinction between the online and offline  formulations is in fact not significant,   and the results are generally mutually transferable.}  to the noiseless special case of our model, by setting $\varepsilon=0$. 

In light of the results in a noiseless setting, it is natural to ask whether only minor policy modifications are needed in order to achieve desirable results in the presence of uncertainty. The answer depends on the level of uncertainty. As one might expect, a naive adaptation of strategies for the deterministic model would perform reasonably well under low levels of uncertainty.  However, as we illustrate below,  its performance deteriorates quickly as the level of uncertainty increases, necessitating a substantial redesign of the policy.

It is shown in \cite{tsitsiklis2018delay} that in a deterministic routing setting, the following family of meta-routing policies will achieve the optimal delay-anonymity trade-off: upon seeing the goal vertex $D$, the agent generates a {random} path, $S$, and proceeds to traverse the entirety of the path. Importantly, the path $S$ includes, but not necessarily terminates at, $D$, and the random mapping from $D$ to $S$ is carefully designed in such a way that conditional on seeing the realized path $S$, the adversary has little idea as to which one of the vertices therein is the true goal. Let us now take intrinsic uncertainty into account. It is easy to show that because the agent can no longer exactly execute her routing actions, it is not possible to implement the type of random path selection that was available in a deterministic world. Instead, we can consider a natural adaptation of the policy: first, the agent would randomly sample an ``intended'' path, $S$, as she would in a deterministic model. Then, the agent would aim to traverse successive vertices in $S$ using a certain stochastic-shortest-path sub-routine. That is, the agent will try to first reach $S_1$, and once that's achieved, try to reach $S_2$, and so on. 

One can show that this adapted algorithm will provide the same level of anonymity guarantee as the original, because the vertices the agent touches between successive visits to vertices in $S$ are purely a result of the intrinsic uncertainty and are independent from the goal vertex. Therefore, they do not provide additional information to the adversary. Unfortunately, the bad news is that the delay cost incurred in executing the path $S$ is now substantially worse: it is not difficult to show that, due to uncertainty, it can take an average of $\mathbf{\Omega}\left(\frac{1}{1-\varepsilon} \right)$ to $\mathbf{\Omega}\left(\frac{1}{1-2\varepsilon} \right)$ steps to execute just one step in the path $S$, depending on the graph topology. In other words, the presence of uncertainty leads to a \emph{multiplicative} increase in delay, leading to a delay-anonymity trade-off of the form:  
\begin{equation}
\mbox{prediction risk}  = \mcal{O} \left(\frac{1}{\mbox{delay} \times \mbox{cost of uncertainty}} \right), 
\end{equation}
which, compared to Eq.~\eqref{eq:high_lel_trade-off}, is significantly worse than the additive delay overhead under an optimal policy when the noise level is high. 

\emph{Harnessing intrinsic uncertainty, not fighting it}. The above example illustrates that naively enforcing a routing policy designed for a deterministic system can lead to orders of magnitude worse performance in the presence of high levels of intrinsic uncertainty. A closer examination of the example reveals a deeper root-cause of the inefficiencies: the agent can spend a substantial fraction of the route in long excursions between reaching consecutive vertices on the pre-planned path $S$, and since the adversary is aware of this fact, these excursions do not meaningfully contribute to lowering the agent's prediction risk. As a result, intrinsic uncertainty only leads to an increase in delay, but not to better obfuscation. 

A central insight of our work addresses this conundrum. Instead of pre-committing to a path $S$ and subsequently fighting against intrinsic uncertainty to implement it, we argue that an efficient policy should leverage the randomness already present in the environment and make it an integral part of the overall randomization. More concretely, this means that the steps that the agent intentionally chooses and those that are a result of the intrinsic uncertainty should be provably indistinguishable to the adversary; otherwise, the adversary would be able to ignore the steps due to intrinsic uncertainty in his analysis and effectively increase the prediction risk. Furthermore, because the realizations of the random shocks are experienced by the agent in a sequential manner, it also suggests that the agent should employ any artificial randomization in a sequential and adaptive manner that reacts to the realizations of the random shocks, rather than deciding from the get-go when and how to reach the target. The family of Water-Filling strategies that we propose in this paper (Section \ref{sec:WaterFill}) exploit the ideas of noise harnessing and sequential randomization. All of our achievability results (i.e., upper bounds) are derived from variants of the Water-Filling family. 

\subsection{Summary of Main Results} 

We now give an informal preview of our main results. The formal statements will be given in Section \ref{sec:main_res}. We will use $n$ to denote the number of vertices in the network, and we fix the noise level of the environment, $\varepsilon$, to be a value in $(0,1)$. We are mostly interested in the regime where the network is large ($n\to \infty$) and the noise level is high ($\varepsilon \approx 1$).%We will use $n$ to denote the number of vertices in the network, and are mostly interested in the regime where the network is large ($n\to \infty$). Throughout, we fix the noise level of the environment, $\varepsilon$, to be a value in $(0,1)$. 

\noindent {\bf 1. Optimal trade-off in complete graphs} (Theorem \ref{thm:wf_thm_infinite}) We first consider the case where the graph $G$ is a complete graph. 
%While the complete graph is a seemingly simple topology, it contains a surprising amount of strategic richness, and illustrates well the core difficulty in designing anonymous routing policies under intrinsic uncertainty. \
We provide a characterization of the trade-off between delay and prediction risk, which is shown to be {asymptotically tight} as the graph size tends to infinity. Recall that delay is the average number of steps it takes for the agent to reach her goal. Denote by $\mathcal{Q}(w)$ the minimal prediction risk that can be achieved across all agent strategies with a delay of at most $w$.  Fix a delay target $w > \frac{\varepsilon}{2(1-\varepsilon)^2}+1$. We show that, as the graph size $n\to \infty$, 
\begin{align}
\mathcal{Q}(w) = \frac{1}{2w-1 - \frac{1}{2w-1}\cdot \frac{\varepsilon}{(1-\varepsilon)^2} - \beta^\varepsilon(w)} + o(1), 
\label{compresult}
\end{align}	 
where $\beta^\varepsilon(w) $ is a discrepancy term with $|\beta^\varepsilon(w)| = O((\varepsilon/w)^2)$, and the term $o(1)$ tends to $0$ as $n\to \infty$. 

Some important observations can be made from Eq.~\eqref{compresult}.  On the positive side,  with a carefully designed agent strategy, the intrinsic uncertainty leads only to an additive, as opposed to multiplicative, delay overhead,  whose magnitude is given by 
\begin{equation}
\mbox{cost of uncertainty} = \frac{1}{2w -1}\cdot \frac{\varepsilon}{1-\varepsilon^2} + O((\varepsilon/w)^2). 
\end{equation}
As a result, the overhead  becomes increasingly negligible as $w$ grows. On the negative side however, the lower bound portion of Eq.~\eqref{compresult} shows that this cost of uncertainty can never be fully eliminated or ``absorbed'', even if the delay budget {significantly surpasses the (stochastic) diameter of the graph}. The presence of uncertainty therefore always adversely affects the agent compared to the noiseless setting. 

\iffalse
In summary, Eq.~\eqref{compresult} establishes that, at a qualitative level, 
\begin{align*}
\mbox{prediction risk} = \Theta \left(\frac{1}{\mbox{delay} - \mbox{cost of uncertainty}} \right).
\end{align*}
\fi

\noindent {\bf 2. Non-Complete graphs} (Theorem \ref{thm:reg}) In many practical settings, the underlying network topology may not be a complete graph. Fixing an average degree parameter $p \in (0,1)$, our next result establishes that there exists a family of non-complete graphs, $\calG_n$, with average degree $pn$, over which the intrinsic uncertainty still leads to an additive delay overhead. In particular, we focus on graphs for which (1) pairs of connected nodes share a large fraction of overlapping neighbors, and (2) the degree distribution exhibits concentration as the graph grows. These two properties will enable us to establish a minimal prediction risk of the form: 
\begin{equation}\label{eq:reg_upper}
\frac{1}{2w+1} \leq \calQ(w) \leq \frac{1}{2w - \lt(\frac{1}{p} +\frac{\varepsilon}{(1-\varepsilon)^2p} \rt) + o(1) } + o(1),
%\leq \frac{1}{2w - \lt(2 +\frac{1}{p} +\frac{\varepsilon}{(1-\varepsilon)^2p} \rt) } + o(1), 
\end{equation}
for any graph in $\calG_n$, as $n \rightarrow \infty$, whenever $w> \frac{\varepsilon}{2p(1-\varepsilon)^2} + \frac{1}{p} $.  Furthermore, we show that the family $\calG_n$ is ``large,'' in the sense that a graph generated according to the Erd\H{o}s-R\'{e}nyi model, $G(n,p)$, belongs to $\calG_n$ with high probability as $n\to \infty$. 

%This result can be extended to sparse graphs by letting $p$ approach zero as the graph size grows; the delay overhead due to uncertainty will remain additive (as a function of $w$) but unfortunately the magnitude of the overhead will grow as $p$ decreases. 

\noindent {\bf 3. Designing sparse networks for anonymity} (Theorem \ref{thm:networkdesign}) The results so far focus on instances where the network topology is treated as given. Unfortunately, the prediction risk upper bound in Eq.~\eqref{eq:reg_upper} deteriorates as the graph becomes more sparse ($p\to 0$). To understand what one can do in a sparse network topology, our next result considers the scenario where the network structure can be designed, subject to an average degree constraint. Here, we show that there exists a  sparse network structure that will offer a similar delay vs.~prediction risk trade-off as that in a complete graph. For any degree sequence $\{\bar{p}_n\}_{n\in \N}$, where $\sqrt{n} \ll \bar{p}_n \leq n$,\footnote{We write $f(n) \ll g(n)$ if ${\lim\limits_{n \rightarrow \infty}}{f(n)/g(n) =0}$.} we show that there exists a family of graphs, $\mathcal{\bar{G}}(n,\bar{p}_n)$, where each $G \in \mathcal{\bar{G}}(n,\bar{p}_n)$ has $n$ vertices and average degree at most $\bar{p}_n$, over which the minimal prediction risk satisfies 
\begin{equation} \label{eq:network_des_upper}
\frac{1}{2w+1} \leq \calQ(w) \leq \frac{1}{2w - 1 -\frac{\varepsilon}{(1-\varepsilon)^2}} + o(1), 
\end{equation}
as $n\to \infty$, whenever $w > \frac{\varepsilon}{2(1-\varepsilon)^2} + 1$. In the absence of the global neighborhood overlap property of complete graphs and graphs in the family $\calG_n$, the proposed network topology here will achieve the desired trade-off in Eq.~\eqref{eq:network_des_upper} by exploiting certain \emph{local connectivity} properties. Specifically, we describe a family of graphs that consist of several highly connected components that are linked to each other via a small number of edges. With this design, each highly connected component essentially functions as a complete graph and enables a low prediction risk, while the edges across different components together with an appropriately modified greedy routing strategy allow the delay to remain small. Thanks to the ability in choosing the network topology, the prediction risk guarantee in Eq.~\eqref{eq:network_des_upper} is stronger than that in Eq.~\eqref{eq:reg_upper}, for networks with the same level of sparsity, where asymptotically both the requirement on $w$ and the additive delay overhead are now independent of the average degree parameter $\bar{p}_n$. 

\subsection{Motivating Examples}\label{sec:motExamp}

We examine two application areas that motivate the formulation of the Anonymous Stochastic Routing model proposed in this paper. While our model is clearly stylized and simplified for direct application, we believe it captures key features of these applications and the theoretical findings reported in this work may provide valuable insights and design guidelines. 

%We further discuss several gaps between our model and these applications in Conclusions. 

\subsubsection{Anonymous Messaging} 

The anonymous routing model we propose can serve as a topology-aware framework for studying anonymous  messaging protocols, which enable Internet users to send messages without revealing the identity of the receiver. A number of existing protocols, such as Bitmessage (\cite{warren2012bitmessage}), Riposte (\cite{corrigan2015riposte}), and Herbivore (\cite{goel2003herbivore}),  ensure anonymity by flooding encrypted messages to a large number of nodes in the network, while only the recipient node has the private key to decode the message. The intention of the flooding is to ensure that any adversary who observes the network traffic will have no way of knowing the intended recipient. However, uniformly flooding the network with each message is not ideal: it can result in both excessive traffic congestion when the network is large or sparse, and can create additional privacy risks by making it easy for any malicious node to obtain a copy of all communications, even if encrypted.

To apply our model in this context, the agent would correspond to an encrypted data packet that is being routed over a communication network. The vertices would represent participating users and the edges the communication links between them. Our routing policies can be adapted into messaging protocols to achieve near-optimal congestion-anonymity trade-off, in a manner that takes advantage of the underlying network structure and avoids unnecessary flooding. Furthermore, our theoretical results would provide rigorous guarantees for the performance of these protocols, which to the best of our knowledge, has not been established for anonymous messaging protocols except for the case of uniform flooding.  Intrinsic uncertainty can manifest in anonymous messaging systems in two ways. First, due to the inherent peer-to-peer nature of these systems, the communication links are subject to random failures (cf.~\cite{perkins2001ad, kurose2013computer}), and participating users may randomly become offline \citep{warren2012bitmessage}. Second, a random fraction of the participating users may be subject to malicious attacks \citep{vu2009peer} in which case they may not follow the routing instruction prescribed by the protocol.  The uncertainty model we propose here can thus serve as a first-order approximation to the type of disruptions the data package may encounter. 

%In Appendix \ref{ap:implementation}, we will use the Bitmessage protocol as an example and explore more in detail how our algorithms might potentially be implemented there. 

\subsubsection{Obfuscation in Anonymity-Aware Logistics}
Concerns over anonymity in routing can also arise in the physical world. With the advent of machine learning, it has become increasingly easier for firms to collect and process data on their competitors' operations. For instance, investment firms have been able to use satellite images of parking lots to estimate a retailer's sales performance before such information is made public \citep{economist2019}.  Such technologies are indeed not unique to the financial industry; other AI-powered systems have also shown remarkable  tracking abilities for cars and ships \citep{rodriguez2018, allioux2018}.

Against this technological backdrop, we speculate that our model can potentially be used to create interesting anonymity-aware logistic services. Imagine a firm, A, that regularly receives shipment from a supplier for a certain vital material or component for manufacturing its products. By tracking the timing, frequency and origination of the delivery trucks to the firm's plants, a competitor firm, B, may be able to infer the nature of the product or its demand, both of which may be sensitive strategic information that A would want to protect. To address this concern, a logistics company could conceivably offer an anonymity-aware service to its clients: instead of sending a truck specifically designated for A each time a shipment is ordered, the truck could potentially visit multiple recipients with similar corporate profiles on a single delivery trip. From the perspective of firm A, over time, some of the visits of the delivery truck indeed carry vital materials, while on other visits the truck may simply be used for non-urgent supplies or pickups. When deployed strategically, such obfuscation can make it difficult or impossible for a competitor to extract valuable information from analyzing the truck movement patterns. 

In this context, the agent in our routing model would correspond to the delivery truck,  the network the physical road network, and  the goal vertex the firm who is the true recipient of the vital material for a particular delivery trip. Intrinsic uncertainty in this application can be a result of unexpected traffic congestion or road disruptions due to weather conditions, political demonstrations, or accidents. Admittedly, there are still large gaps to be bridged in order for our model to be useful for logistical services in the real world. Nevertheless, it could serve as a first step towards a conceptual framework for addressing anonymity issues in logistics and transportation.

%\subsection{Organization} \tr{TBD} The remainder of the paper is organized as follows. We formally describe our model in Section \ref{sec:model} and present our main results in \ref{sec:main_res}. Section \ref{sec:lit} surveys the related literature. The main family of strategies we use, the Water-Filling Strategies, will be introduced in Section \ref{sec:WaterFill}. Sections \ref{sec:XXX} through \ref{sec:YYY} are devoted to the proofs of our main results. We conclude the paper with some discussion in Section \ref{sec:conc}.
%Section \ref{sec:num} presents numerical studies and simulations on finite, real-world networks.

\section{Model: Anonymous Stochastic Routing}\label{sec:model}

In this section, we formally define our Anonymous Stochastic Routing model. 

%The model consists of an agent, who traverses the edges of a graph $G$ in order to reach a goal vertex, and an adversary who observes the trajectory of the agent and tries to predict the goal vertex $D$.

{\bf Model primitives.} Let $G = (\calV, \calE)$ be an undirected graph with $n$ vertices. We consider an agent who operates on the graph $G$ in discrete time $t \in \{1, \ldots, K \}$, where $K \in \N$ is the time horizon. We denote the goal vertex by $D$, which is sampled uniformly at random from $\calV$. In particular, the set of vertices $\calV$ represents the states of the agent, and the set of edges $\calE$ the allowed state transitions in each period. We will denote by $\neig(v)$ the \emph{neighborhood} of a vertex $v$: $\neig(v) = \{ v' \in \calV: (v,v') \in \calE \}$. The initial state of the agent is given by $x_0 \in \calV$. Our results will not depend on the distribution of $x_0$ and without loss of generality, we may assume that $x_0$ is fixed. 

We denote by $\varepsilon$ the noise level in the system. Then, define a collection of i.i.d.~Bernoulli random variables $\{B_t\}_{t=1}^K$ with success probability $1-\varepsilon$. The random variables $\{B_t\}_{t =1}^K$ capture the \emph{intrinsic uncertainty} in the system, where $B_t$ indicates whether the agent's chosen action at time $t$ will be fulfilled.

Finally, we define two mutually independent sequences of independent random variables $\mathcal{R}_A$ and $\mathcal{R}_D$. Specifically, $\mathcal{R}_A$ and $\mathcal{R}_D$ are independent random variables that can be used for the purpose of randomization, by the agent or the adversary, respectively.

{\bf Agent trajectory and strategy.} An agent \emph{trajectory} is a sequence of random variables $\{X_t\}_{t =1}^K$ where $X_t \in \calV$ denotes the {state} of the agent at time $t$. 
%We require that $(X_t, X_{t+1}) \in \calE$ with probability $1$ for all $t \geq 1$. That is, the agent can only travel to a neighboring vertex in each time step. 
We will denote by $a_t$ the \emph{action} undertaken by the agent at time $t$, where $a_t$ can be a neighboring vertex of the agent's current state $X_t$. Given $a_t$ and $X_t$, the agent's next state $X_{t+1}$ is generated as follows. If $B_t=1$, then $X_{t+1}$ is equal to $a_t$, and if $B_t=0$, then $X_{t-1}$ is sampled uniformly at random from the neighboring vertices. This dynamics is equivalent to the following transition probabilities: 
\begin{align*}
\bP(X_{t+1} = v \abs X_t=x_t, a_t) = \begin{cases}
1- \varepsilon + \frac{\varepsilon}{|\neig(x_t)|}, & \mbox{if } v = a_t,\\
\frac{\varepsilon}{|\neig(x_t)|}, & \mbox{if } v \in \neig(x_t) \setminus \{a_t\}, \\
0, & \mbox{otherwise.}
\end{cases}
\end{align*} 
More formally,  the actions will be generated by an {\bf agent strategy}, $\psi$, which is a sequence of mappings that  sequentially generates the agent's actions,  $a_t = \psi_t (H_t, \mathcal{R}_A, D)$, where $H_t = \{ (X_i, a_i, B_i) \}_{i=1}^{t-1} \cup \{X_t\} $ is the agent's history up to time $t$.

Under any agent strategy $\psi$, the time at which the agent reaches the goal is referred to as the \emph{goal-hitting time}, and is given by
\begin{align}
T^{\psi} = \inf \{ t \geq 1: X_t =D \}.
\label{eq:goal_hitting}
\end{align}
Then, the {\bf delay} of agent strategy $\psi$ is defined as $\bE(T^{\psi})$, where the expectation is taken with respect to the randomness in $D$, $\mathcal{R}_A$ and $\{B_t\}_{t =1}^K$.

If by the end of the {time horizon} $K$ the agent has not yet reached her goal, then at time $K+1$, she is automatically sent to $D$ and we say that the adversary succeeds. Specifically, we set $X_{K+1} = D$ and assume $B_{K}=1$. The use of a horizon is a primarily a technical construction of the model to ensure the integrability of $T^{\psi}$. For the most part, our analysis will focus on cases where $K$ is sufficiently large and has a negligible effect on the overall system dynamics.

{\bf Adversary strategies and prediction risk.} The adversary has access to $G$, $x_0$, $\mathcal{R}_D$, $\psi$ and the agent's entire trajectory $\{X_t \}_{t =1}^K$. He does not observe the agent's actual actions $\{a_t\}_{t =1}^K$ and does not know whether the agent has reached her goal. Based on this information, the adversary's strategy, $\chi$, produces a prediction $\hat{D}_\chi$ in period $K$. Appendix \ref{ap:example} provides an illustrative instance of the interaction between the agent and the adversary. 

We say that the adversary succeeds if his prediction correctly matches the goal or if the agent fails to reach her goal vertex by the end of the horizon. We define \emph{prediction risk} of a given pair of agent and adversary strategies $(\psi, \chi)$ as the probability 
\begin{align}
q(\psi, \chi) = \bP(\hat{D}_\chi= D  ) + \bP(T^{\psi} >K),
\end{align}
where the probability is measured with respect to the randomness in $D, \mathcal{R}_A, \mathcal{R}_D$, and $\{B_t\}_{t =1}^K$. Note that, for any fixed agent strategy $\psi$, the probability that the adversary succeeds due to the agent's failing to reach the goal by the end of the horizon diminishes as the length of the horizon increases, i.e., $\lim\limits_{K \rightarrow \infty} \bP(T^{\psi} > K) = 0$. Hence, the main reason the adversary succeeds will be due to his correct prediction of the goal vertex. 

{\bf Minimax prediction risk.} Given an agent strategy $\psi$, we define the \emph{maximal prediction risk}:
\begin{align}
q^*(\psi) = \sup_{\chi} q(\psi, \chi).
\end{align}
For any given time budget $w \in \mathbb{R}_{+}$, $\Psi_w$ is defined to be the set of all agent strategies for which the delay is at most $w$: $\Psi_w = \{ \psi: \bE(T^{\psi}) \leq w \}.$ 

Finally, we define our main metric of interest, the {\bf minimax prediction risk}: 
\begin{align}
\calQ(w) = \inf_{\psi \in \Psi_w} q^*(\psi) = \inf_{\psi \in \Psi_w} \sup_{\chi} q(\psi, \chi).
\end{align}
In words, $\calQ(w)$ measures the least amount of prediction risk that any agent strategy will be able to guarantee subject to a delay of at most $w$. Characterizing $\calQ(w)$ will allow us to design efficient agent policies with limited resources, under uncertainty.

\subsection{Notation}
Let $f, g: \bN \rightarrow \mathbb{R}$ be two functions. We use the following asymptotic notation: $f(n) \ll g(n)$ if ${\lim\limits_{n \rightarrow \infty}}{f(n)/g(n) =0}$; $f(n) \sim g(n)$ if ${\lim\limits_{n \rightarrow \infty}}{f(n)/g(n) =1}$; $f(n) \preceq g(n)$ if ${\limsup\limits_{n \rightarrow \infty}}{f(n)/g(n) < \infty}$; and $f(n) \lleq1 g(n)$ if ${\limsup\limits_{n \rightarrow \infty}}{f(n)/g(n) \leq 1}$. 

\section{Main Results}\label{sec:main_res}

We state the main results in this section. The proofs will be given in Sections \ref{sec:pf_upper} and \ref{sec:pf_lower}, with an overview in Section \ref{sec:pf_over}. In the remainder, we fix a noise level $\varepsilon \in \lt[0,1\rt)$ and a sequence of horizons $\{K_n\}_{n \in \N}$ such that $1\ll K_n \ll {n}$. 

The first theorem gives an asymptotically tight characterization of the prediction risk in complete graphs. We focus on the large-graph asymptotic regime in which the number of vertices, $n$, is let to approach infinity.

\begin{theorem}\label{thm:wf_thm_infinite}
Fix delay budget $w$ such that $w > \frac{\varepsilon}{2(1-\varepsilon)^2} + 1$. Fix $n \in \N$, and let $G = (\calV,\calE)$ be a complete graph with $n$ vertices. Then, the minimax prediction risk satisfies
\begin{align*}
\frac{1}{2w-1-\alpha^{\varepsilon}(w)} -\delta_n \leq \calQ(w) \leq & \frac{1}{2w-1 - \alpha^{\varepsilon}(w) - \beta^{\varepsilon}(w)} + \delta_n,
\end{align*}
where $\lim\limits_{n \rightarrow \infty} \delta_n = 0$, and $\alpha^{\varepsilon}(w)$ and $\beta^{\varepsilon}(w)$ are defined by
\begin{align*}
\alpha^{\varepsilon}(w)=  \frac{\varepsilon}{(2w-1)(1-\varepsilon)^2} >0, \quad \mbox{ and } \quad \beta^{\varepsilon}(w)=  \frac{\varepsilon^2}{2 \left(w-\frac{1}{2} \right)^3 (1-\varepsilon)^4}>0.
 \end{align*}

\end{theorem}

Notably, for large complete graphs, Theorem \ref{thm:wf_thm_infinite} establishes that the delay overhead due to intrinsic uncertainty is only additive as a function of the noise level. Further, the overhead is always strictly positive regardless the size of the delay budget.

%Theorems \ref{thm:reg} and \ref{thm:networkdesign} examine non-complete graphs when the network topology is given exogenously and when it can be designed by the agent, respectively. 

Theorem \ref{thm:reg} shows that there exists a family of non-complete graphs, $\mcal{G}_n$, with average degree $pn$, on which an adaptive agent strategy guarantees an additive delay overhead.
% Our analysis relies on the design of an agent strategy that can be implemented on any undirected graph. 
We further prove that the family $\mcal{G}_n$ is ``large,'' in the sense that an Erd\H{o}s-R\'{e}nyi random graph with $n$ vertices and edge probability $p$ belongs to $\mcal{G}_n$ with high probability, as $n \to \infty$.

\begin{theorem}\label{thm:reg}
Fix edge density $p \in [0,1]$ and delay budget $w > \frac{1}{p}  \lt( \frac{\varepsilon}{2(1-\varepsilon)^2} + 1 \rt)$. Then, there exists a sequence of  families of graphs, $\{\calG_n \}_{n\in \N}$, such that the following are true.

\noindent (a)  Suppose that $\tilde{G}$ is a graph drawn from the Erd\H{o}s-R\'{e}nyi random graph model with $n$ vertices and edge probability $p$. Then, 
\begin{align*}
\bP(\tilde{G} \in \calG_n ) \geq 1-\theta_n, 
 \end{align*} 
 where $\lim\limits_{n \rightarrow \infty} \theta_n = 0$.
 
\noindent (b)  Fix $n\in \N$, and suppose that $G\in \calG_n$. Then, the minimax prediction risk satisfies
		\begin{align*}
		\frac{1}{2w+1} \leq \calQ(w) \leq \frac{1}{ 2 w - p^{-1} (1 + c^\varepsilon) - \lambda_n  } + \delta_n,
		\end{align*}		
where $\delta_n$ and $\lambda_n$ are constants that do not depend on $G$ such that $\lim\limits_{n \rightarrow \infty} \delta_n = 0$, and $\lim\limits_{n \rightarrow \infty} \lambda_n = 0$ hold, and $c^{\varepsilon} = \frac{\varepsilon}{(1-\varepsilon)^2} $.

\end{theorem}

Theorem \ref{thm:reg} demonstrates that additive overhead on delay can still be achieved on a ``typical" non-complete graph,  i.e., one that is generated by a random graph model with linear edge density. Essentially, the low minimax prediction risk in Theorem \ref{thm:reg} for the family $\mcal{G}_n$ is enabled by two topological properties: (1) concentrated degree distribution, and (2) neighborhood overlap property. In Section \ref{sec:pf_upper}, we exploit these properties to characterize the prediction risk and the delay respectively.

It is not difficult to formally extend Theorem \ref{thm:reg} to a sparse regime and allow the density of the graph, $p$, to diminish as $n$ grows. Unfortunately, in this limit, the additive overhead in the upper bound will tend to infinity at rate $\Omega(1/p)$, rendering the characterization rather weak. This limitation raises the question of whether we can still obtain desirable results in sparse graphs, albeit with more structure. For instance, one could be in a setting where the agent can design the network topology, subject to a constraint on the average degree. The next theorem accomplishes this, showing that we can design networks with average degree as low as $\sqrt{n}$, while maintaining a desirable prediction risk-delay trade-off that is competitive even when compared to a complete graph; in Section \ref{sec:proof-net}, we explain how to design these graphs explicitly. At a high level, the proposed topology aims to balance the prediction risk and delay by building a graph with several highly connected components that are also linked to each other by a small number of edges. We illustrate that the connections across different components, together with a variant of the Water-Filling Strategy that uses greedy routing subroutines, drive the delay to be low; whereas the prediction risk analysis follows by realizing that each individual component behaves as a complete graph.
\begin{theorem}
\label{thm:networkdesign}
%Fix noise level $\varepsilon \in \lt[0,1\rt)$. Fix a sequence of horizons $\{K_n\}_{n \in \N}$ such that $1\ll K_n \ll {n}$. 
Fix a degree sequence $\{\bar{p}_n\}_{n \in \bN}$ such that $\bar{p}_n \leq n$ and $\bar{p}_n \gg \sqrt{n}$. Fix $n \in \bN$. Then, there exists a family of graphs, $\mathcal{\bar{G}}(n, \bar{p}_n)$, consisting of graphs with $n$ vertices and average degree at most $\bar{p}_n$. Fix $w$ such that $w > \frac{1}{\rho_n^\varepsilon}  \lt( \frac{\varepsilon}{2(1-\varepsilon)^2} + 1 \rt)$. Then, the minimax prediction risk for this family satisfies
	\begin{align*}
	\frac{1}{2w+1} \leq \calQ(w) \leq \frac{1}{2w \rho_n^\varepsilon -1- c^{\varepsilon} } + \delta_n,
	\end{align*}
where $\lim\limits_{n \rightarrow \infty} \rho_n^\varepsilon = 1 $ and $\lim\limits_{n \rightarrow \infty} \delta_n = 0$, and $c^{\varepsilon} = \frac{\varepsilon}{(1-\varepsilon)^2} $.
\end{theorem}

Contrasting Theorems \ref{thm:reg} and \ref{thm:networkdesign}, we see that the ability to design the topology allows the agent to achieve superior anonymity guarantees on graphs of similar edge density. First, the leading constant in front of $w$ in the upper bound of Theorem \ref{thm:networkdesign} matches that of the lower bound, whereas that of Theorem \ref{thm:reg} does not. More interestingly, while both results demonstrate an additive delay overhead due to uncertainty, the overhead in Theorems \ref{thm:networkdesign} is independent of the graph's average degree $\bar{p}_n$ (provided that $\bar{p}_n\gg \sqrt{n}$), whereas the overhead in Theorem \ref{thm:reg} deteriorates as the average degree decreases.

\section{Related Work}\label{sec:lit}

The Anonymous Stochastic Routing model is inspired by, and generalizes, the Goal Prediction Game proposed by \cite{tsitsiklis2018delay}, who showed that, in a deterministic anonymous  path-planning model, the agent's prediction risk is inversely proportional to her delay in undirected graphs with a small diameter. Our model diverges from \cite{tsitsiklis2018delay} in two primary ways. First and foremost,  our formulation overcomes a crucial limitation of the model in \cite{tsitsiklis2018delay} by allowing the agent's actual trajectory to be perturbed by noise and uncertainty. This distinction is significant, since as explained in Section \ref{sec:why_uncertainty}, the incorporation of intrinsic uncertainty requires a fundamentally different approach to policy design than in the deterministic setting. Secondly, the model in \cite{tsitsiklis2018delay} focuses more on applications in sequential planning and thus considers an online adversary who makes prediction in real-time and aims to predict the goal before the agent reaches it, whereas we are mostly interested in modeling a passive data collector and allow the adversary to make offline predictions, after the agent's routing actions have already been completed. In this way, the adversary model in our formulation is more powerful. However, we do not expect this second difference to be significant, and most of our results and those of \cite{tsitsiklis2018delay} can be derived in both settings, with possible changes in the leading constant of the prediction risk bounds. 

Anonymity in routing or messaging has been extensively studied in the computer science literature \citep{reiter1998crowds, corrigan2010dissent, fanti2015spy, luo2016infection, kwon2017atom}. Models in this literature vary depending on whether the algorithm protects the anonymity of the sender or receiver, and on the strength of adversarial assumption (e.g., local versus global adversary). In the context of this literature, we focus on receiver anonymity, under a global adversary who can observe all routing traffic in the system, and provide information theoretic anonymity guarantees that do not depend on cryptographic or computational hardness assumptions. In this way, our work is distinct from those that study sender anonymity (e.g., \cite{fanti2015spy, luo2016infection}), assume local adversaries who only observe the traffic over a subset of the network (e.g., \cite{reiter1998crowds,tang2020privacy}) or require encryption to prevent the adversary from eavesdropping on the message's destination (e.g., \cite{dingledine2004tor}). There have been systems proposed that, like ours, also provide receiver anonymity against a global adversary, but they require the presence of a central server to collect and shuffle messages (e.g., \cite{corrigan2015riposte}). As mentioned in Section \ref{sec:motExamp}, existing peer-to-peer protocols that do not require a central server mostly rely on broadcasting each (encrypted) message to all members of a group \citep{goel2003herbivore, corrigan2010dissent, warren2012bitmessage}, leading to sub-optimal network congestion, especially in sparse networks. Finally, we are not aware of any existing model or protocol that provides formal anonymity guarantees in the presence of intrinsic uncertainty as in the present paper. 

At a higher level, our work relates to a growing literature on anonymity-preserving mechanisms in sequential (rather than static) decision-making, including those in  operations research (cf.~\cite{cummings2016empirical, tsitsiklis2018delay}), game theory (cf.~\cite{blum2015privacy, gradwohl2017perception, augenblick2018reveal}), computer science (cf.~\cite{fanti2015spy, lindell2009secure}) and learning theory (cf.~\cite{calmon2015fundamental, shokri2015privacy, tossou2016algorithms, xu2018query, tsitsiklis2018private}). Similar to our model, most of the existing algorithms rely on injecting artificial randomness into the agent's decision-making policy to achieve obfuscation. However, unlike in our model, to the best of our knowledge, almost all existing models consider the noiseless setting where the agent has full control of the effects of her actions. In contrast, the agent's actions in our model are subject to random shocks and perturbations, which makes efficient obfuscation substantially more difficult. 

More broadly, our model is related, in spirit, to the literature on differential privacy \citep{dwork2008differential, dwork2014algorithmic}, in that both are interested in providing information-theoretic guarantees on protecting sensitive information against a sophisticated adversary. However, our model is not comparable to the differential privacy paradigm: they are mostly concerned with privatizing datasets that consist of \emph{multiple} individuals, and the main goal there is to make sure that an algorithm running on two datasets that differ by one individual entry does not lead to obviously diverging outputs (i.e., a ``differential'' perturbation). This being said, there could be interesting variants of the anonymity metric in our model that are inspired by differential privacy: for instance, the routing strategy might want to ensure that neighboring goal vertices are associated with similar distributions of routing trajectory. We will leave these avenues of investigation for future work.

\section{Water-Filling Strategies}
\label{sec:WaterFill}

We describe in this section the family of Water-Filling Strategies which will form the foundation of our analysis. All upper bounds in our theorems are proved by analyzing an agent that employs a variant of the Water-Filling Strategy, and the lower bound in Theorem \ref{thm:wf_thm_infinite} also relies on insights derived from this family.  
To simplify terminology, we define the following two meta actions to serve as a short-hand encoding for the agent's actual actions: 
\begin{enumerate}
	\item[(1)] We say that the agent chooses the \textbf{Random-Step} meta action ($\bar a^{R}$), if she samples a vertex uniformly at random from the set of neighboring vertices of her current state, i.e. $a_t \sim \mbox{Unif}(\neig(X_t)) $.
	\item[(2)] We say that the agent chooses the \textbf{Goal-Attempt} meta action ($\bar a^{G}$), if the agent chooses the goal vertex $D$ as her action, i.e., $a_t = D$. The agent can choose this action only if the goal vertex is in her immediate neighborhood. 
\end{enumerate}
 We will denote by $\bar{\mathcal{A}} = \{\bar a^{R}, \bar a^{G}\}$ the set of meta actions, and by $\bar{a}_t \in \bar{\mathcal{A}}$ the meta action taken at time $t$. 

We will make repeated use of the concept of \emph{intentional goal-hitting time}, defined as the first time the goal is reached by the agent intentionally via  a successful Goal-Attempt meta action, as opposed to reaching the goal coincidentally via a random shock: \footnote{Note that $T_{\mbox{\tiny IH}}\leq K+1$: if the agent has not yet reached her goal by the end of the horizon, then we have $T_{\mbox{\tiny IH}}=K+1$ since we set $X_{K+1} = D$ and $B_{K}=1$. }
	\begin{align}
	\tih = \inf \{ t \geq 1: \mbox{agent chooses Goal-Attempt at $t-1$ and } B_{t-1} =1 \}. 
	\end{align}
	The intentional goal-hitting time serves as a natural upper bound on the actual goal-hitting time, but is more amenable to analysis.
 
We are now ready to define the Water-Filling Strategies. Define the counter $L(t)$ to be the number of times the state $X$ has visited the neighborhood of the goal vertex, $\neig(D)$, by time $t$:

\begin{equation}
L(t) = \sum_{s=1}^t \mathbb{I}(X_s \in \neig(D)).
\label{eq:counter}
\end{equation}
We will denote by $\tih^{wf}$ the intentional goal-hitting time under this family of strategies.

\begin{definition}[Water-Filling Strategy] \label{defn:wf_strategy}
	Fix a target risk level $\bar{q} \in [0,1]$. Let $G=(\calV,\calE)$ be a connected undirected graph with $n$ vertices. Define a threshold\footnote{To simplify notation and to avoid floor and ceiling functions, we will henceforth assume the parameters are such that $\frac{1}{\bar q} - \frac{\varepsilon}{1-\varepsilon}$ is an integer.} 
		\begin{equation*}
	t^* = {\bigg \lceil}{ \frac{1}{\bar q} - \frac{\varepsilon}{1-\varepsilon} } {\bigg \rceil}. 
	\end{equation*}
	and a sequence of probabilities indexed by $t$: 
	\begin{align}
	{p}_{t} =
	\begin{cases}
	\frac{\bar{q}}{1-\varepsilon} \left(1- t \bar{q}\right)^{-1} , & \mbox{ if } 0 \leq t < t^*-1, \\
	1 , & \mbox{ otherwise. }
	\end{cases}
	\label{eq:ptdef}
	\end{align}
	The Water-Filling Strategy, $\psi\wf_{\bar{q}}$, is defined as follows: for $t\in \N$
	\begin{enumerate}
	\item If $X_t \in \neig(D)$ (agent in the neighborhood of goal vertex) and $\tih^{wf}>t$ (no intentional goal-hitting so far):
		\begin{enumerate}
			\item[(1)] with probability $p_{L(t)}$, the agent chooses the Goal-Attempt meta action, where $L(t)$ and $\{p_t\}_{t\in \N}$ were defined in Eqs.~\eqref{eq:counter} and\eqref{eq:ptdef}, respectively; 
		\item[(2)] otherwise, the agent chooses the Random-Step  meta action. 
		\end{enumerate}
	\item  If $X_t \notin \neig(D)$ or $\tih^{wf} \leq t$, the agent always chooses the Random-Step meta action.
	
\end{enumerate}
\end{definition}

In words, under the Water-Filling Strategy, the agent always performs a Random-Step meta action when she is not in the immediate neighborhood of the goal vertex. Otherwise, she will attempt to reach the goal with a probability that depends on the level of intrinsic uncertainty, $\varepsilon$, as well as how many times she has already visited the goal vertex's neighborhood, $L(t)$. 

To deploy the Water-Filling Strategy for different delay budgets, it suffices to vary the risk level $\bar{q}$ until a desirable delay is achieved: as we will formally show in the subsequent sections, larger $\bar{q}$ leads to shorter delays, and \emph{vice versa}. 

\subsection{Interpretation of Water-Filling Strategy}
\label{sec:intuition}

We now explain the intuition behind the Water-Filling Strategy, and how it addresses  the design principles outlined in Section \ref{sec:why_uncertainty}. At a high level, the Water-Filling Strategy aims to achieve the following objectives: 
\begin{enumerate}
\item \emph{Independence between trajectory and goal-hitting time}. Since the adversary observes the entire trajectory, predicting the identity of the goal vertex is equivalent to correctly predicting the goal-hitting time (Eq.~\eqref{eq:goal_hitting}). Therefore, to be able to accurately control the adversary's ability to predict the goal-hitting time, the Water-Filling Strategy ensures that the goal-hitting time is (approximately) independent from the agent's trajectory. This independence, if true,  reduces the analysis of the prediction risk to that of the probability mass function (PMF) of the goal-hitting time, thus simplifying the analysis. 

The Water-Filling Strategy achieves this independence via carefully chosen meta actions. It is essential that the strategy only chooses to approach the goal vertex when $X_t$ is in the goal's immediate neighborhood. This ensures the needed symmetry in the trajectory, since the adversary has no way of knowing whether a single move is reaching the goal vertex or just a random step. Moreover, besides this Goal-Attempt meta action, the agent's action in all other times, Random-Step, is clearly independent from the goal vertex. 

\item \emph{Flattening the PMF of goal-hitting time, while minimizing delay}: Once the above-mentioned independence is established, the next task is to ensure that the maximum value of the PMF of the goal-hitting time is small. Moreover, we also need to ensure that while doing so, delay is minimized subject to a level of prediction risk. 

This is the part where the Water-Filling Strategy employs {sequential randomization}, as eluded to in Section \ref{sec:why_uncertainty}. To see how this works, it is easiest to consider the special case where the network $G$ is a complete graph. In this case, $X_t$ is essentially always in $\neig(D)$ prior to the intentional goal-hitting time, and the Water-Filling Strategy will direct the agent to choose Goal-Attempt with progressively smaller probability, $p_t$, until the goal vertex is reached, in the form of Eq.~\eqref{eq:ptdef}. The particular values of $p_t$ are chosen in such a way that 
\begin{enumerate}
\item[$(a)$] the PMF of the intentional goal-hitting time $\tih^{wf}$ never exceeds the risk level $\bar{q}$, and
\item[$(b)$] given $(a)$ is satisfied, they minimize the resulting expected value of $\tih^{wf}$. 
\end{enumerate}
In particular,  $(b)$ is achieved by greedily shifting the probability mass of $\tih^{wf}$ towards small $t$. This is what leads to the name-sake ``Water-Filling'': treating each entry of the PMF of $\tih^{wf}$ as a bucket, the strategy is essentially trying to fill the buckets with small $t$ up to the risk level $\bar{q}$, until it's no longer possible to do so due to the presence of intrinsic uncertainty (beyond the threshold $t^*$).\footnote{In the special case of a deterministic system with $\varepsilon=0$, this water-filling procedure in fact leads to a sequential construction of the random variable $T_{{IH}}^{wf}$ that is uniformly distributed between $1$ and $t^*$.} Figure \ref{fig:Twf} provides an illustration of this type of intentional goal-hitting time PMF induced by the Water-Filling Strategy on a complete graph. 
\begin{figure}[h]
\centering
	\includegraphics[width=0.55\textwidth]{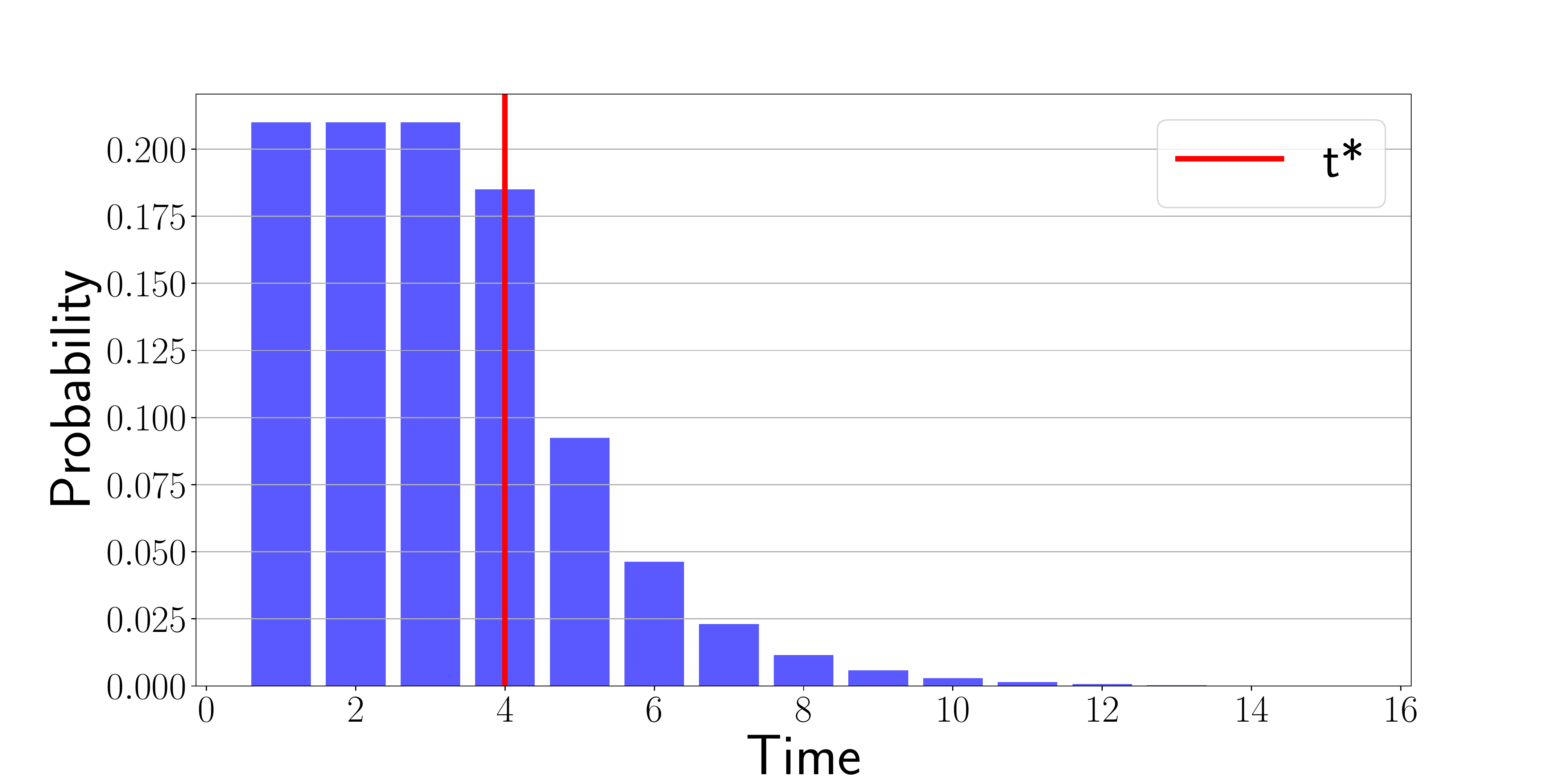}
	\caption{Distribution of $\tih\wf$ under the Water-Filling Strategy, with $\bar{q}=0.21$ and $\varepsilon=0.5$ on a complete graph. The vertical line depicts the threshold $t^*$.  }
	\label{fig:Twf}
\end{figure} 

When the graph is not complete, the agent is clearly not able to choose the Goal-Attempt meta action if she is not in the goal vertex's neighborhood, and in this case, the Water-Filling Strategy simply lets the agent take a Random-Step meta action. Our choice to employ the Random-Step is based on the following insight: the adversary does not observe the agent's actions, and as a result, under a carefully designed policy, he cannot distinguish a Random-Step from a Goal-Attempt purely based on the resulting trajectory. This feature ensures that the Water-Filling Strategy can continue to achieve a near-optimal prediction risk for a given level of delay, though, unfortunately, the minimum delay that can be implemented under a Water-Filling Strategy will be higher when the network is sparser.  
\end{enumerate}

The sequential randomization employed by the Water-Filling Strategy also highlights the distinction between environments with and without intrinsic uncertainty: in a deterministic system, there is little benefit to sequential randomization because the agent can perform randomization before routing begins and she will be sure as to when and how the goal vertex will be reached. Under intrinsic uncertainty, however, because the agent could not have foreseen with certainty when the goal vertex could be reached, the Water-Filling Strategy must tune the probability of Goal-Attempt in an adaptive manner, depending on whether a Goal-Attempt action has already succeeded in the past. For instance, it is easy to show that if the agent does not promptly switch to Random-Step after $\tih^{wf}$ but continues to choose Goal-Attempt, then the goal vertex $D$ would tend to be over-represented in the final trajectory, leading to a sub-optimal delay vs.~prediction risk trade-off. Note also that the magnitudes of $p_t$ (or rather $1-p_t$), which correspond to the degree of artificial randomization the agent injects, depend on the noise level $\varepsilon$ from the intrinsic uncertainty. This is how the Water-Filling Strategy harnesses intrinsic uncertainty, and thus avoids unnecessarily large delays. 

\section{Numerical Experiments}\label{sec:num}
While our theoretical results focus on the anonymity vs.~delay trade-off in the large-graph asymptotic, in this section we present some numerical examples and simulations on finite graphs with a modest size. We simulate an agent running a Water-Filling Strategy with $\bar{q} = \frac{1}{2w-1-c^{\varepsilon}}$, where $c^{\varepsilon}=\frac{\varepsilon}{(1-\varepsilon)^2}$, for a range of delay budget values, $w$. For simplicity, we will assume that the adversary predicts the goal vertex to be the first vertex after the initial vertex along the trajectory. This adversary estimator is an optimal MAP estimator against the Water-Filling strategy in a complete graph, but not necessarily optimal for non-complete graphs. Nevertheless, since the trajectory under Water-Filling Strategy has been shown to be independent of the goal, and the distribution of its goal-hitting time is non-increasing in time by design, this estimator should give us a reasonable approximation of the optimal prediction risk. We should acknowledge that the attack of the adversary could potentially be strengthened on non-complete graphs by using the actual posterior probability of the goal, or by endowing the adversary  with additional knowledge such as the distance between the destination and the initial vertex. However, doing so over a non-complete graph tends to greatly increase the complexity in evaluating the prediction risk. Assessing the performance of these more sophisticated adversary estimators could be an interesting direction for future work.

\begin{figure}[h]
\centering
		\includegraphics[scale=0.4]{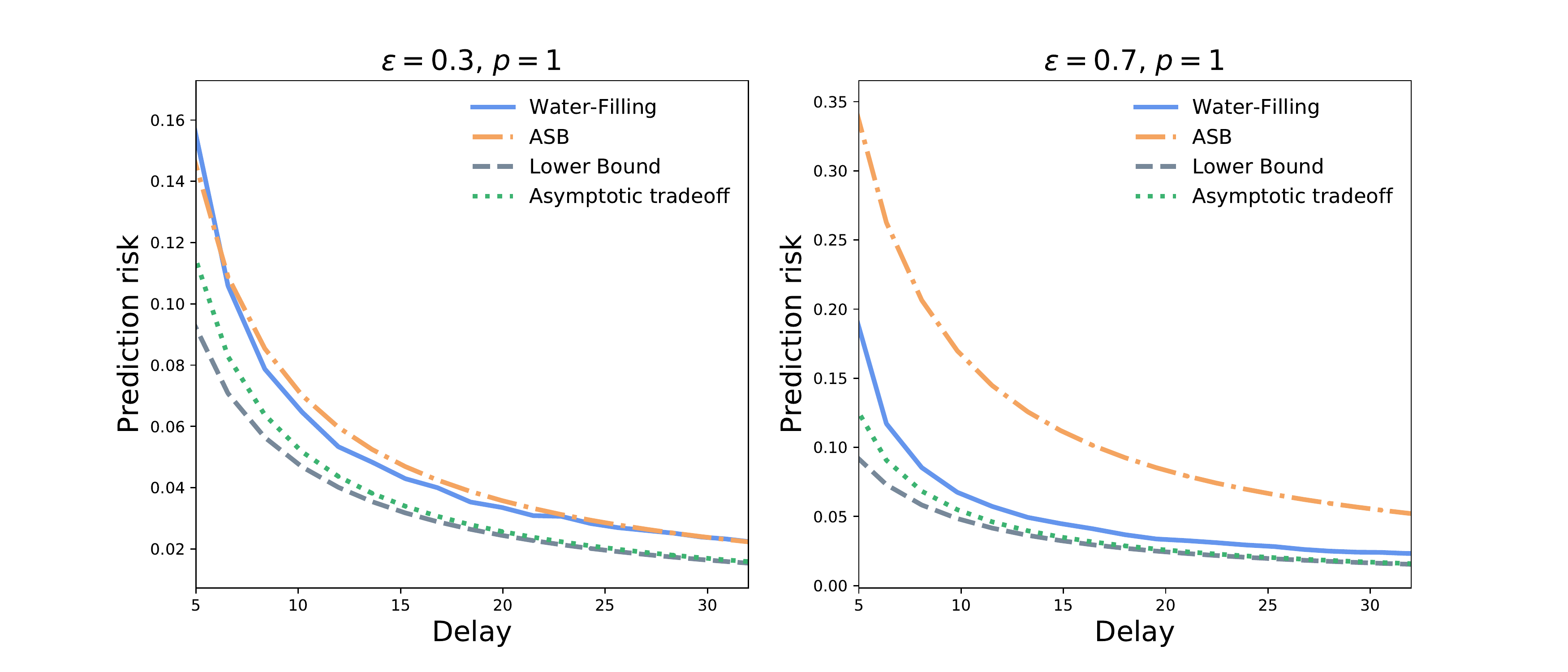}
	\caption{An illustration of the prediction risk as a function of the delay on a complete graph with $n=100$ vertices, under noise level $\varepsilon=0.3$ and $0.7$, respectively.  }
	\label{fig:comp}
\end{figure}

We first study a complete graph with $n=100$ vertices (Figure \ref{fig:comp}). For each delay budget value, we compute the prediction risk and the delay by averaging over $10^5$ trajectory realizations. Finally, we plot the asymptotic characterization of the delay-prediction trade-off in Theorem \ref{thm:wf_thm_infinite}, which holds as $n\to \infty$, as well as the looser lower bound in Theorem \ref{thm:reg} which holds for any graph of any size.  For comparison, we also include an estimated prediction risk for a more naive agent strategy, described in Section \ref{sec:why_uncertainty}. The strategy, dubbed \emph{adapted segment-based strategy} (ASB), serves as a proxy of what one may have obtained had they tried to implement a routing strategy designed for a deterministic model in an environment with uncertainty. Specifically, we consider the family of segment-based strategies in \cite{tsitsiklis2018delay} and assume that the agent first randomly selects an intended path, and subsequently traverses adjacent vertices along this path using a stochastic-shortest-path sub-routine. As discussed in Section \ref{sec:why_uncertainty}, in a large complete graph, the overhead for traversing any particular pre-selected vertices takes on average $\frac{1}{1-\varepsilon}$ steps, and the prediction risk can be analytically computed accordingly.  

\begin{figure}[h]
\centering
	\includegraphics[scale=0.4]{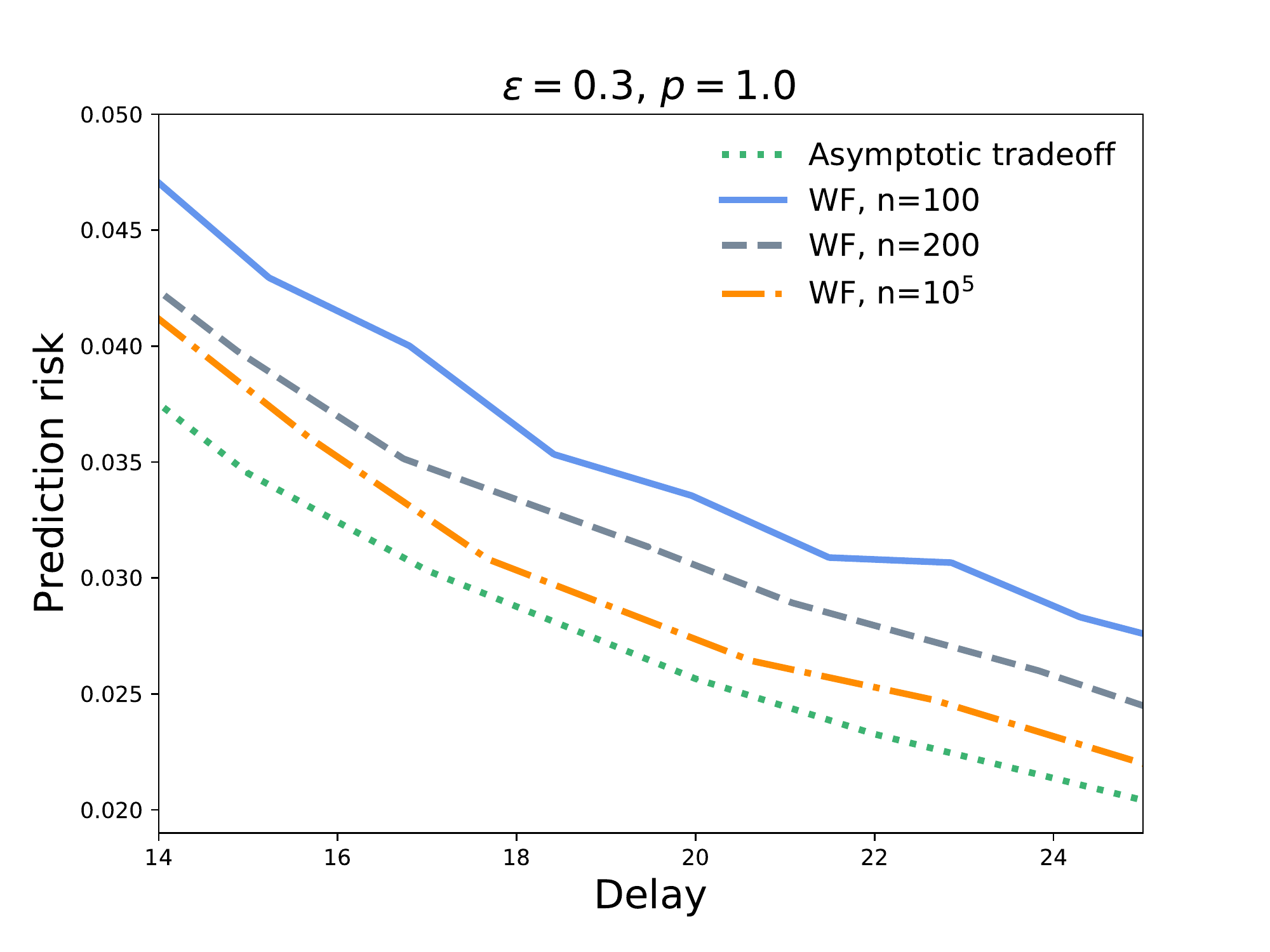}
	\caption{Comparison of the prediction risks on a complete graph at different network sizes, with $\varepsilon=0.3$. Here we zoom into a smaller section of the delays for better visual clarity. }
	\label{fig:changen}
\end{figure}

The figures show that our bounds offer reasonably good approximates for a finite graph with moderate size. Furthermore, the discrepancy between the performance of ASB and the Water-Filling Strategy is relatively small at a low noise level, but becomes much more pronounced as the noise level increases. This is to be expected since the ASB suffers a multiplicative overhead in delay due to increased uncertainty and hence its prediction risk is more sensitive to the changes in the noise level. Similarly, we observe that as $\varepsilon$ increases, the prediction risk also increases for any delay budget since the agent is required to choose higher $\bar{q}$ values to ensure similar delay values. Finally, there persists a small gap between the simulated curve as the asymptotic trade-off predicted by Theroem \ref{thm:wf_thm_infinite}. We expect this to be largely due to the finite size of the graph. Figure \ref{fig:changen} shows that, as we increase the network size, the simulated prediction risk tends to converge towards that of the asymptotic trade-off curve. 

\begin{figure}[h]
\centering
	\includegraphics[scale=0.4]{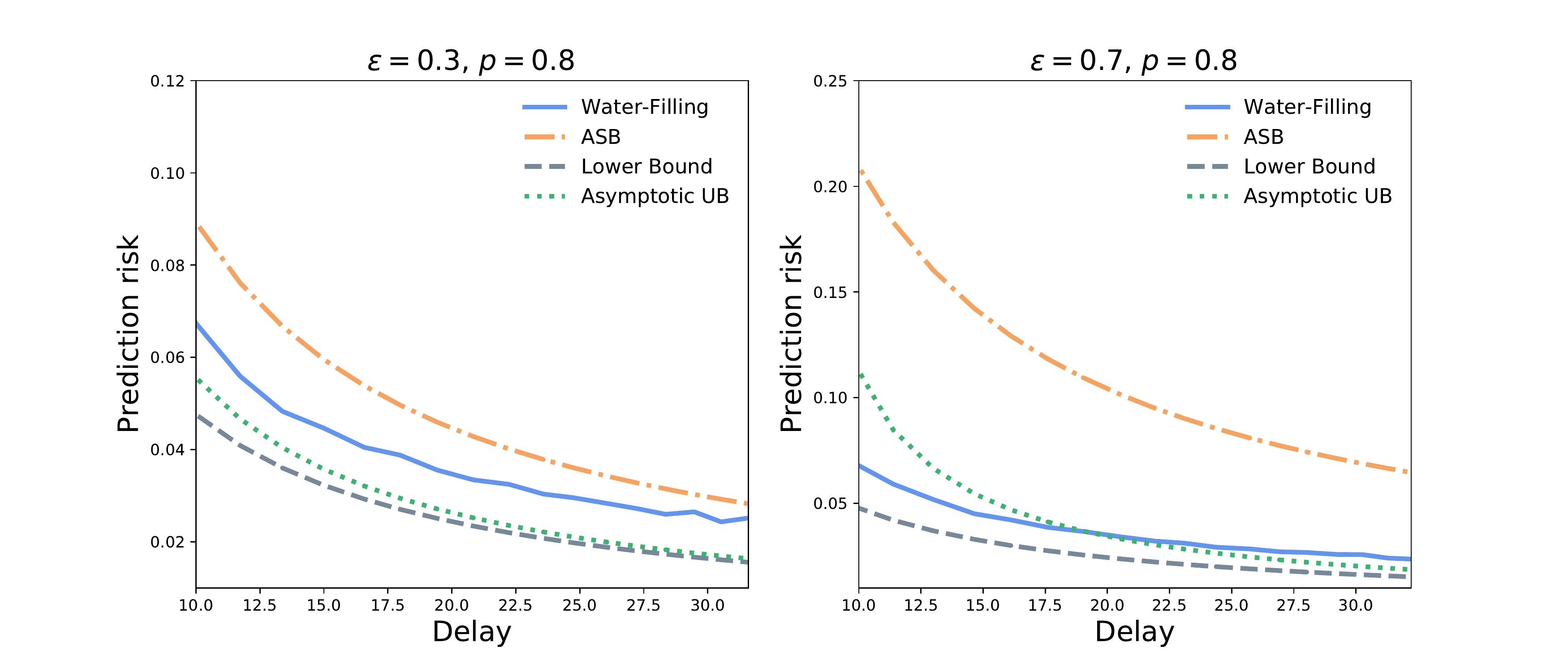}
	\caption{An illustration of the prediction risk as a function of the delay on Erd\H{o}s-R\'{e}nyi random graphs with $n=100$ vertices and edge density $p=0.8$, under noise level $\varepsilon=0.3$ and $0.7$, respectively. }
	\label{fig:er}
\end{figure}

In Figure \ref{fig:er}, we turn to non-complete graphs and repeat our analysis on Erd\H{o}s-R\'{e}nyi random graphs with $100$ vertices and edge density $p=0.8$. For each delay budget value, we compute the prediction risk and the delay by averaging over $100$ trajectory realizations, $100$ graph instances and $100$ goal vertices. Since we do not have a matching asymptotic lower bound for this case, we will display only the upper bound from Theorem \ref{thm:reg}, which is likely not tight. Since the delay under an ASB strategy is more difficult to estimate here, we will use some more optimistic estimates (in favor of ASB). We will assume that each step takes on average $\frac{1}{p(1-\varepsilon)} $ steps to traverse between two adjacent vertices on the pre-selected path. This is based on the calculation that for a typical vertex in such a graph, the agent  has roughly a probability of $p$ of being directly connected to the vertex she wishes to reach, and each attempt to reach it will succeed with probability $1-\varepsilon$.  

Figure \ref{fig:er} shows that, compared to the simulations on the complete graph, the bounds in these non-complete graphs are less tight, and with an increased noise level, the upper bound appears to be more pessimistic in the regime of small delays. On the positive side, we see that the performance improvement of Water-Filling compared to the naive ASB strategy widens even more in this setting; even at a modest noise level of $\varepsilon=0.3$, the Water-Filling Strategy can achieve up to a 30\% reduction in prediction risk. This appears to be due to the fact that the multiplicative delay overhead associated with ASB increases as $p$ decreases. 

\section{Discussion}\label{sec:conc}
%We proposed in this work the {Anonymous Stochastic Routing} model to analyze recipient-anonymous routing in a stochastic environment with intrinsic uncertainty. Our main result establishes near-optimal characterization of the delay-anonymity trade-off in a number of network topologies. We also propose a novel family of routing strategies, the Water-Filling Strategies, that achieve near-optimal trade-offs in these settings. 

There is a number of interesting directions for future research. A limitation of the Water-Filling Strategy is that it cannot enforce a small delay when the network is very sparse, which is due to the fact that the agent does not try to intentionally move closer to the goal unless she is already in its immediate neighborhood. While a variant of the strategy used in Theorem \ref{thm:networkdesign} does address this issue in sparse networks by exploiting a clique structure, this puts additional constraints on the type of networks. An interesting direction would be to understand whether one can design a routing strategy that simultaneously delivers low delay and optimal anonymity-delay trade-off in sparser networks. Our analysis suggests that in the absence of strong connectivity, an agent wanting to achieve low delay would have to attempt to reach the goal sooner and more frequently. One potential approach would be to generalize the notion of the ``goal-attempt'' action in the Water-Filling strategy from a single step to multiple steps: the agent may attempt to reach the goal as soon as she is sufficiently close to the goal vertex. The challenge here would be to ensure that the resulting trajectory of the multi-step goal-attempt can be seamlessly blended with the rest of the route and does not overtly reveal the identity of the goal.

In another direction, it would be interesting to extend this model to the setting with multiple rounds of routing tasks: our strategy may fail if the recipients across rounds can be correlated, if, for instance, the adversary focuses on the vertices that are shared across different rounds of routing trajectories. A solution to this problem may require some form of buffering and shuffling of goal vertices across the rounds, so as to erase any correlation structure, which is similar to the approaches of centralized anonymous messaging systems such as \cite{corrigan2015riposte}. 

Our investigation also reveals an intriguing conceptual puzzle concerning the interplay between the sparsity of the network and the ease of obfuscation. While in many centralized routing problems the performance of the system only improves with denser networks, this is not so obvious in the case of stochastic anonymous routing. The lower bounds in Theorems \ref{thm:reg} and \ref{thm:networkdesign} turn out to be weaker than the one we have for the complete graph in Theorem \ref{thm:wf_thm_infinite}. This is counter-intuitive, since one would expect that having more edges in the network helps the agent, and therefore, at the very least, the lower bound for the complete graph should extend to graphs that are strictly more sparse. However, this apparent monotonicity turns out to be far from obvious: on the one hand, adding edges should make it easier for the agent to traverse the graph, but at the same time it also makes it possible for random shocks to throw the agent off in ways that were not previously possible. Note that in a deterministic setting with $\varepsilon=0$, a denser graph always helps the agent (since she can simply choose to ignore certain edges), which suggests that the puzzle is a direct consequence of intrinsic uncertainty. To better understand the impact of sparsity on anonymity, a potential approach could be to derive a tighter lower bound for non-complete graphs that explicitly depends on the intrinsic uncertainty level $\varepsilon$. Resolving this puzzle around the interplay between sparsity and the ease of obfuscation is an interesting future direction and it is still an open question whether such a lower bound can be obtained.

Finally, it would be interesting to consider more complex models of intrinsic uncertainty. We currently assume that the random shocks do not depend on the agent's current state or chosen action. To generalize, one could imagine a full ``Markovian'' noise model, where the random shock in each step may depend both on the agent's state and her chosen action. The state dependence appears to be easier to handle, since the adversary already observes the agent's past trajectory, and the main insights should carry through as long as the noise distribution is approximately uniform across the neighboring vertices. The independence of the shocks from the intended action appears more difficult to relax. We expect that our results can be extended to cases where the noise distributions vary mildly as a function of the intended action, though the analysis will be non-trivial. It is less clear, however, what strategies would be desirable if the noise distributions depend very heavily on the intended action. In an extreme case, the adversary may be able to effectively observe the agent's chosen actions by looking at the resulting trajectory. Curiously, the policy adapted from the deterministic setting (Section \ref{sec:why_uncertainty}) does not require concealing the agent's intended actions, so an interesting question here is whether, as the noise distribution becomes less uniform and depends more heavily on the action, the prediction risk overhead due to uncertainty would become multiplicative instead of additive. 

\section{Proof Overview}\label{sec:pf_over}
The next two sections are devoted to the proofs of the main theorems. The upper bounds are proved in Section \ref{sec:pf_upper}, and our proofs will be based on analyzing the delay and prediction risk experienced by an agent who employs a Water-Filling type strategy: the bounds in Theorems \ref{thm:wf_thm_infinite} and \ref{thm:reg} are directly based on the strategy outlined in Definition \ref{defn:wf_strategy}. The proof of the upper bound in Theorem \ref{thm:networkdesign} uses a variant of the Water-Filling Strategy that is adapted to exploit a clique-like structure in the network to obtain even smaller delays. The primary techniques used here are concentration inequalities for random graph asymptotics, combined with the independence property outlined in Section \ref{sec:intuition}.  

The lower bounds are proved in Section \ref{sec:pf_lower}. For the lower bound in Theorem \ref{thm:wf_thm_infinite} (complete graphs), we leverage the ``water-filling''  analogy in Section \ref{sec:intuition} and show that the Water-Filling Strategy by construction induces an intentional goal-hitting time with the minimum expected delay, across all possible strategies that yield a maximum risk level of $\bar{q}$. We then use the large-graph asymptotics to show that the expected intentional goal-hitting time coincides with the true delay as $n \to \infty$. Unfortunately, this more delicate line of analysis does not extend to more complex topologies of the graphs in Theorems \ref{thm:reg} and \ref{thm:networkdesign}. In those cases, we will use a reduction argument to extend a prediction risk lower bound proved in \cite{tsitsiklis2018delay} for a noiseless model to our setting.

\section{Proofs of Upper Bounds}\label{sec:pf_upper}

\subsection{Upper Bounds in Theorems \ref{thm:wf_thm_infinite} and \ref{thm:reg}}
We prove in this subsection the upper bounds in Theorems \ref{thm:wf_thm_infinite} and \ref{thm:reg}, using the Water-Filling Strategies. We begin by introducing the two main technical results which will form the foundation of the proofs. Both results concern the performance of an agent that uses the Water-Filling Strategy outlined in Definition \ref{defn:wf_strategy}: 
\begin{enumerate}
\item Proposition \ref{prop:delay_general} establishes an upper bound on the delay as a function of the risk level $\bar{q}$,  noise level $\varepsilon$, and a parameter $p$ that captures the edge density of the network $G$. 
\item Proposition \ref{prop:risk_general} establishes an upper bound on the maximal prediction risk that only depends on the target risk level $\bar{q}$, as well as the maximum and minimum degrees of $G$. In particular, it does not depend on $\varepsilon$, nor any other topological properties of $G$. 
\end{enumerate} 
We will then combine these technical results to get the upper bounds in Theorems \ref{thm:wf_thm_infinite} and \ref{thm:reg}. 

To ensure that the delay under the Water-Filling Strategy is not too large, we will rely on the following Neighborhood Overlap property, which essentially states that pairs of connected nodes share a large fraction of overlapping neighbors. Intuitively,  this condition ensures that a random walk over the graph will frequently visit \emph{some} neighboring vertex of the goal vertex, though not necessarily the goal vertex itself. 
\begin{definition}[Neighborhood Overlap]
Fix $p \in [0,1]$. $G= (\calV, \calE)$ satisfies the Neighborhood Overlap property with parameter $p$, if for any pair $u,v \in \calV$, at least $p$ fraction of the neighborhood of $v$ intersects the neighborhood of $u$, i.e., ${\abs \neig(u) \cap \neig(v) \abs}/{\abs \neig(v) \abs} \geq p$.
\end{definition}

Now, we state Proposition \ref{prop:delay_general} which derives an upper bound on the expected delay under the Water-Filling Strategy on any graph with the Neighborhood Overlap property. Notably, we can see that the impact of the noise level $\varepsilon$ on the delay is additive. Proposition \ref{prop:delay_general} is proven in Section \ref{sec:delay_general_proof}, using the expected intentional goal-hitting time as a natural upper bound on the actual goal-hitting time.
\begin{proposition}\label{prop:delay_general}
Fix graph size $n \in \N$, and target risk level $\bar{q} \in [0,1]$. Let $G = (\calV,\calE)$ be a graph with $\abs \calV \abs = n$. Suppose $G$ satisfies the Neighborhood Overlap property with parameter $p$. Under the Water-Filling Strategy, $\psi\wf_{\bar{q}}$, the expected delay satisfies 
\begin{align*}
\bE(T) \leq \bE(\tih) \leq \frac{1}{p} \left( \frac{1}{2\bar{q}} + \frac{1}{2} + \frac{\bar{q} \varepsilon}{2(1-\varepsilon)^2} \right).
\end{align*}	
\end{proposition}

Next, we state an upper bound on the maximal prediction risk under the Water-Filling Strategy on any undirected connected graph. Proposition \ref{prop:risk_general} expresses the maximal prediction risk in terms of the target risk level, $\bar{q}$. The result shows that the maximal prediction risk under the Water-Filling Strategy depends on the graph topology only through the minimum and maximum degrees of the vertices. The proof is given in Section \ref{sec:risk_general_proof}.
\begin{proposition}\label{prop:risk_general}
%Fix a sequence of horizons $\{K_n\}_{n \in \N}$ such that $1\ll K_n \ll {n}$. 
Fix graph size $n \in \bN$, and target risk level $\bar{q} \in [0,1]$. Let $G = (\calV,\calE)$ be a graph with $| \calV | = n $. Under the Water-Filling Strategy, $\psi\wf_{\bar{q}}$, the maximal prediction risk satisfies
\begin{align*}
q^*(\psi\wf_{\bar{q}}) \leq \frac{\bar{q} \overline{\Delta}_G}{n} + \delta_n'
\end{align*}
where $\delta_n' = \frac{w}{K_n}+ 1 - \lt( 1- \frac{1}{\underline{\Delta}_G} \rt)^{K_n}$, $\lim\limits_{n \rightarrow \infty} \delta_n' = 0$, and $\underline{\Delta}_G$ and $\overline{\Delta}_G$ are the minimum and maximum degrees of $G$, i.e., $\underline{\Delta}_G = \min\limits_{ v \in \calV}  | \neig(v)|$ and $\overline{\Delta}_G = \max\limits_{ v \in \calV}  | \neig(v)|$. 
\end{proposition}

We now combine Propositions \ref{prop:delay_general} and \ref{prop:risk_general} to derive an upper bound on the minimax prediction risk, as a function of the delay target, $w$. Let 
\begin{equation}
 c^\varepsilon = \frac{\varepsilon}{(1-\varepsilon)^2},
 \end{equation} 
and suppose $ w > \frac{c^\varepsilon}{2} + 1 $ holds. By Proposition \ref{prop:delay_general}, to obtain $\bE(T) \leq w$ under the Water-Filling Strategy with target risk level $\bar{q}$, it suffices to set
\begin{align}\label{eq:barq}
\bar{q} = \frac{1}{2 w p - 1 - c^\varepsilon}.
\end{align}

Next, we note that the minimax prediction risk $\calQ(w)$ is upper bounded by the maximal prediction risk under any strategy whose expected delay is at most $w$. Specifically, we can look at the Water-Filling Strategy whose target risk level is given by Eq.~\eqref{eq:barq}. Then, Proposition \ref{prop:risk_general} implies
\begin{align}\label{eq:minimax_general}
\calQ(w) & \leq q^*(\psi\wf_{\bar{q}}) \leq \frac{\bar{q} \overline{\Delta}_G}{n} + \delta_n' = \frac{\frac{\overline{\Delta}_G}{n}}{2wp - 1- c^\varepsilon} + \delta_n'.
\end{align}
We can then substitute $\overline{\Delta}_G$, $\underline{\Delta}_G$, and $p$ to obtain an upper bound on the minimax prediction risk on different network topologies. We present the derivation of these upper bounds in Theorems \ref{thm:wf_thm_infinite} and \ref{thm:reg} in the following subsections.

\subsubsection{Complete Graphs: Theorem \ref{thm:wf_thm_infinite}}
We first note that on a complete graph we have $\underline{\Delta}_G = \overline{\Delta}_G = n$ and the Neighborhood Overlap property holds with parameter $p=1$. Using Eq.~\eqref{eq:minimax_general}, this immediately implies that for any delay budget $w$ such that $w > \frac{\varepsilon}{2(1-\varepsilon)^2} + 1$ holds, the minimax prediction risk satisfies
\begin{align}\label{eq:minimax_comp_loose}
\calQ(w) \leq \frac{1}{2w - 1 - c^\varepsilon}  + \delta_n'.
\end{align}
This already shows that a delay overhead that is only additive in $\varepsilon$ can be achieved via the Water-Filling Strategy. 

To obtain the exact additive overhead, we can conduct a more refined analysis. In what follows, we will derive a prediction risk upper bound that improves upon  Eq.~\eqref{eq:minimax_comp_loose}, leading to the upper bound in Theorem \ref{thm:wf_thm_infinite}. Fix a delay budget $w$ and consider the class of strategies, $\Phi_w$, under which the expected intentional goal-hitting time is at most $w$, i.e., $\Phi_w= \{ \psi: \bE(\tih^{\psi}) \leq w \}$.  Define $\bar{q}(w)$ to be the minimum target risk level $\bar{q}$ that can be attained by a Water-Filling Strategy with expected intentional goal-hitting time at most $w$, that is, $\bar{q}(w) = \inf \{ \bar{q}: \exists \psi^{wf}_{\bar{q}} \in \Phi_w \} $. Suppose $w > \frac{\varepsilon}{2(1-\varepsilon)^2} + 1$ holds. Proposition \ref{prop:delay_general} implies that the expected intentional goal-hitting time on a complete graph is at most $\frac{1}{2\bar{q}} + \frac{1}{2} + \frac{\bar{q} \varepsilon}{2(1-\varepsilon)^2}  $. This shows that $\bar{q}(w)$ satisfies:
\begin{align}\label{eq:barq_derive}
\bar{q}(w) \leq \frac{1}{\left(w-\frac{1}{2} \right) \left(1+ \sqrt{ 1 - \frac{\varepsilon}{\left(w-\frac{1}{2} \right)^2 (1-\varepsilon)^2 }} \right)}.
\end{align}
Let $x=\frac{\varepsilon}{\left(w-\frac{1}{2} \right)^2 (1-\varepsilon)^2} $. By the Taylor expansion of $\sqrt{1-x}$ for $x <1$, we have that
\begin{align}\label{eqn:qbar}
\bar{q}(w) \leq \frac{1}{2w-1 - \alpha^{\varepsilon}(w) - \beta^{\varepsilon}(w)},
\end{align}
where $\alpha^{\varepsilon}(w) = \frac{\varepsilon}{(2w-1)(1-\varepsilon)^2}$ and $\beta^{\varepsilon}(w) = \frac{\varepsilon^2}{2 \left(w-\frac{1}{2} \right)^3 (1-\varepsilon)^4}$. 

Finally, we recall that setting $\bar{q} = \bar{q}(w)$ implies $\bE(T\wf) \leq \bE(\tih\wf) \leq w$. Hence, we use Proposition \ref{prop:risk_general} and Eq.~\eqref{eqn:qbar} to obtain
\begin{align}\label{eq:comp_upper_tay}
\calQ(w) \leq \frac{1}{2w-1 - \alpha^{\varepsilon}(w) - \beta^{\varepsilon}(w)} + \delta_n'.
\end{align}
This concludes the proof of the upper bound in Theorem \ref{thm:wf_thm_infinite}. \hfill \qed

\subsubsection{Family $\calG_n$: Theorem \ref{thm:reg}}\label{sec:er_graphs}
We now turn to the proof of the upper bound in Theorem \ref{thm:reg} and show that an additive delay overhead can still be attained on a family of non-complete graphs. In Definition \ref{def:fam}, we formally define a parameterized family of graphs $\calF_n(p,\gamma)$ with concentrated degree distribution and Neighborhood Overlap property, which ensure that the prediction risk is low and the delay is small. Then, we prove that an Erd\H{o}s-R\'{e}nyi random graph with $n$ vertices and edge density $p$ belongs to the family $\calF_n(p,\gamma_n)$ with high probability as $n \rightarrow \infty$. In a way, this suggests that ``typical" graphs with average degree $pn$ under the Erd\H{o}s-R\'{e}nyi model still enjoy low minimax prediction risk as was the case in complete graphs. Finally, we use Eq.~\eqref{eq:minimax_general} to conclude the upper bound.

\iffalse
The properties we need are given in Definition \ref{def:fam}. Intuitively, the first property implies that the degrees of vertices are concentrated around $pn$. In light of Proposition \ref{prop:risk_general}, this observation allows us to conclude that the prediction risk under the Water-Filling Strategy with target risk level $\bar{q}$ is upper bounded by $p \bar{q}$. On the other hand, as we have seen previously, the Neighborhood Overlap property guarantees that the delay is small since a random walk frequently visits the neighborhood of the goal vertex. Compared to a complete graph, the prediction risk and the delay on a graph in the family $\calF_n$ are scaled by $p$ and $1/p$, respectively. We will shortly show that these factors cancel each other, hence, a similar performance to that on complete graphs can be achieved on graphs from $\calF_n$, as well. 
\fi
\begin{definition}[Family $ \calF_n  (p, \gamma)$ of Graphs] \label{def:fam}
Fix $p \in [0,1]$ and $\gamma \in [0,1] $. For each $n \in \bN$, we define $ \calF_n(p, \gamma)$ as the set of all graphs $G=(\calV,\calE)$ with $n$ vertices such that (1) the degrees of all vertices are upper- and lower-bounded by $pn(1+\gamma)$ and $pn(1-\gamma)$ respectively, and (2) the Neighborhood Overlap property holds with parameter $p(1-\gamma)$.
\end{definition}

Next, we will set the value of $\gamma$ in $\calF_n(p,\gamma)$ to define our main object of interest, the sequence of families $\{ \calG_n(p) \}_{n \in \N}$.

\begin{definition}[Family $\calG_n(p)$ of Graphs]\label{def:gn}
Fix $p\in [0,1]$. Let $\{\gamma_n\}_{n \in \bN}$ be a sequence such that $\gamma_n \in [0,\frac{1}{2}] $, $\gamma_n \gg \sqrt{{\log n}/{n}} $ and $\lim\limits_{n \rightarrow \infty} \gamma_n =0$. For each $n \in \N$, we define $\calG_n(p) = \calF_n(p, \gamma_n)$.
\end{definition}

\iffalse
\begin{remark}
Fix $p=1$ and set $\gamma_n = 0$ for all $n \in \N$. Then, the family $\calG_n(p)$ reduces to a sequence of complete graphs with increasing graph size $n$.
\end{remark}
\fi

In the remainder, we will restrict our focus on the family $\calG_n(p)$ where $\gamma_n$ is defined as in Definition \ref{def:gn}. Recall that under the Erd\H{o}s-R\'{e}nyi random graph model, a graph with $n$ vertices is constructed by independently connecting each pair of vertices by an edge with probability $p$. Lemma \ref{lem:erdos} shows that with high probability, a graph generated using the Erd\H{o}s-R\'{e}nyi random graph model with edge probability $p$ belongs to the family $\calG_n(p)$. The proof relies on an application of concentration inequalities and is given in Appendix \ref{ap:erdos}. We note that Lemma \ref{lem:erdos} proves Theorem \ref{thm:reg}(a).

\begin{lemma}\label{lem:erdos}
Fix edge density $p \in [0,1]$. Fix graph size $n \in \bN$ and consider the family of graphs $\calG_n(p)$. Let $G$ be an Erd\H{o}s-R\'{e}nyi random graph with $n$ vertices and edge probability $p$. Then, there exists a sequence $\{\theta_n\}_{n \in \N}$, with $\lim\limits_{n \rightarrow \infty} \theta_n = 0 $, such that for every $n \in \N$, $ \bP(G \in \calG_n(p)) \geq 1 -\theta_n$ holds.
\end{lemma}

We now turn to the proof of Theorem \ref{thm:reg}(b) and use Eq.~\eqref{eq:minimax_general}. Assume that $w > \frac{1}{p} + \frac{\varepsilon}{2p(1-\varepsilon)^2} $ holds. We then substitute the Neighborhood Overlap parameter $p = p(1-\gamma_n) $, and the minimum and maximum degrees $\underline{\Delta}_G=pn(1-\gamma_n)$ and $\overline{\Delta}_G=pn(1+\gamma_n)$. Thus, we obtain
\begin{align*}
\calQ(w) & \leq \frac{p(1+\gamma_n)}{2wp(1-\gamma_n ) - 1- c^\varepsilon} +  \delta_n' = \frac{1}{2w- \frac{1}{p} (1 + c^\varepsilon) - 2\gamma_n w } + \frac{\gamma_n}{2wp(1-\gamma_n ) - 1- c^\varepsilon} + \delta_n',
\end{align*}
where $\delta_n' = 1- \left( 1- \frac{1}{pn(1-\gamma_n)} \right)^{K_n} + \frac{w}{K_n}$. Let $\delta_n =  \delta_n' + \frac{\gamma_n}{2wp(1-\gamma_n ) - 1- c^\varepsilon}$ and $\lambda_n = 2 \gamma_n n $. Recall that for the family $\calG_n(p)$ with $\gamma=\gamma_n$, we have $\lim\limits_{n \rightarrow \infty} \gamma_n =0$. Therefore, the second term in $\delta_n$ will approach $0$ as $n \rightarrow \infty$. Similarly, we have $\lim\limits_{n \rightarrow \infty} \lambda_n = 0$. Next, let us consider $\delta_n'$. Since $w$ is fixed and does not depend on $n$, and since $K_n \ll n$ by assumption, we also have $\lim\limits_{n \rightarrow \infty} \delta_n' = 0$. Thus, we conclude $\lim\limits_{n \rightarrow \infty} \delta_n = 0$. Finally, we get $\calQ(w) \leq \frac{1}{2w - p^{-1} (1 + c^\varepsilon) - \lambda_n } + \delta_n$ where $\lim\limits_{n \rightarrow \infty} \delta_n = 0$. This completes the proof of Theorem \ref{thm:reg}. 
%$\lambda_n = 2 \gamma_n n \geq 2 \gamma_n w $ since w<K_n \ll n, defining \lambda_n with n instead of w makes it independent from the delay budget
\qed

Note that as we let $n$ and $K$ grow, the upper bound becomes 
\begin{align*}
\calQ(w) \leq \frac{1}{2w-\frac{1}{p} - \frac{\varepsilon}{(1-\varepsilon)^2p}},
\end{align*}
showing that an additive uncertainty overhead is still achievable, even when the graph is generated from a random graph model. However, the performance deteriorates as $p \to 0$.

%In Appendix \ref{sec:reg_n}, we prove Theorem \ref{thm:reg_n} which extends the analysis to sparse non-complete graphs where the edge density $p_n$ may approach zero as $n \rightarrow \infty$. 

\subsection{Upper Bound in Theorem \ref{thm:networkdesign}}\label{sec:proof-net}

In this section, we prove the upper bound in Theorem \ref{thm:networkdesign} and analyze what one can achieve by selecting a sparse network topology. The proof of Theorem \ref{thm:networkdesign} consists of three main steps. First, we describe a family of networks $\mathcal{\bar{G}}(n, \bar{p}_n)$, having $n$ vertices and average degree at most $\bar{p}_n$. 

The main insight is to build a graph composed of several highly connected components that are connected to each other via a small number of edges; as such, each connected component functions as a complete graph, and the time to travel between different components is small. Next, we design the Clique Water-Filling Strategy to be applied on $\mathcal{\bar{G}}(n, \bar{p}_n)$ and quantify its delay as a function of the delay of $\psi\wf_{\bar{q}}$ on a complete graph. To do so, we use a coupling argument in which we ``freeze" time whenever the agent leaves the component containing her goal. Finally, we derive an upper bound on the prediction risk using a reduction argument to the complete graph setting. Specifically, we observe that if the agent announces which component contains her goal, then the adversary can ignore the rest of the graph and focus only on the announced component. Since this component is itself a complete graph, the maximum prediction risk under $\psi\wf_{\bar{q}}$ applies as an upper bound. We note that the graphs and the strategy developed in this subsection generalize the complete graphs and the Water-Filling Strategy, respectively. 

We start with the following definition that constructs the graph $G$ with the desired properties: We first create $k$ fully connected cliques with size $m$ each. Then, we connect the $l$th vertex in the cliques to one another, for each $l=1, \ldots, m$. A sample $k$-clique graph is illustrated in Appendix \ref{ap:clique_fig}.
\begin{definition}[$k$-Clique Graph]
Fix $n, k \in \bN$, and denote $m= \frac{n}{k}$. Let $G=(\calV, \calE)$ be a $k$-clique graph with $n$ vertices. Then, $G$ is constructed as follows.
\begin{enumerate}
	\item[(1)] For each $i \in \{ 1,...,k \}$, construct a complete graph $G_i = (\calV_i, \calE_i)$ where $\calV_i = \{v_1^i, \ldots, v_m^i\}$. 
	\item[(2)] Define $\tilde{\calE} = \{ (v_l^i, v_l^j) : 1 \leq i<j \leq k, 1\leq l \leq m\}$. 
	\item[(3)] Let $\calV = \bigcup_{i=1}^{k} \calV_i$ and $\calE = \tilde{\calE} \cup ( \bigcup_{i=1}^{k} \calE_i)$. 
	\item[(4)] Set $G = (\calV, \calE)$ so that $|\calV|=n$ and $| \calE | =\frac{n}{2}\left( \frac{n}{k}+k-1 \right)$. 
\end{enumerate}
\end{definition}

Under this construction, each vertex has degree $\frac{n}{k}+k-1$, with $\frac{n}{k}$ within the clique and $k-1$ across cliques. Note that if the average degree of the network is constrained to be at most $\bar{p}_n$, then one can choose the number of cliques, $k$, so as to have $\frac{n}{k}+k-1 \leq \bar{p}_n$. Therefore, we let $\mathcal{\bar{G}}(n, \bar{p}_n)$ be the family of $k$-clique graphs where $k$ satisfies $\frac{n}{k}+k-1 \leq \bar{p}_n$. %Note that a complete graph is a 1-clique graph.

Next, adapting the Water-Filling Strategy to leverage the network structure, we design the Clique Water-Filling Strategy, $\psi\k $, which can be implemented on a $k$-clique graph $G$. The main idea is to apply the Water-Filling Strategy whenever the state is inside the same clique with the goal and to freeze time whenever the state is outside. Define the counter $L(t)$ to be the number of times the state $X$ has been in the clique that contains the goal, $\calV_D$, by time $t$:
\begin{equation}
L(t) = \sum_{s=1}^t \mathbb{I}(X_s \in \calV_D).
\label{eq:counter_clique}
\end{equation} 

\begin{definition}[Clique Water-Filling Strategy]
	Fix target risk level $\bar{q} \in [0,1]$. Let $G=(\calV,\calE)$ be a $k$-clique graph with $n$ vertices and let $\calV_D$ denote the set of vertices in the clique containing the goal. 
	The Clique Water-Filling Strategy, $\psi\k_{\bar{q}}$, is defined as follows: for $t\in \N$
	\begin{enumerate}
	\item If $X_t \notin \calV_D$ (agent not in the same clique as the goal), set $a_t = v_l^{\calV_D}$ to go back to $\calV_D$, where $X_t = v^i_l$. 
	\item If $X_t \in \calV_D$, and $\tih\k > t$ (no intentional goal-hitting so far):
	\begin{enumerate}
			\item[(1)] with probability $p_{L(t)}$, the agent chooses the Goal-Attempt meta action, where $L(t)$ and $\{p_t\}_{t\in \N}$ were defined in Eqs.~\eqref{eq:counter_clique} and\eqref{eq:ptdef}, respectively; 
		\item[(2)] otherwise, the agent chooses the Random-Step meta action. 
	\end{enumerate}
	\item If $\tih\k \leq t$, the agent always chooses the Random-Step action.
\end{enumerate}
\end{definition}

Now, we quantify the delay under $\psi\k$ in terms of the expected intentional goal-hitting time under the Water-Filling Strategy on a complete graph with $\frac{n}{k}$ vertices. The proof is given in Appendix \ref{ap:cliquedelay}. Intuitively, the delay under $\psi\k$ is the sum of two terms: the expected time spent in the clique containing the goal and the expected time spent outside. To prove Proposition \ref{cliquedelay}, we begin by providing an upper bound on the expected total time the agent spends outside $\calV_D$. Then, we observe that the time spent inside $\calV_D$ is less than the expected intentional goal-hitting time under $\psi\wf$, where the time is indexed by the counter $L$ and hence, conclude the proof.

\begin{proposition}\label{cliquedelay}
Fix target risk level $\bar{q} \in [0,1]$. Fix a degree sequence $\{\bar{p}_n\}_{n \in \bN}$ such that $\bar{p}_n \leq n$ and $\bar{p}_n \gg \sqrt{n}$. Fix graph size $n \in \bN$ and clique number $k \in \N$ such that $\frac{n}{k}+k-1 \leq \bar{p}_n$. Let $G=(\calV,\calE)$ be a $k$-clique graph with $| \calV | = n$. Under the Clique Water-Filling Strategy, $\psi\k_{\bar{q}}$, the expected delay satisfies
\begin{align*}
\bE(T\k) \leq \bE(\tih\wf) \left(1+ \frac{k^2}{(1-\varepsilon) (n+(k-1)k) + \varepsilon k} \right),
\end{align*}
where $\tih\wf$ is the intentional goal-hitting time under the Water-Filling Strategy $\psi\wf_{\bar{q}}$, on a complete graph with $\frac{n}{k}$ vertices.
\end{proposition}

Suppose the adversary knows the clique containing the goal, $\calV_D$. Then, the adversary can simply ignore whenever the agent is outside $\calV_D$. Consequently, from the perspective of the adversary, the game reduces to predicting $D$ of an agent using $\psi\wf$ on a complete graph with $\frac{n}{k}$ vertices. This implies that the maximal prediction risk under $\psi\k_{\bar{q}}$ can be upper bounded by the maximal prediction risk under $\psi\wf_{\bar{q}}$ on a complete graph with $\frac{n}{k}$ vertices. Hence, we obtain the following result.
\begin{corollary}\label{cor:cliquerisk}
Fix a degree sequence $\{\bar{p}_n\}_{n \in \bN}$ such that $\bar{p}_n \leq n$ and $\bar{p}_n \gg \sqrt{n}$. Fix graph size $n \in \bN$ and clique number $k \in \N$ such that $\frac{n}{k}+k-1 \leq \bar{p}_n$. Let $G=(\calV,\calE)$ be a $k$-clique graph with $| \calV | = n$. Under the Clique Water-Filling Strategy, $\psi\k_{\bar{q}}$, the maximal prediction risk satisfies
\begin{align*}
q^*(\psi\k_{\bar{q}}) \leq \bar{q} + \delta_n,
\end{align*}	
where $\delta_n = \frac{w}{K_n} + 1- \left( 1- \frac{k}{n} \right)^{K_n} $, and $\lim\limits_{n \rightarrow \infty} \delta_n = 0$.
\end{corollary}

We are now ready to prove Theorem \ref{thm:networkdesign} by combining Proposition \ref{cliquedelay} and Corollary \ref{cor:cliquerisk}.

\emph{Proof of Theorem \ref{thm:networkdesign}.} Let $G \in \mathcal{\bar{G}}(n,\bar{p}_n)$ be a $k$-clique graph with $n$ vertices, for some $k \in \N$ such that $\frac{n}{k}+k -1 \leq \bar{p}_n$. Then, consider the agent strategy $\psi\k_{\bar{q}}$, that is, the Clique Water-Filling Strategy with target risk level $\bar{q} \in [0,1]$. By Proposition \ref{cliquedelay}, the expected delay under strategy $\psi\k$ satisfies
	\begin{align*}
		\bE(T\k) \leq \bE(\tih\wf) \left(1+ \frac{k^2}{(1-\varepsilon) (n+(k-1)k) + \varepsilon k} \right) := \bE(\tih\wf) (1+x).
	\end{align*}
	Then, to obtain $\bE(T\k) \leq w$, it suffices to have $\bE(\tih\wf) (1+x) \leq w $. By Lemma \ref{lem:twf}, for $\bE(\tih\wf) \leq y$, it suffices to set $\bar{q} = \frac{1}{2y -1 - c^{\varepsilon}}$ where $c^{\varepsilon} = \frac{\varepsilon}{(1-\varepsilon)^2}$. Thus, letting $y = \frac{w}{1+x}$, Corollary \ref{cor:cliquerisk} implies
	\begin{align*}
		\calQ(w) \leq q^*(\psi\k) \leq \bar{q}+ \delta_n = \frac{1}{2w \rho_n^\varepsilon(k) - 1 - c^{\varepsilon} } + \delta_n ,
	\end{align*}
	where $\rho_n^\varepsilon(k) = \frac{1}{1+x} = \frac{(1-\varepsilon) (n+ k^2-k) + \varepsilon k }{ n+2k^2-k-\varepsilon(n+k^2-2k) } $. Finally, by the assumption $\bar{p}_n \gg \sqrt{n}$, for each $k \in \N$, we have $\lim\limits_{n \rightarrow \infty} \rho_n^\varepsilon(k) = 1 $. This completes the proof of Theorem \ref{thm:networkdesign}.
\qed

Note that as $k$ increases, the average degree of the network with design parameter $k$ decreases while the prediction risk grows. Therefore, by tuning the value of $k$, one can design a private network with the most efficient cost and privacy trade-off within this family. Finally, letting $n \rightarrow \infty$, we obtain the following upper bound on the minimax prediction risk
\begin{align*}
\calQ(w) \leq \frac{1}{2w-1-c^{\varepsilon}},
\end{align*}
implying that the performance on complete graphs can be asymptotically achieved by a careful design of the network topology.

%In conclusion, for any degree sequence $\{ \bar{p}_n \}_{n \in \bN}$, where $\sqrt{n} \gg \bar{p}_n \leq n$, subject to the constraint that the average degree of the network is at most $\bar{p}_n$, there exists a family of private networks $\mathcal{G}(p)$ with $n$ vertices, indexed by the design parameter $k_n$ such that $\frac{n}{k_n}+k_n -2 \leq \bar{p}_n $ holds. Moreover, on this family of graphs, the minimax prediction risk only bears an additive overhead. Crucially, observe that it is possible to design a private network topology subject to an average degree constraint down to the threshold $p_n \gg \sqrt{n}$, allowing relatively sparse private networks. 

\subsection{Proofs of Propositions \ref{prop:delay_general} and \ref{prop:risk_general}}
\subsubsection{Delay: Proof of Proposition \ref{prop:delay_general}}\label{sec:delay_general_proof}
In this section, we will characterize the agent's delay under the Water-Filling Strategies and prove Proposition \ref{prop:delay_general}. We will begin by observing that the expected delay can be broken into two components: time spent in the neighborhood of $D$ and the total time spent outside. To analyze the first term, we will use the concept of intentional goal-hitting time. Next, using the Neighborhood Overlap property we will show that an agent using a random walk will ``stumble'' upon the neighborhood of her goal every $\frac{1}{p}$ time steps on average. We will subsequently combine this observation with a ``time-stretching" argument to establish that the delay on a graph satisfying the Neighborhood Overlap property is upper-bounded by the delay on a complete graph, but slowed down by a factor of $\frac{1}{p}$. 

Recall the counter $L(t)$ that tracks the total time spent in the neighborhood of $D$. For all $t \geq 1$, let $C^o_t = t - L(t)$ so that $C^o_t$ records the total time spent outside $\neig(D)$ up to time $t$. Then, we can write the expected delay as the sum of the two counters at time $T$:
\begin{align}\label{eq:delay_counters}
\bE(T) = \bE(L(T)) + \bE(C^{o}_T).
\end{align}
%In the remainder, for notational simplicity we will refer to $L(T)$ and $C^o_T$ as $T\wf$ and $T^o$, respectively. 

Suppose $G$ is a complete graph. Since every vertex of $G$ is in the neighborhood of $D$, we will have $C^o_T = 0$. This suggests that we can interpret $L(T)$ as the goal-hitting time on a complete graph under the Water-Filling Strategy, $T\wf$. Using this observation, Lemma \ref{lem:twf} derives an upper bound on $\bE(T\wf) = \bE(L(T))$ in terms of the intentional goal-hitting time on a complete graph under the Water-Filling Strategy, $\tih\wf$. The proof is in Appendix \ref{ap:lem_twf}.\footnote{In the remainder, we may suppress the argument $\bar{q}$ from $\psi^{wf}$ for notational convenience, when the context is clear.}
\begin{lemma}\label{lem:twf}
Fix graph size $n \in \N$, and target risk level $\bar{q} \in [0,1]$. Let $G=(\calV,\calE)$ be a complete graph with $| \calV | = n$. Under the Water-Filling Strategy, $\psi\wf_{\bar{q}}$, the expected goal-hitting and intentional goal-hitting times satisfy
\begin{align*}
\bE(T\wf) \leq \bE(\tih\wf) \leq \frac{1}{2\bar{q}} + \frac{1}{2} + \frac{\bar{q} \varepsilon}{2(1-\varepsilon)^2}.
\end{align*}
\end{lemma}

Hence, on any undirected connected graph $G$ under the Water-Filling Strategy we will have
\begin{align*}
\bE(L(T)) \leq \frac{1}{2\bar{q}} + \frac{1}{2} + \frac{\bar{q} \varepsilon}{2(1-\varepsilon)^2}.
\end{align*}
Crucially, the noise level $\varepsilon$ has an additive effect on the delay, in contrast to the per step overhead under forced pre-commitment.

To control the expected delay, we need to ensure that time spent outside $\neig(D)$ is small. For this purpose, we now use the Neighborhood Overlap property and derive an upper bound on the expected delay. The following result establishes that the delay under $\psi\wf$ is upper-bounded by the expected intentional goal-hitting time under the Water-Filling Strategy on a complete graph, scaled by $\frac{1}{p}$. Intuitively, the vertices of the graph $G$ are on average connected to $p$ fraction of the vertices so that under $\psi\wf$, the agent has an opportunity to attempt her goal $p$ fraction of the time. Using a coupling argument we show that $\bE(\tih\wf)$ slowed down by $\frac{1}{p}$ gives an upper bound for $\bE(T)$. The proof is given in Appendix \ref{ap:lem_nhb}.
\begin{lemma}\label{lem:nhb}
Fix graph size $n \in \N$, and target risk level $\bar{q} \in [0,1]$. Let $G = (\calV,\calE)$ be a graph with $\abs \calV \abs = n$. Suppose $G$ satisfies the Neighborhood Overlap property with parameter $p$. Under the Water-Filling Strategy, $\psi\wf_{\bar{q}}$, the expected delay satisfies 
\begin{align*}
\bE(T) \leq \frac{\bE(\tih\wf)}{p},
\end{align*}	
where $\tih\wf$ is the intentional goal-hitting time under the Water-Filling Strategy, $\psi\wf_{\bar{q}}$, on a complete graph with $n$ vertices.
\end{lemma}

Finally, we combine Lemmas \ref{lem:twf} and \ref{lem:nhb} to obtain 
\begin{align*}
\bE(T) \leq \frac{1}{p} \left( \frac{1}{2\bar{q}} + \frac{1}{2} + \frac{\bar{q} \varepsilon}{2(1-\varepsilon)^2} \right).
\end{align*}	
This concludes the proof of Proposition \ref{prop:delay_general}. \qed

\subsubsection{Prediction Risk: Proof of Proposition \ref{prop:risk_general}}\label{sec:risk_general_proof}
We now turn to the agent's prediction risk under the Water-Filling Strategies and prove Proposition \ref{prop:risk_general}. Recall that the maximal prediction risk, $q^*(\psi)$, is given by the following expression
\begin{align*}
	\sup_{\chi} q(\psi, \chi) = &  \sup_{\chi} \bP(\hat{D}_{\chi} = D ) + \bP(T > K).
\end{align*}

\iffalse
\begin{proposition}\label{prop:reg_risk_general1}
Fix a sequence of horizons $\{K_n\}_{n \in \N}$ such that $1\ll K_n \ll {n}$. Fix graph size $n \in \bN$, and target risk level $\bar{q} \in [0,1]$. Let $G = (\calV,\calE)$ be a graph with $| \calV | = n $. Under the Water-Filling Strategy, $\psi\wf_{\bar{q}}$, the maximal prediction risk satisfies
\begin{align*}
q^*(\psi\wf_{\bar{q}}) \leq \frac{\bar{q} \overline{\Delta}_G}{n} + \delta_n'
\end{align*}
where $\delta_n' = \frac{w}{K_n}+ 1 - \lt( 1- \frac{1}{\underline{\Delta}_G} \rt)^{K_n}$, $\lim\limits_{n \rightarrow \infty} \delta_n' = 0$, and $\underline{\Delta}_G$ and $\overline{\Delta}_G$ are the minimum and maximum degrees of $G$, i.e., $\underline{\Delta}_G = \min\limits_{ v \in \calV}  | \neig(v)|$ and $\overline{\Delta}_G = \max\limits_{ v \in \calV}  | \neig(v)|$. 
\end{proposition}
\fi

We first note that the probability with which the adversary succeeds due to the agent failing to reach the goal by the end of the horizon diminishes as the length of the horizon increases, i.e., $\lim\limits_{K \rightarrow \infty} \bP(T > K) = 0$. Specifically, for any agent strategy $\psi$, we have
\begin{align*}
	\bP(T > K ) \leq \frac{\bE(T)}{K} \leq \frac{w}{K},
\end{align*}
by the Markov's Inequality and the definition of the delay budget $w$. Then, for fixed $\psi$ and $w$, we conclude $\lim\limits_{K \rightarrow \infty} \bP(T > K) = 0$ since $\lim\limits_{K \rightarrow \infty} \frac{w}{K} = 0$. 
	
Next, we analyze the probability that the adversary makes a correct prediction, $\sup_{\chi} \bP(\hat{D}_{\chi} = D )$. Specifically, we will first show that $\sup_{\chi} \bP(\hat{D}_{\chi} = D )$ is upper bounded by the sum of two terms: (1) the probability that adversary predicts the intentional goal-hitting time correctly, and (2) the probability that the intentional goal-hitting time differs from the actual goal-hitting time. Then, we will prove Lemmas \ref{lem:pred_tih1} and \ref{lem:tih_t_diff1} to provide upper bounds for these terms, respectively.  

Let $\what{T}\ih$ denote the optimal Bayes estimator for $\tih$. That is, for a given trajectory realization $\bx$, let $\what{T}\ih(\bx) = \argmax_{t \geq 1} \bP(\tih=t \abs \mathbf{X}=\bx)$. Using this definition, we will first analyze the probability that the adversary succeeds by predicting correctly. Let $\mathcal{X}$ be the set of all trajectories under horizon $K$. Then, we can write
	\begin{align} \label{eq:predrisk_reg}
	\sup_{\chi} \bP(\hat{D}_{\chi} = D) = & \sum_{\bx \in \mathcal{X}} \max_{t \geq 1} \bP(T = t \abs \bold{X}=\bx ) \bP(\bold{X}=\bx) \nln
	\overset{(a)}{\leq} & \sum_{\bx \in \mathcal{X}} \max_{t \geq 1} [\bP(\tih=t \abs \bold{X}=\bx) + \bP(\tih \ne T \abs \bold{X}=\bx)] \bP(\bold{X}=\bx) \nln
	= & \left( \sum_{\bx \in \mathcal{X}} \max_{t \geq 1} \bP(\bold{X}=\bx, \tih=t) \right)  + \bP(\tih \ne T) \nln
	\overset{(b)}{=} & \bP(\what{T}\ih = \tih ) + \bP(\tih \ne T),
	\end{align}
	where $(a)$ follows from the observation that for each $t \geq 1$, the following holds
	\begin{align*}
	\bP(T = t \abs \bold{X}=\bx ) = & \bP(T=t, \tih = T \abs \bold{X}=\bx) + \bP(T=t, \tih \ne T \abs \bold{X}=\bx) \\
	\leq &  \bP(\tih=t, \tih = T \abs \bold{X}=\bx) + \bP(\tih \ne T \abs \bold{X}=\bx) \\
	\leq & \bP(\tih=t \abs \bold{X}=\bx) + \bP(\tih \ne T \abs \bold{X}=\bx),
	\end{align*}
and $(b)$ from the definition of $\what{T}\ih$ as the optimal Bayes estimator for $\tih$.

Note that the two terms on the right-hand side of \eqref{eq:predrisk_reg} correspond to the adversary's success in predicting the intentional goal-hitting time, $\bP(\what{T}\ih = \tih ) $, and the probability that the intentional goal-hitting time differs from the actual goal-hitting time, $\bP(\tih \ne T) $. 

We first state Lemma \ref{lem:pred_tih1} which will provide an upper bound on the success probability of the optimal Bayes estimator for $\tih$. To prove Lemma \ref{lem:pred_tih1}, we leverage a path counting argument and analyze the trajectories that can be generated under the Water-Filling Strategies. The proof is given in Appendix \ref{ap:pred_what}.
\begin{lemma}\label{lem:pred_tih1}
Fix graph size $n \in \N$, and target risk level $\bar{q} \in [0,1]$. Let $G = (\calV,\calE)$ be a graph with $| \calV | = n$. Under the Water-Filling Strategy, $\psi\wf_{\bar{q}}$, 
\begin{align*}
\pb(\what{T}\ih = \tih) \leq \frac{\bar{q} \overline{\Delta}_G}{n},
\end{align*}
where $\overline{\Delta}_G$ is the maximum degree of $G$, i.e., $\overline{\Delta}_G = \max\limits_{ v \in \calV}  | \neig(v)|$. 
\end{lemma}

Next, we state Lemma \ref{lem:tih_t_diff1} quantifying the probability that $\tih$ and $T$ are different under $\psi\wf$. The proof relies on a trajectory based analysis of $T$ and explicitly uses the design of the strategy $\psi\wf$. The proof is provided in Appendix \ref{ap:pred_diff}.
\begin{lemma}\label{lem:tih_t_diff1}
Fix graph size $n \in \N$, and target risk level $\bar{q} \in [0,1]$. Let $G = (\calV,\calE)$ be a graph with $| \calV | = n$. Under the Water-Filling Strategy, $\psi\wf_{\bar{q}}$, 
\begin{align*}
\pb(\tih \ne T) \leq 1-\left( 1-\frac{1}{\underline{\Delta}_G} \right)^K,
\end{align*}
where $\underline{\Delta}_G$ is the minimum degree of $G$, i.e., $\underline{\Delta}_G = \min\limits_{ v \in \calV}  | \neig(v)|$. 
\end{lemma}

We now apply Lemmas \ref{lem:pred_tih1} and \ref{lem:tih_t_diff1} to the terms in \eqref{eq:predrisk_reg}, respectively. Thus, we obtain
	\begin{align*}
	\sup_{\chi} \bP(\hat{D}_{\chi} = D) \leq & \bP(\what{T}\ih = \tih) + \bP(\tih \ne T) {\leq}  \frac{\bar{q} \overline{\Delta}_G}{n} + 1- \left( 1- \frac{1}{\underline{\Delta}_G} \right)^{K},
	\end{align*}
and, we conclude $q^*(\psi\wf) \leq \frac{\bar{q} \overline{\Delta}_G}{n} + 1- \left( 1- \frac{1}{\underline{\Delta}_G} \right)^{K} + \frac{w}{K}$. Note that we can replace $K$ with $K_n$ since we have $K_n \rightarrow \infty$ as $n \rightarrow \infty$. Finally, we define $\delta_n' = \frac{w}{K_n}+ 1 - \lt( 1- \frac{1}{\underline{\Delta}_G} \rt)^{K_n}$ and observe that $\lim\limits_{n \rightarrow \infty} \delta_n' = 0$ holds. This completes the proof of Proposition \ref{prop:risk_general}. \qed

\section{Proofs of Lower Bounds}\label{sec:pf_lower}

\subsection{Lower Bound in Theorem \ref{thm:wf_thm_infinite}} \label{sec:lb_comp}

So far, we have demonstrated that it is possible for the agent to harness the intrinsic uncertainty to achieve an additive delay overhead. In this section, we will focus on the lower bound and show that the agent always has to bear this strictly positive overhead due to uncertainty, even if the delay budget $w$ is large. That is, the effect of even a mild amount of uncertainty can never be fully naturalized.

To prove the lower bound in Theorem \ref{thm:wf_thm_infinite}, we follow three main steps. We begin by showing that when the number of vertices is sufficiently large, $\bE(T^{\psi})$ and $\bE(\tih^{\psi})$ coincide. Second, for any strategy $\psi$, we define its maximal time-only prediction risk, $\tilde{q}_{\psi}$, as its maximal prediction risk when the adversary strategy only uses the distribution of $\tih^{\psi}$ to make a prediction, and not the actual trajectory. Consequently, we conclude that the maximal prediction risk is asymptotically lower bounded by $\tilde{q}_{\psi}$. Third, we establish that among all agent strategies that can be defined on a complete graph $G$ with $n$ vertices, the strategy $\psi\wf_{\bar{q}} $ minimizes $\bE(\tih^{\psi})$ subject to the constraint that $\tilde{q}_{\psi}$ is at most $\bar{q}$. Thus, for any delay budget, we can assert that the minimum target risk level that can be attained by a Water-Filling Strategy with this budget is a lower bound on the maximal time-only prediction risk for any strategy with that budget. Note that we can also conclude the Water-Filling Strategies are optimal. Precisely, given any target risk level they minimize the delay and given any delay budget they achieve minimal prediction risk.

Lemma \ref{lem3} states that for sufficiently large complete graphs, $\bE(T^{\psi})$ and $\bE(\tih^{\psi})$ are equivalent so that we can use $\bE(\tih^{\psi})$ as a proxy for $\bE(T^{\psi})$. The proof is in Appendix \ref{ap:lem3}. 

\begin{lemma}\label{lem3}
Fix graph size $n \in \N$. Let $G=(\calV,\calE)$ be a complete graph with $| \calV | = n$. Let $\psi$ be an agent strategy on $G$. Under $\psi$, the expected goal-hitting and intentional goal-hitting times satisfy
	\begin{align*}
	\bE(\tih^{\psi}) \leq \bE(T^{\psi}) + \sigma_n,
	\end{align*}
	where $\sigma_n =\left( 1 - \left( 1- \frac{1}{n} \right)^{K_n} \right) \frac{K_n(K_n+1)}{2}  $, and $\lim\limits_{n \rightarrow \infty} \sigma_n = 0$.
\end{lemma}
Note that Lemma \ref{lem3} allows us to state $\bE(T^{\psi}) \leq \bE(\tih^{\psi}) \leq \bE(T^{\psi})+\sigma_n$, since the first inequality always holds. Thus, for sufficiently large graphs $\bE(\tih^{\psi})$ is a good approximation for $\bE(T^{\psi})$.

Now, we analyze the maximal prediction risk. For any strategy $\psi$, define its maximal time-only prediction risk,
\begin{align*}
\tilde{q}_{\psi} = \max_{t \in \bN} \bP(\tih^{\psi} = t).
\end{align*} 
The next lemma shows that the maximal prediction risk under any strategy $\psi$ is lower-bounded by its maximal time-only prediction risk, $\tilde{q}_{\psi}$, as the number of vertices increases. For sufficiently large graphs, this allows us to use the maximal time-only prediction risk, $\tilde{q}_{\psi}$ as a lower bound on $q^*(\psi)$. The proof is given in Appendix \ref{ap:lem1}. 
\begin{lemma}\label{lem1}
Fix graph size $n \in \N$. Let $G=(\calV,\calE)$ be a complete graph with $| \calV | = n$. Let $\psi$ be an agent strategy on $G$. Under $\psi$, the maximal prediction risk satisfies
	\begin{align*}
	q^*(\psi) \geq (1-\bar{\delta}_n) \tilde{q}_{\psi},
	\end{align*}
	where $\bar{\delta}_n = 1- \lt( 1- \frac{1}{n} \rt)^{K_n}$, and $\lim\limits_{n \rightarrow \infty} \bar{\delta}_n = 0$.
\end{lemma}
We now use Lemmas \ref{lem3} and \ref{lem1} to derive the lower bound. Recall that $\Phi_w$ denotes the class of strategies under which the expected intentional goal-hitting time is at most $w$. Define $\hat{q}(w)$ to be the minimum $\tilde{q}_{\psi}$ value attained by some strategy in this class, where $\tilde{q}_{\psi} =\max\limits_{t \in \bN} \bP(\tih^{\psi} = t) $ as before. That is, let
\begin{align*}
\hat{q}(w) = \inf_{\psi \in \Phi_w} \tilde{q}_{\psi}.
\end{align*}

For any $\psi$ such that $\bE(T^{\psi}) \leq w$, Lemma \ref{lem3} implies $\bE(\tih^{\psi}) \leq w + \sigma_n$ and consequently, $\Psi_w \subseteq \Phi_{w+\sigma_n}$. From this observation, for any $\psi$, we can write
\begin{align}\label{hatq}
\inf_{\psi \in \Psi_w} \tilde{q}_{\psi} \geq \inf_{\psi \in \Phi_{w+\sigma_n}} \tilde{q}_{\psi} = \hat{q}(w+\sigma_n) .
\end{align}
Then, we obtain
\begin{align}\label{eqnstar}
\calQ(w) \overset{(a)}{=} \inf_{\psi \in \Psi_w} q^*(\psi) \overset{(b)}{\geq} \inf_{\psi \in \Psi_w} (1-\bar{\delta}_n) \tilde{q}_{\psi} \overset{(c)}{\geq} (1-\bar{\delta}_n) \inf_{\psi \in \Psi_w} \tilde{q}_{\psi} \overset{(d)}{\geq} (1-\bar{\delta}_n) \hat{q}(w + \sigma_n) ,
\end{align}
where $(a)$, $(b)$ and $(d)$ follow from the definition of $\calQ(w)$, Lemma \ref{lem1}, and Eq.~\eqref{hatq}, respectively. Finally, $(c)$ is due to the observation that $1-\bar{\delta}_n \geq 0$ and is independent of $\psi$.

To conclude the lower bound from Eq.~\eqref{eqnstar}, we need the following result on the optimality of $\psi\wf$. Lemma \ref{wf_optimal} establishes that the Water-Filling Strategy actually achieves minimal expected intentional goal-hitting time subject to the maximal time-only prediction risk being upper-bounded by the target risk level. The proof can be found in Appendix \ref{app:optimal}.
\begin{lemma}\label{wf_optimal}
	Fix graph size $n \in \bN$ and $\bar{q} \in [0,1]$. Let $G=(\calV,\calE)$ be a complete graph with $|\calV|=n$. The Water-Filling Strategy with target risk level $\bar{q}$ solves the optimization problem,
	\begin{align*}
	\min_{\psi \in \Psi } \mbox{ } \bE(\tih^{\psi}) \mbox{ s.t. } \max_{t \in \bN} \mbox{ } \bP(\tih^{\psi}=t) \leq \bar{q},
	\end{align*}
	where $\Psi$ denotes the set of all agent strategies defined on $G$.
\end{lemma}

Recall that $\bar{q}(w)$ is the minimum target risk level $\bar{q}$ that can be attained by a Water-Filling Strategy with expected intentional goal-hitting time at most $w$. Then, by Lemma \ref{lem3}, we have that
\begin{align*}
\bar{q}(w) = \inf \{ \bar{q}: \exists \psi^{wf}_{\bar{q}} \in \Phi_w \} \leq \inf\{ \bar{q}: \exists \psi^{wf}_{\bar{q}} \in \Psi_{w-\sigma_n} \},
\end{align*}
since $\Psi_{w-\sigma_n} \subseteq \Phi_w$. In addition, by Lemma \ref{wf_optimal}, if under any strategy $\psi$ we have $\bE(\tih^{\psi}) < \bE(\tih\wf)$, then it must be the case that $\tilde{q}_{\psi} > \tilde{q}_{\psi\wf}$. Therefore, using Lemma \ref{wf_optimal} and Eq.~\eqref{eqnstar}, we obtain
\begin{align}\label{lower_wf}
\calQ(w) \geq (1-\bar{\delta}_n) \bar{q}(w+\sigma_n).
\end{align}

To complete the lower bound, we recall the derivation of $\bar{q}(w)$ and again use the Taylor expansion of $\sqrt{1-x}$ for $x <1$. Consequently, letting $\alpha^{\varepsilon}(w) = \frac{\varepsilon}{(2w-1)(1-\varepsilon)^2}$, we obtain
\begin{align}\label{eqn:qbar-}
\frac{1}{2w-1-\alpha^{\varepsilon}(w)} \leq \bar{q}(w).
\end{align}
Then, combining Eqs.~\eqref{lower_wf} and \eqref{eqn:qbar-} gives 
\begin{align*}
\calQ(w) \geq \frac{1-\bar{\delta}_n}{2w-1-\alpha^{\varepsilon}(w+\sigma_n)+\sigma_n} \overset{(a)}{\geq} \frac{1-\bar{\delta}_n}{2w-1-\alpha^{\varepsilon}(w)+\sigma_n' } \overset{(b)}{\geq} \frac{1}{2w-1-\alpha^{\varepsilon}(w)} - \tilde{\delta}_n,
\end{align*}
where $(a)$ and $(b)$ are obtained by the Taylor expansion of $\alpha^\varepsilon(w)$ and $\frac{1}{x}$, respectively, and we let
\begin{align*}
\sigma'_n = \sigma_n \lt( 1+ \frac{2}{(2w-1)^2} \cdot \frac{\varepsilon}{(1-\varepsilon)^2} \rt), \mbox{and } \tilde{\delta}_n =  \frac{\bar{\delta}_n}{2w-1-\alpha^{\varepsilon}(w)+\sigma_n' } + \frac{\sigma_n'}{(2w-1-\alpha^\varepsilon(w))^2}.
\end{align*}
Hence, we observe $\lim\limits_{n \rightarrow \infty} \tilde{\delta}_n =0$ and conclude the lower bound:
\begin{align}\label{eq:comp_lower_tay}
\calQ(w) \geq  \frac{1}{2w-1-\alpha^{\varepsilon}(w)} - \tilde{\delta}_n.
\end{align}
Finally, we let $\delta_n = \max \{ \tilde{\delta}_n, \delta_n' \}$. Then, combining Eqs.~\eqref{eq:comp_upper_tay} and \eqref{eq:comp_lower_tay}, we get
\begin{align}\label{eq:comp_bds}
\frac{1}{2w-1-\alpha^{\varepsilon}(w)} - {\delta}_n \leq \calQ(w) \leq  \frac{1}{2w-1 - \alpha^{\varepsilon}(w) - \beta^{\varepsilon}(w)} + \delta_n.
\end{align}
This completes the proof of Theorem \ref{thm:wf_thm_infinite}. \hfill \qed

Observe that Eq.~\eqref{eq:comp_bds} yields an asymptotically tight characterization of the minimax prediction risk in the context of complete graphs. Specifically, letting $n \rightarrow \infty$, we have
\begin{align*}
 \frac{1}{2w-1-\alpha^{\varepsilon}(w)} \leq \calQ(w) \leq \frac{1}{2w-1 - \alpha^{\varepsilon}(w) - \beta^{\varepsilon}(w)}.
\end{align*}

\subsection{Lower Bounds in Theorems \ref{thm:reg} and \ref{thm:networkdesign}}

To prove the lower bounds in Theorems \ref{thm:reg} and \ref{thm:networkdesign}, we apply the lower bound in Theorem 1 of \cite{tsitsiklis2018delay}. We first note that the addition of noise makes the strategy space of the agent more restrictive than the deterministic setting $(\varepsilon = 0)$, since in the latter case the agent can also simulate the noise in her state transition if necessary. Moreover, we model an offline adversary rather than an online adversary who needs to make a correct prediction before the agent reaches her goal vertex. Since the offline adversary is more powerful than its online counterpart, the prediction risk against an offline adversary will be greater than that against an online adversary. As a result, the prediction risk lower bound in \cite{tsitsiklis2018delay} for the deterministic setting with an online adversary will still hold in our case. The lower bounds in Theorems \ref{thm:reg} and \ref{thm:networkdesign} thus follow from the lower bound in Theorem 1 of \cite{tsitsiklis2018delay} and we conclude the proof.

For completeness, we state Theorem 1 of \cite{tsitsiklis2018delay} below and provide its proof in Appendix \ref{ap:lower_bound}. The proof examines a simple adversary strategy that guarantees a prediction risk of at least $\frac{1}{2w+1}$. Intuitively, for the delay to be at most $w$, the distribution of the goal-hitting time $T$ has to be concentrated. This implies that, under any agent strategy $\psi$ with delay budget $w$, there exists a time $t(\psi) \in \N$ such that the goal-hitting time $T$ is equal to $t(\psi)$ with probability at least $\frac{1}{2w+1}$. Therefore, the adversary can predict the goal vertex to be the agent state at time $t(\psi)$ and achieve a prediction risk of at least $\frac{1}{2w+1}$. 
\begin{theorem}[Theorem 1 of \cite{tsitsiklis2018delay}] \label{thm:tx_lowerbound}
Fix $n \in \N$, $\varepsilon = 0 $ and let $G = (\calV, \calE)$ be a connected undirected graph with $n$ vertices. Fix $x_0 \in \calV$ and $w \in \{1, \cdots, n \}$. Then, the minimax prediction risk under a time budget of $w$ satisfies
\begin{align}
\frac{1}{2w+1} \leq \calQ(w).
\end{align}
\end{theorem}

%\tr{KX: For completeness, can we add a precise statment of the result we are using here? Also, the model is a bit different from Tsitsiklis 2018, since we do offline prediction. Might be good to include a short proof in the appendix, and point out some minor changes from the prior work. State that the proof is essentially the same, but we include it here for completeness. }

%\bibliographystyle{abbrv}
\bibliographystyle{apa}
\bibliography{references.bib}

\newpage 

\ifx \useplain\undefined

\begin{center}
\large {\bf Online Companion for ``Anonymous Stochastic Routing''}
\end{center}

\begin{APPENDICES}
\else
\appendix
\fi

\normalsize

\section{Example Instance}
\label{ap:example}
	Suppose that $G$ is as in Figure \ref{fig:example}. The agent starts at vertex $1$ and her secret goal is $D=5$. The agent's trajectory is $1 \rightarrow 2 \rightarrow 3 \rightarrow 5$ whereas her strategy is given by the sequence of actions $a_1=3$, $a_2=3$, $a_3=5$. The sequence of Bernoulli trials is given by $B_1=B_3=1$ and $B_2=0$. Note that the state of the agent at time $t=2$ being $2$ even though her action was vertex $3$ exemplifies a case in which the agent ends up at a different vertex due to noise. Assume that the adversary predicts that the goal vertex is $3$, i.e., $\hat{D}=3$. Since $\hat{D} \ne D$, the agent succeeds with a goal-hitting time of $T=4$. Nonetheless, if the adversary had made a prediction $\hat{D} =5$, then he would have succeeded.

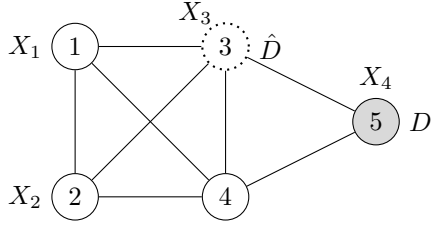
\begin{figure}[h]
	\vspace{-1em}
	\centering
	\begin{tikzpicture}[node distance=2cm]
	\node (v1) [fcn, label=left:{$X_1$}] {1};
	\node (v2) [fcn, below of=v1, label=left:{$X_2$}] {2};
	\node [dotted, thick] (v3) [fcn, right of=v1, label={[xshift=-0.4cm, yshift=-0.15cm]$X_3$}, label={right:$\hat{D}$}] {3};
	\node (v4) [fcn, right of=v2] {4};
	\node (v5) [fcn, right of=v3, yshift=-1cm, fill=gray!30!white,, label=above:{$X_4$}, label=right:{$D$}] {5};
	
	\draw (v1) -- (v3);
	\draw (v2) -- (v4);
	\draw (v1) -- (v4);
	\draw (v3) -- (v5);
	\draw (v1) -- (v2);
	\draw (v3) -- (v2);
	\draw (v3) -- (v4);
	\draw (v4) -- (v5);
	\end{tikzpicture}
	\caption{An example of the Anonymous Stochastic Routing model.}
	\label{fig:example}
\vspace{-1em}
\end{figure}

\section{Sample $k$-Clique Graph}
\label{ap:clique_fig}
\begin{figure}[H]
\centering
		\includegraphics[scale=0.5]{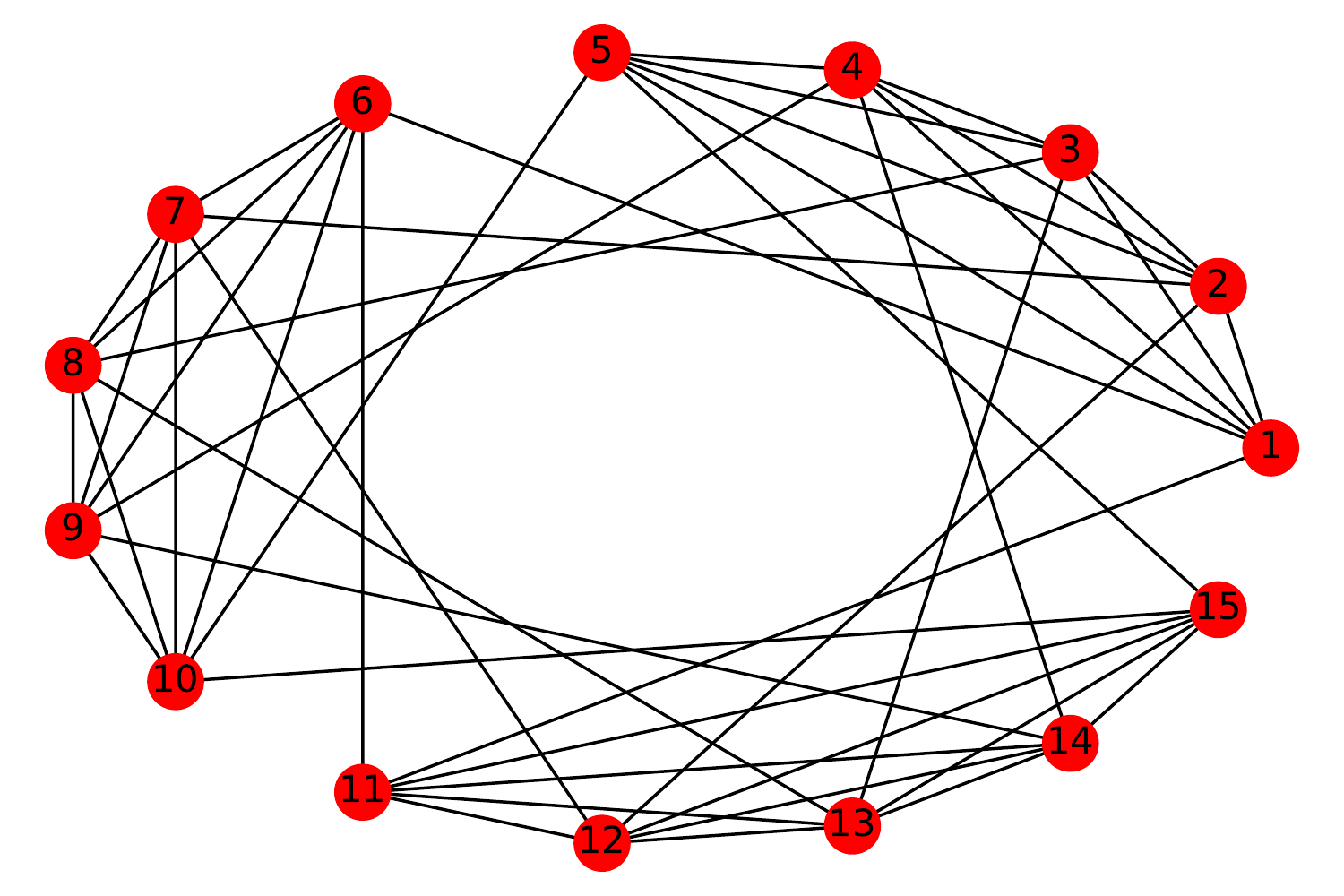}
	\caption{A $3$-clique graph with $5$ vertices in each clique. For simplicity, self edges are not drawn. Sets of vertices $\{1,2,3,4,5\}$, $\{6,7,8,9,10\}$ and $\{11,12,13,14,15\}$ correspond to the three cliques.}
	\label{fig:clique}
\end{figure}

\section{Proofs}
\subsection{Proof of Lemma \ref{lem:erdos}} \label{ap:erdos} %changed
\bpf To prove that a graph generated using the Erd\H{o}s-R\'{e}nyi random graph model belongs to the family $\calG_n(p)$ with high probability, we will examine each property in Definition \ref{def:fam} separately. We will begin by verifying the upper bound on the degree using the Chernoff bound. Then, we will repeat similar arguments to prove the lower bound on the degree and the Neighborhood Overlap property.
	
	\textit{Degree upper bound.} For each vertex $v \in \calV$, define the event $A_v = \{ |\neig(v) | \geq pn(1+\gamma_n) \}$ to represent whether the degree of $v$ is greater than $pn(1+\gamma_n)$. Since $|\neig(v)| \sim \mbox{Binomial}(n,p)$ under the Erd\H{o}s-R\'{e}nyi random graph model, using the Chernoff bound with $\gamma_n \geq 0$ we obtain
	\begin{align}\label{eq:cher_max}
	\bP(A_v) = \bP( |\neig(v) | \geq pn(1+\gamma_n)) \leq \exp\left( {\frac{-\gamma_n^2 pn}{2+\gamma_n}} \right).
	\end{align}
	Then, the probability that the maximum degree is at most $pn(1+\gamma_n)$ can be written as
	\begin{align}
	\bP \left( \max_{v \in \calV}  |\neig(v)| \leq pn(1+\gamma_n) \right) = & 1- \bP \left( \bigcup_{v \in \calV} A_v \right) \nln
	 \geq & 1- \sum_{v \in \calV} \bP(A_v)  \nln
	 \overset{(a)}{\geq} & 1 - n \exp\left( {\frac{-\gamma_n^2 pn}{2+\gamma_n}} \right), \label{eq:erdos_up}
	\end{align}
	where $(a)$ uses \eqref{eq:cher_max}. Recall that $\gamma_n$ satisfies $\gamma_n \gg \sqrt{{\log n}/{n}} $ and $\lim\limits_{n \rightarrow \infty} \gamma_n =0$. This implies that $\exp(n \gamma_n^2) \gg n $ and consequently we obtain $\lim\limits_{n \rightarrow \infty} n \exp\left( {\frac{-\gamma_n^2 pn}{2+\gamma_n}} \right) = 0$. Hence, we conclude that the degrees of all vertices are upper-bounded by $pn(1+\gamma_n) $ with high probability as $n \rightarrow \infty$, i.e., $\lim\limits_{n \rightarrow \infty} \bP \left( \max\limits_{v \in \calV}  |\neig(v)| \leq pn(1+\gamma_n) \right) = 1$.
	
	\textit{Degree lower bound.} As in the proof for the upper bound, for any vertex $v \in \calV$ define the event $B_v = \{ |\neig(v) | \leq pn(1-\gamma_n) \}$ to represent whether the degree of $v$ is less than $pn(1-\gamma_n)$. Using the Chernoff bound with $\gamma_n \in [0,1]$ we get
	\begin{align}\label{eq:cher_min}
	\bP(B_v) = \bP( |\neig(v) | \leq pn(1-\gamma_n)) \leq \exp\left( {\frac{-\gamma_n^2 pn}{2}} \right).
	\end{align}
	Similarly, we write the probability that the minimum degree is at least $pn(1-\gamma_n)$ as
	\begin{align}
	\bP \left( \min_{v \in \calV}  |\neig(v)| \geq pn(1-\gamma_n) \right) = & 1- \bP \left( \bigcup_{v \in \calV} B_v \right) \nln
	 \geq & 1- \sum_{v \in \calV} \bP(B_v) \nln
	  \overset{(a)}{\geq} & 1 - n \exp\left( {\frac{-\gamma_n^2 pn}{2}} \right), \label{eq:erdos_low}
	\end{align}
	where $(a)$ follows from \eqref{eq:cher_min}. 
	
	Once again we recall $\gamma_n \gg \sqrt{{\log n}/{n}} $ and $\lim\limits_{n \rightarrow \infty} \gamma_n =0$. Then, we can write $\exp(n \gamma_n^2) \gg n $ and obtain $\lim\limits_{n \rightarrow \infty} n \exp\left( {\frac{-\gamma_n^2 pn}{2}} \right) = 0$. Thus, we conclude that the degrees of all vertices are lower-bounded by $pn(1-\gamma_n) $ with high probability as $n \rightarrow \infty$, i.e., 
	\begin{equation}
	\lim\limits_{n \rightarrow \infty} \bP \left( \min\limits_{v \in \calV}  |\neig(v)| \geq pn(1-\gamma_n) \right) = 1.
	\end{equation}
	
\textit{Neighborhood Overlap.} For each pair of vertices $u, v \in \calV$, define the event $C_{uv} = \left\lbrace  \frac{| \neig(u) \cap \neig(v) |}{| \neig(v) |} \leq p(1-\gamma_n) \right\rbrace$. For simplicity of notation, denote the random variables $\abs \neig(u) \cap \neig(v) \abs$ and $\abs \neig(v) \abs$ by $S_{uv}$ and $S_v$, respectively. Note that $S_v \sim \mbox{Binomial}(n,p)$, and $S_{uv} \sim \mbox{Binomial}(n,p^2)$. The probability that the Neighborhood Overlap property holds can be written as 
\begin{align}
	& \bP \left(\frac{\abs \neig(u) \cap \neig(v) \abs}{\abs \neig(v) \abs} \geq p(1-\gamma_n), \forall u,v \in \calV \right) =  1- \bP \left( \bigcup_{u,v \in \calV} C_{uv}  \right)  \geq  1- \sum_{u,v \in \calV} \bP(C_{uv}).  \label{eq:nhb_prob}
	\end{align}
Next, we will derive an upper bound on $\bP(C_{uv})$. To this end, let $\eta \in [0, 1]$ and for each vertex $v \in \calV$, define the event $D_{v} = \{ S_v < np(1+\eta) \}$. Applying the Chernoff bound with $\eta \geq 0$  for $S_v$, we obtain $\bP(D_v^c) = 1- \bP(D_v) \leq \exp \lt( \frac{-\eta^2 p n}{2 +\eta} \rt)$.

Similarly, for each pair of vertices $u,v \in \calV$, define the event $D_{uv} = \{ S_{uv} > np^2(1-\eta) \}$. Then, the Chernoff bound applied to $S_{uv}$ with $\eta \in [0, 1]$ implies $\bP(D_{uv}^c) = 1- \bP(D_{uv}) \leq \exp \lt( \frac{-\eta^2 np^2}{2} \rt)$.

Now, observe that $D_v \cap D_{uv} \subseteq \left\lbrace \frac{S_{uv}}{S_v} > p \lt( \frac{1-\eta}{1+\eta} \rt) \right\rbrace$ holds so that we write $\bP(D_v \cap D_{uv}) \leq \bP \lt(\frac{S_{uv}}{S_v} > p \lt(\frac{1-\eta}{1+\eta}\rt) \rt) $. Equivalently, we have
\begin{align*}
\bP \lt(\frac{S_{uv}}{S_v} \leq p \lt(\frac{1-\eta}{1+\eta}\rt) \rt)  \leq & 1- \bP(D_v \cap D_{uv}) = \bP(D_v^c \cup D_{uv}^c) \leq \bP(D_v^c) + \bP(D_{uv}^c).
\end{align*}
Since $\eta \in [0,1]$, we further have 
\begin{align*}
\bP\lt(\frac{S_{uv}}{S_v} \leq p \lt(1-2\eta\rt) \rt) \leq \bP \lt(\frac{S_{uv}}{S_v} \leq p \lt(\frac{1-\eta}{1+\eta}\rt) \rt).
\end{align*}
Hence, choosing $\eta=\frac{\gamma_n}{2}$, we obtain
\begin{align}
\bP(C_{uv}) \leq \bP(D_v^c) + \bP(D_{uv}^c) \leq \exp \lt( \frac{-\gamma_n^2 p n}{8 + 2\gamma_n} \rt) +  \exp \lt( \frac{-\gamma_n^2 np^2}{8} \rt). \label{eq:nhb_bound}
\end{align}
Lastly, combining Eqs.~\eqref{eq:nhb_prob} and \eqref{eq:nhb_bound} yields
\begin{align}
& \bP \left(\frac{\abs \neig(u) \cap \neig(v) \abs}{\abs \neig(v) \abs} \geq p(1-\gamma_n), \forall u,v \in \calV \right) \geq 1- \sum_{u,v \in \calV} \bP(C_{uv}) \nln
& \geq 1- n^2 \lt( \exp \lt( \frac{-\gamma_n^2 p n}{8 + 2\gamma_n} \rt) +  \exp \lt( \frac{-\gamma_n^2 np^2}{8} \rt) \rt). \label{eq:erdos_neig}
\end{align}

We now observe that the assumptions $\gamma_n \gg \sqrt{{\log n}/{n}} $ and $\lim\limits_{n \rightarrow \infty} \gamma_n =0$ imply $\exp(n \gamma_n^2) \gg n^2$. Consequently, we can conclude $\lim\limits_{n \rightarrow \infty} n^2 \exp \lt( \frac{-\gamma_n^2 p n}{8 + 2\gamma_n} \rt) =0$ and $\lim\limits_{n \rightarrow \infty} n^2  \exp \lt( \frac{-\gamma_n^2 np^2}{8} \rt)   =0$. Therefore, we obtain $\lim\limits_{n \rightarrow \infty} \bP \left( \frac{| \neig(u) \cap \neig(v) |}{| \neig(v) |} \geq p(1-\gamma_n), \forall u,v \in \calV  \right) = 1$ and conclude that Neighborhood Overlap property holds with high probability as $n \rightarrow \infty$. 
Finally, we combine \eqref{eq:erdos_up}, \eqref{eq:erdos_low}, and \eqref{eq:erdos_neig} to obtain
\begin{align*}
& \bP(G \in \calG_n) \geq 1-\bP \left( \bigcup_{v \in \calV} A_{v}+ B_v  \right)- \bP \left( \bigcup_{u,v \in \calV} C_{uv}  \right) \nln
& \geq 1-  n \left( \exp\left( {\frac{-\gamma_n^2 pn}{2+\gamma_n}} \right) +\exp\left( {\frac{-\gamma_n^2 pn}{2}} \right) \right) -n^2 \lt( \exp \lt( \frac{-\gamma_n^2 p n}{8 + 2\gamma_n} \rt) +  \exp \lt( \frac{-\gamma_n^2 np^2}{8} \rt) \rt) \nln
&= 1- \theta_n.
\end{align*}	
	We observe that $\lim\limits_{n \rightarrow \infty} \theta_n =0$ follows. This completes the proof that under the Erd\H{o}s-R\'{e}nyi model, the degrees of vertices are upper- and lower-bounded by $pn(1+\gamma_n) $ and $pn(1-\gamma_n)$, and the Neighborhood Overlap property holds with high probability as $n \rightarrow \infty$.
\qed

\subsection{Proof of Lemma \ref{lem:twf}}
\label{ap:lem_twf}
\bpf
Suppose $G$ is a complete graph. We will compute the expected intentional goal-hitting time on $G$ under the assumption that the horizon is infinite. Then, when the horizon is finite, i.e., $K<\infty$, we can simply truncate the goal-hitting time at time $K$ which will result in a smaller expected delay.

	For simplicity of notation, let us define $q_t = \mathbb{P}(\tih=t)$ and note that $q_0=0$. In order to calculate the expected value of the intentional goal-hitting time, we condition on the event $\{\tih \geq t \}$ and obtain a recursive equation on $q_t$:
	\begin{align*}
	q_t = &  \mathbb{P}(\tih = t \,\big{|}\, \tih \geq t ) \mathbb{P}(\tih \geq t) + \mathbb{P}(\tih= t \,\big{|}\, \tih< t ) \mathbb{P}(\tih< t) \\
	= & \mathbb{P}(\tih=t \,\big{|}\, \tih \geq t ) \left( 1 - \sum_{i=1}^{t-1} {q}_i \right) + 0 \left(\sum_{i=1}^{t-1} {q}_i \right) \\
	\overset{(a)}{=} & {p}_{t-1} (1- \varepsilon) \left( 1 - \sum_{i=1}^{t-1} {q}_i \right),
	\end{align*}
	where $(a)$ follows from the following observation:
	\begin{align*}
	\mathbb{P}(\tih=t \,\big{|}\, \tih \geq t ) = \mathbb{P} (\bar{F}_t=1 \,\big{|}\, \tih \geq t) = \mathbb{P} (\bar{a}_{t-1}= \bar{a}^G , X_t=D, B_{t-1}=1 \,\big{|}\, \tih \geq t) = p_{t-1}(1-\varepsilon).
	\end{align*}
	Solving recursively, we obtain:
	\begin{align*}
	q_t = \begin{cases}
	\bar{q}, & \mbox{ if } 1 \leq t < {t}^*, \\
	\varepsilon^{t-{t}^*+1} \left( 1- \varepsilon \right) \left( 1 - (t^* -1)\bar{q} \right), & \mbox{ otherwise.}
	\end{cases}
	\end{align*}
	Finally, we compute the expected intentional goal-hitting time under $\psi^{wf}_{\bar{q}}$:
	\begin{align*}
	\bE( \tih ) = & \mathbb{E}(\tih \,\big{|}\, \tih < {t}^* ) \mathbb{P}(\tih < {t}^*) + \mathbb{E}(\tih \,\big{|}\, \tih \geq {t}^* ) \mathbb{P}(\tih \geq {t}^*) \\
	= & \frac{{t}^* -1 + 1 }{2} \left( ({t}^*-1)\bar{q}\right) + \left( {t}^* -1 + \frac{1}{1-\varepsilon} \right) \left( 1- ({t}^*-1)\bar{q} \right) \\
	= & \frac{\bar{q}}{2} \left( \frac{1}{\bar{q}} - \frac{\varepsilon}{1-\varepsilon} \right) \left( \frac{1}{\bar{q}} - \frac{1}{1-\varepsilon} \right) + \frac{1}{\bar{q}} \left( 1- \bar{q} \left( \frac{1}{\bar{q}} - \frac{1}{1-\varepsilon} \right) \right) \\
	= & \frac{1}{2\bar{q}} + \frac{1}{2} + \frac{\bar{q} \varepsilon}{2(1-\varepsilon)^2} .
	\end{align*}	
	When $K< \infty$, we will have $\bE(T\wf) \leq \bE(\tih\wf) \leq \frac{1}{2\bar{q}} + \frac{1}{2} + \frac{\bar{q} \varepsilon}{2(1-\varepsilon)^2} $. This completes the proof of Lemma \ref{lem:twf}. 
 \qed

\subsection{Proof of Lemma \ref{lem:nhb}} \label{ap:lem_nhb}
\bpf
Throughout the proof we will use the superscripts p and wf to denote quantities on a graph $G$ such that Neighborhood Overlap property holds with parameter $p$ and on a complete graph, respectively.

To prove Lemma \ref{lem:nhb}, we will derive the expected time between two consecutive trials and express the index of the successful trial in terms of $\tih\wf$ on a complete graph. For any $t \geq 1$, define $N_t = \sum_{i=1}^{t} \mathbb{I}(X_i \in \neig(D))$ to be the number of times the agent has been in the neighborhood of her goal up to time $t$. Let $U_n = \sup \{ t\geq 1: N_t < n \}$ denote the time of the $n^{th}$ occurrence of this event. Under the Water-Filling Strategy, the system dynamics can equivalently be generated as follows:
	\begin{enumerate}
		\item[(1)] Draw the goal, i.e., $D \sim \mbox{Unif}(\calV)$.
		\item[(2)] Draw the number of trials needed before the agent succeeds for the first time, $R$, where the attempt probabilities are given by those under the Water-Filling Strategy on a complete graph with $n$ vertices, $\psi\wf_{\bar{q}}$. Note that $R = \tih\wf-1$.
		\item[(3)] Generate a random walk of length $K$ on $G$, $\{X_1, ..., X_K\}$. Then, set $X_{K+1} = D$.
		\item[(4)] If $U_R+1 \leq K$, set $X_{U_R+1}=D$ so that the agent indeed succeeds at her $R^{\mbox{\scriptsize th}}$ attempt, i.e., $\tih\p = U_R+1$. Replace $\{X_{U_R+2, ..., X_K}\}$ with a random walk starting at $D$.
	\end{enumerate}
	
	Then, the observation $\tih\p = U_R+1$ implies
	\begin{align}\label{eq:tih_u}
	\bE(\tih\p-1) = \sum_{r=1}^{\infty} \bE(U_r \abs R=r) \bP(R=r) \overset{(a)}{=} \sum_{r=1}^{\infty} \bE(U_r) \bP(R=r) ,
	\end{align}
	where $(a)$ follows from the independence of $U_r$ and $R$.
	
	Letting $Y_i = U_i-U_{i-1}$ and $U_0=0$, we can now write $U_r = \sum_{i=1}^{r} Y_i$. To analyze $\bE(U_r)$, we state Lemma \ref{lem:y_coupling} whose proof is given in Appendix \ref{ap:lem_returntime}.
	
	\begin{lemma}\label{lem:y_coupling}
		Let $N_t = \sum_{i=1}^{t}  \mathbb{I}(X_i \in \neig(D))$ where $\{X_t\}_{t \in \bN}$ is the trajectory generated under $\psi\wf$ on $G$. Define $U_i = \sup\{t \geq 1: N_t < i\}$ and $Y_i = U_i - U_{i-1}$ for $i \geq 1$. Then, $\bE(Y_i) \leq \frac{1}{p}$.
	\end{lemma}
	
	Using Eq.~\eqref{eq:tih_u} and applying Lemma \ref{lem:y_coupling} in step $(a)$, we obtain:
	\begin{align*}
	\bE(\tih\p-1) = \sum_{r=1}^{\infty} \bE \left(\sum_{i=1}^r Y_i \right) \bP(R=r)  \overset{(a)}{\leq} \sum_{r=1}^{\infty} \frac{r\bP(R=r)}{p}  = \frac{\bE(R)}{p} \overset{(b)}{=} \frac{\bE(\tih\wf)-1}{p} ,
	\end{align*}
	where $(b)$ is due to $\bE(R) = \bE(\tih\wf) -1$. Finally, since $p \leq 1$, we get
	\begin{align*}
 \bE(T\p) \leq \bE(\tih\p) \leq \frac{\bE(\tih\wf)}{p} - \frac{1}{p} + 1 \leq \frac{\bE(\tih\wf)}{p}.
	\end{align*}
	Thus, we conclude the proof of Lemma \ref{lem:nhb}. 
\qed

\subsubsection{Proof of Lemma \ref{lem:y_coupling}} \label{ap:lem_returntime}
\bpf
	Let $\tilde{Y}$ be a geometric random variable with success probability $p$, drawn independently from the rest of the system. We will now show that for all $t \geq 1$, $\bP(Y_1 > t) \leq \bP(\tilde{Y} > t)$ holds so that $Y_1$ is stochastically dominated by $\tilde{Y}$. This will give the upper bound on $\bE(Y_1)$.
	
	First, note that $Y_1$ corresponds to the first hitting time of the random walk to the subset $\neig(D)$ and $Y_i$ the $i^{th}$ return time to $\neig(D)$. We begin with $t=1$. Write $Z_t = \mathbb{I}(X_t \in \neig(D))$ for $t \geq 1$. Then,
	\begin{align*}
	\bP(Y_1 > 1) = \bP(Z_1 =0) = 1- \frac{\abs \neig(X_0) \cap \neig(D) \abs}{\abs \neig(X_0) \abs} \leq 1- p = \bP(\tilde{Y} > 1),
	\end{align*}
	by the Neighborhood Overlap property of the graph $G$. Next, suppose $\bP(Y_1 > t) \leq \bP(\tilde{Y} >t)$ holds for some $t \in \bN$. Then, we have
	\begin{align*}
	 \bP(Y_1 > t+1) = & \bP(Y_1 > t) \bP(Y_1 >t+1 \abs Y_1 >t) \\
	= & \bP(Y_1 > t) \bP(Z_{t+1} =0 \abs \{Z_i\}_{i=1}^{t} =0) \\
	= & \bP(Y_1 > t) \left( \sum_{v \in \neig(X_{t-1})} \bP(Z_{t+1} = 0 \abs \{Z_i\}_{i=1}^t =0, X_t=v) \bP(X_t =v \abs \{Z_i\}_{i=1}^t =0) \right) \\
	\overset{(a)}{=} & \bP(Y_1 >t) \left( \sum_{v \in \neig(X_{t-1})} \bP(Z_{t+1}= 0 \abs X_t=v) \bP(X_t =v \abs  \{Z_i\}_{i=1}^t =0) \right) \\
	\leq & \bP(Y_1 >t) \left( \sum_{v \in \neig(X_{t-1})} \left( \max_{v \in \neig(X_{t-1})} \bP(Z_{t+1} = 0 \abs X_t=v) \right) \bP(X_t =v \abs  \{Z_i\}_{i=1}^t =0) \right) \\
	\leq & \bP(Y_1>t) \left( \max_{v \in \neig(X_{t-1})} \bP(Z_{t+1} = 0 \abs X_t=v) \right)  \\
	\overset{(b)}{=} & \bP(Y_1>t) \left( 1- \frac{\abs \neig(\bar{v}) \cap \neig(D) \abs}{\abs \neig(\bar{v}) \abs} \right) \\
	 \overset{(c)}{\leq} & (1- p )^{t+1} \\
	 = & \bP(\tilde{Y} > t+1),
	\end{align*}
	where $(a)$ is due to the Markov property, and $(b)$ uses the definition $\bar{v} = \argmax\{v \in \neig(X_{t-1}): \bP(Z_{t+1} = 0 \abs X_t=v) \}$. Step $(c)$ is again by the Neighborhood Overlap property and induction. 
	
	Since $\bP(Y_1 > t) \leq \bP(\tilde{Y} > t)$ holds for all $t \geq 1$ as claimed, we obtain that $Y_1$ is stochastically dominated by $\tilde{Y}$. Hence, we conclude $\bE(Y_1) \leq \bE(\tilde{Y}) = 1/p$. 
	
	Finally, note that for $Y_i$ with $i>1$, the following still holds 
	\begin{align*}
	\bP(Y_i > 1) = \bP(Z_{U_i+1} =0) = 1- \frac{\abs \neig(X_{U_i}) \cap \neig(D) \abs}{\abs \neig(X_{U_i}) \abs} \leq 1- p = \bP(\tilde{Y} > 1),
	\end{align*}
	and a similar induction argument will conclude that $\bP(Y_i >t) \leq \bP(\tilde{Y}>t)$ for all $t \geq 1$. Note that the only difference in the argument is in the first step since $X_{U_i} \in \neig(D)$ rather than being chosen uniformly at random from $\calV$. Since the Neighborhood Overlap property of the graph $G$ holds for all pairs of vertices, we can conclude $\bE(Y_i) \leq \bE(\tilde{Y})$ as before. This completes the proof. 
\qed

\subsection{Proof of Lemma \ref{lem:pred_tih1}}
\label{ap:pred_what}
\bpf
Let $\mathcal{X}$ denote the set of all trajectories under horizon $K$. Recall that for a given trajectory realization $\bx \in \mathcal{X}$, the optimal Bayes estimator for $\tih$ can be expressed as $\what{T}\ih(\bx) = \argmax_{t \geq 1} \bP(\tih=t \abs \bold{X}=\bx)$. Then, we can write
\begin{align}
\bP(\what{T}\ih = \tih) = & \sum_{\bx \in \mathcal{X}} \bP(\bold{X}=\bx, \what{T}\ih(\bx) = \tih) \nln
= & \sum_{\bx \in \mathcal{X}} \bP( \what{T}\ih(\bx) = \tih \abs \bold{X}=\bx) \bP(\bold{X}=\bx) \nln
= & \sum_{\bx \in \mathcal{X}} \max_{t \geq 1} \bP( \tih = t \abs \bold{X}=\bx) \bP(\bold{X}=\bx) \nln
= & \sum_{\bx \in \mathcal{X}} \max_{t \geq 1} \bP(\bold{X}=\bx, \tih=t). \label{eq:what_tih_overall}
\end{align}
Now, we will compute the joint probability, $\bP(\bold{X}=\bx,\tih=t)$. For this purpose, we will first fix a vertex $v \in \calV$. Then, we will condition on this fixed $v$ being the goal and compute $\bP(\bold{X}=\bx,\tih=t \abs D=v)$. Finally, we will let $v = x_t$ and complete the derivation of $\bP(\bold{X}=\bx,\tih=t)$.

To find $\bP(\bold{X}=\bx,\tih=t \abs D=v)$, we now define a sequence of indicators $\{J_i\}_{i=1}^K$ where for each $i$, we write $J_i = 1$ only if a successful goal-attempt occurs at time $i$ under the Water-Filling Strategy. Similarly, for each $i \geq 1$, we let $A_i$ denote the event $\{X_i = x_i \}$ and define the events $\tilde{A}_i$ as follows:
	\begin{align*}
	\tilde{A}_i = & 
	\begin{cases}
	\{ J_i = 1 \} & \mbox{if } i=t, \\
	\{ J_i = 0 \} & \mbox{otherwise}.
	\end{cases}
	\end{align*}
We can then write
\begin{align}\label{eq:traj_tih_givenD}
	& \bP(\bold{X}=\bx , \tih =t \abs D=v)  = \bP \lt( \bigcap_{i=1}^K ( A_i \cap \tilde{A}_i ) \abs D=v \rt) \nln
	& = \bP \lt( \bigcap_{i=2}^K ( A_i \cap \tilde{A}_i ) \abs A_1 \cap \tilde{A}_1, D=v \rt) \bP \left( A_1 \cap \tilde{A}_1 \abs D=v \right) \nln
	% & = \bP \lt( \bigcap_{i=3}^K ( A_i \cap \tilde{A}_i ) \abs \bigcap_{i=1}^2 A_i \cap \tilde{A}_i, D=v \rt) \bP \left( A_2 \cap \tilde{A}_2 \abs A_1 \cap \tilde{A}_1, D=v \right) \bP \left( A_1 \cap \tilde{A}_1 \abs D=v \right) \nln
	& \overset{(a)}{=} \lt[ \prod_{i=2}^K \bP \lt( A_i \cap \tilde{A}_i \abs \bigcap_{j=1}^{i-1} A_j \cap \tilde{A}_j, D=v \rt) \rt] \bP \left( A_1 \cap \tilde{A}_1 \abs D=v \right),
\end{align}
where $(a)$ is obtained by recursively conditioning on the events $A_i \cap \tilde{A}_i$ for each $i$.

For every $t \geq 1$, let $C_t^v$ denote the number of times a vertex which is a neighbor of the vertex $v$ appears in the trajectory $\bx$ up to time $t$, i.e., $C_t^v = \sum_{i=1}^{t} \mathbb{I} (x_i \in \neig(v))$.  We now define some useful indicator variables to facilitate our analysis. Recall the random variables $\{B_t\}$ defined in Section \ref{sec:model} that capture intrinsic uncertainty: if $B_t=1$, then the agent's action in period $t$ will be  successfully executed whereas she will be sent to a random neighboring vertex, otherwise. With this notation, we define the following sequence $\{\bar{F}_t\}_{t \in \bN}$ as the indicators for the event that the agent chooses Goal-Attempt at $t$ and is successful: 
\begin{equation}
\bar F_{t+1} = \mathbb{I}\left(\bar a_t = \bar a^G, \, B_t = 1\right). 
\end{equation}
 Using this sequence, the intentional goal-hitting time, that is, the first time the agent chooses the meta action Goal-Attempt ($\bar a_t = \bar a^G$) and actually succeeds, can be written as
 \begin{equation}
 \tih = \inf \{t \geq 1 : \bar{F}_t =1 \}
 \end{equation}
Using these definitions, we next examine $\bP \lt( A_i \cap \tilde{A}_i \abs \bigcap_{j=1}^{i-1} A_j \cap \tilde{A}_j, D=v \rt)$ for each $i$ and observe that one of four cases must hold. 

\textbf{Case 1: $i>t$.} Given a successful attempt has already occurred at $t$, the agent will do a random walk under $\psi\wf$ at every period $i>t$. The Random-Step meta action yields $J_i = 0$ so that we obtain
\begin{align*}
\bP \lt( A_i \cap \tilde{A}_i \abs \bigcap_{j=1}^{i-1} A_j \cap \tilde{A}_j, D=v \rt) = \frac{1}{\abs \neig(x_{i-1}) \abs}.
\end{align*}

\textbf{Case 2: $i<t$ and $x_{i-1} \notin \neig(v)$.} Under $\psi\wf$, for each $i<t$, the agent will choose the Random-Step meta action if she is outside the neighborhood of the goal, $v$. Thus, we will have
\begin{align*}
\bP \lt( A_i \cap \tilde{A}_i \abs \bigcap_{j=1}^{i-1} A_j \cap \tilde{A}_j, D=v \rt) = \frac{1}{\abs \neig(x_{i-1}) \abs}.
\end{align*}

\textbf{Case 3: $i<t$ and $x_{i-1} \in \neig(v)$.} If the agent is currently inside the neighborhood of the goal $v$, and if there has not been a successful goal-attempt yet, the agent will choose the corresponding action under the Water-Filling Strategy for period $C_i^v$. Consequently, the probability that the agent does not have a successful attempt at period $i$ and travels to $x_i$ can be written as
\begin{align*}
\bP \lt( A_i \cap \tilde{A}_i \abs \bigcap_{j=1}^{i-1} A_j \cap \tilde{A}_j, D=v \rt) = \frac{ \bP( \bar{F}_{C_{i-1}^v+1} =0) }{\abs \neig(x_{i-1}) \abs}.
\end{align*}

\textbf{Case 4: $i=t$.} Given that a successful attempt has not occurred up to time $t$, the agent can reach her goal intentionally at $t$ if (a) she is in the neighborhood of $v$ at $t-1$, (b) $x_t$ is equal to $v$, and (c) $\bar{F}_{C_{t-1}^v+1}$ is 1. Therefore, 
\begin{align*}
\bP \lt( A_t \cap \tilde{A}_t \abs \bigcap_{j=1}^{t-1} A_j \cap \tilde{A}_j, D=v \rt) = \bI \{ x_{t-1} \in \neig(v) \} \bI \{ x_{t} =v \} \bP( \bar{F}_{C_{t-1}^v+1} =1) .
\end{align*}

Substituting these cases into \eqref{eq:traj_tih_givenD}, we obtain the following:
\begin{align}
& \bP \lt(\bold{X}=\bx, \tih =t \abs D = v \rt) \nln
& = \lt( \prod_{i=1}^{t-1} \frac{\lt( \bI\{ x_{i-1} \notin \neig(v) \} + \bI \{ x_{i-1} \in \neig(v) \} \bP(\bar{F}_{C_{i-1}^v+1} =0) \rt)}{\abs \neig(x_{i-1}) \abs} \rt)\nln
&\cdot\bI \{ x_{t-1} \in \neig(v) \} \bI \{ x_t =v \} \bP(\bar{F}_{C_{t-1}^v+1} =1) \cdot \prod_{i =t+1}^K \frac{1}{\abs \neig(x_{i-1}) \abs} \nln
& {=}  \left(\prod_{i \neq t} \frac{1}{|\neig(x_{i-1})|}\right)\left[ \prod_{i=1}^{t-1}  \left( \bI\{ x_{i-1} \notin \neig(v) \} \right. \right. \nln
& \left. \left. + \bI \{ x_{i-1} \in \neig(v) \} \bP(\bar{F}_{C_{i-1}^v+1} =0) \right) \right] \nln
& \cdot  \bI \{ x_{t-1} \in \neig(v) \} \bI \{ x_t =v \} \bP(\bar{F}_{C_{t-1}^v+1} =1). 
\label{eq:pXtcondv}
\end{align}

We now state and prove Lemma \ref{lem:prod_deg} to obtain an upper bound on the term $\prod\limits_{i \neq t} \frac{1}{|\neig(x_{i-1})|}$. We will shortly use Lemma \ref{lem:prod_deg}.
\begin{lemma}
\label{lem:prod_deg}
Let $G=(\calV, \calE)$ be a connected undirected graph. Fix $x_0 \in \calV$ and $K\in \N$. Denote by $\tilde{X}^{[K]}$ the (random) trajectory of a $K$-step random walk on $G$ that starts from $x_0$. Fix a $K$-step trajectory $\bx$, and $t\in \{1, 2, \ldots, K\}$. We have
\begin{align}
\prod_{i\in \{1, 2, \ldots, K\}\backslash \{t\}} \frac{1}{|\neig(x_{i-1})|} \leq \overline{\Delta}_G \pb\lt(\tilde{X}^{[K]} = \bx \rt).
\label{eq:prod_deg}
\end{align}
\end{lemma}
\emph{Proof of Lemma \ref{lem:prod_deg}.}
We have that
\begin{align*}
\prod_{i \neq t} \frac{1}{|\neig(x_{i-1})|} \leq&  \max_{1 \leq s \leq K}\prod_{i \neq s} \frac{1}{|\neig(x_{i-1})|}  \nln
= & \lt(\max_{0 \leq s \leq K-1} |\neig(x_s)| \rt) \prod_{i =1}^K \frac{1}{|\neig(x_{i-1})|} \nln
\leq & \lt(\max_{v \in \calV} |\neig(v)| \rt) \prod_{i =1}^K \frac{1}{|\neig(x_{i-1})|} \nln
= & \overline{\Delta}_G \prod_{i =1}^K \frac{1}{|\neig(x_{i-1})|} \nln
= & \overline{\Delta}_G \pb(\tilde{X}^{[K]} = \bx), 
\end{align*}
where the last step follows from the definition of a random walk. 
\qed 

We now evoke Lemma \ref{lem:prod_deg} to obtain an upper bound on the term $\prod\limits_{i \neq t} \frac{1}{|\neig(x_{i-1})|}$.  Substituting \eqref{eq:prod_deg} from Lemma \ref{lem:prod_deg} into \eqref{eq:pXtcondv}, we have that
\begin{align}
& \bP \lt(\bold{X}=\bx, \tih =t \abs D = v \rt)  \leq  \overline{\Delta}_G  \pb(\tilde{X}^{[K]} = \bx) \nln
& \cdot \lt[ \prod_{i=1}^{t-1} {\lt( \bI\{ x_{i-1} \notin \neig(v) \} + \bI \{ x_{i-1} \in \neig(v) \} \bP(\bar{F}_{C_{i-1}^v+1} =0) \rt)}\rt] \nln
& \cdot  \bI \{ x_{t-1} \in \neig(v) \} \bI \{ x_t =v \} \bP(\bar{F}_{C_{t-1}^v+1} =1).
\end{align}

Now, let $\bx$ be a trajectory such that $x_t = v$, since otherwise we will have $\bP \lt(\bold{X}=\bx, \tih =t \abs D = v \rt)=0$. For this trajectory, observing that $\bI \{ x_{t-1} \in \neig(v) \} \bI \{ x_t =v \}=1$ holds, we get:
\begin{align}
& = \overline{\Delta}_G  \pb(\tilde{X}^{[K]} = \bx)  \lt[ \prod_{i=1}^{t-1} \lt( \bI\{ x_{i-1} \notin \neig(v) \} + \bI \{ x_{i-1} \in \neig(v) \} \bP(\bar{F}_{C_{i-1}^v+1} =0) \rt) \rt] \bP(\bar{F}_{C_{t-1}^v+1} =1) \nln
& \overset{(a)}{=} \overline{\Delta}_G  \pb(\tilde{X}^{[K]} = \bx) \lt( \prod_{i=1}^{C_{t-2}^v+1}  \bP(\bar{F}_{i} =0) \rt) \bP(\bar{F}_{C_{t-1}^v+1}=1 ) \nln
& \overset{(b)}{=} \overline{\Delta}_G  \pb(\tilde{X}^{[K]} = \bx) \bP(\tih = C_{t-1}^v+1), \label{eq:pXtcondv_upp}
\end{align}	
where $(a)$ follows after substituting the values of $\bI \{ x_{i-1} \in \neig(v) \}$ and $\bI \{ x_{i-1} \notin \neig(v) \}$. Step $(b)$ is obtained by observing that the event $ \lt( \bigcap_{i=1}^{t-1} \{ \bar{F}_i = 0 \} \rt) \cap \{ \bar{F}_t =1 \}$ is equivalent to $\{ \tih = t\}$. 
	
	Next, we use \eqref{eq:pXtcondv_upp} to characterize $\bP(\bold{X}=\bx, \tih =t)$ and replace $v=x_t$. We get:
\begin{align}\label{eq:joint_x_tih}
	\bP(\bold{X}=\bx, \tih = t) = & \bP(\bold{X}=\bx, \tih=t \abs D = x_t) \bP(D=x_t) \nln
	{\leq} & \overline{\Delta}_G  \pb(\tilde{X}^{[K]} = \bx) \bP(\tih = C_{t-1}^{x_t}+1)  \bP(D=x_t) \nln
	\overset{(a)}{\leq} & \overline{\Delta}_G  \pb(\tilde{X}^{[K]} = \bx) \bar{q} \bP(D=x_t) \nln
	\overset{(b)}{\leq} & \overline{\Delta}_G  \pb(\tilde{X}^{[K]} = \bx) \left( \frac{\bar{q}}{n} \right),
	\end{align}	
where step $(a)$ uses that $\bP(\tih = t) \leq \bar{q}$ holds for all $t\geq1$ by design. Finally, step $(b)$ is obtained by recalling that the goal is drawn uniformly at random from $\calV$. 

Since the right hand side of \eqref{eq:joint_x_tih} is independent of $t$, we further have
\begin{align}\label{eq:max_what_tih}
\max_{t \geq 1} \bP(\bold{X}=\bx, \tih = t) \leq \overline{\Delta}_G  \pb(\tilde{X}^{[K]} = \bx) \left( \frac{\bar{q}}{n} \right).
\end{align}

Now, we substitute \eqref{eq:max_what_tih} into \eqref{eq:what_tih_overall} and obtain
	\begin{align*}
\bP(\what{T}\ih = \tih) \leq & \sum_{\bx \in \mathcal{X}} \max_{t \geq 1} \bP(\bold{X}=\bx, \tih=t)	\nln
	\leq & \sum_{\bx \in \mathcal{X}} \overline{\Delta}_G  \pb(\tilde{X}^{[K]} = \bx) \left( \frac{\bar{q}}{n} \right) \\
	\overset{(a)}{=} & \bar{q} \frac{\overline{\Delta}_G}{n}.
	\end{align*}
	Since $\mathcal{X}$ is the set of all trajectories of length $K$ that can be traversed on $G$, $\tilde{X}^{[K]}$ always takes values in $\mathcal{X}$. Thus, step $(a)$ follows from the observation that $\sum\limits_{\bx \in \mathcal{X}} \bP(\tilde{X}^{[K]}=\bx) =1$. This concludes the proof of Lemma \ref{lem:pred_tih1}.
\qed

\subsection{Proof of Lemma \ref{lem:tih_t_diff1}}
\label{ap:pred_diff}
\bpf
To compute the probability that $T$ and $\tih$ are different under the Water-Filling Strategy, we first condition on the value of $\tih$:
\begin{align}
	\bP(\tih \ne T ) = & \sum_{h=1}^{K+1} \bP(\tih \ne T \abs \tih=h ) \bP(\tih =h) . \label{eq:t_diff_condition_tih}
\end{align}
Recall that if $B_t=1$, then the agent's action in period $t$ has been successful whereas she has been sent to a random vertex, otherwise. For each $1 \leq t \leq K+1$, define the random variable $A_t$ as follows:
\begin{align*}
A_t = \bI \{ X_t = D \mbox{ and } B_{t-1}=0 \},
\end{align*} 
so that $A_t$ denotes whether there has been a goal hit that is \emph{not} intentional, at time $t$. 

Note that under $\psi\wf$, at any period $t$, we can have $A_t=1$ under one the following two scenarios:
\begin{enumerate}
\item[(1)] the agent chooses the Random-Step meta action and is sent to $D$, or
\item[(2)] she chooses the Goal-Attempt meta action, fails with $\varepsilon$ probability (i.e., $B_{t-1}=0$), and nature sends her to $D$. 
\end{enumerate}
Since both the Random-Step meta action and the intrinsic uncertainty sample a vertex uniformly at random, independently from all other sources of randomness in the game, $\{A_t\}_{t=1}^{K+1}$ are independent. Hence, under $\psi\wf$, for any $t$ we have:
\begin{align}\label{eq:pA1}
\bP(A_t=1) = \frac{p_{C_{t-1}} \varepsilon + (1-p_{C_{t-1}})}{\abs \neig(X_{t-1}) \abs} \bI \{ X_{t-1} \in \neig(D) \} \leq \frac{1}{\underline{\Delta}_G},
\end{align}
where $C_t = \sum_{i=1}^t \bI \{X_i \in \neig(D) \}$ and $\underline{\Delta}_G = \min\limits_{v \in \calV} \abs \neig(v) \abs$. 

With this notation, given $\tih=h$, we will have $\tih = T$ if for all $t<h$, $A_t$ is 0. We can then write $\bP(\tih = T \abs \tih=h)$ as follows:
\begin{align}
\bP(\tih = T \abs \tih=h) = & \bP \left( \bigcap_{t=1}^{h-1} \{A_t=0\} \right) \nln
\overset{(a)}{=} & \prod_{t=1}^{h-1} \bP \left( A_t=0 \right) \nln
\overset{(b)}{\geq} & \left(1- \frac{1}{\underline{\Delta}_G} \right)^{h-1}, \label{eq:t_same_given_tih}
\end{align}
where $(a)$ is due to the independence of $\{A_t\}_{t=1}^{K+1}$ and $(b)$ follows from \eqref{eq:pA1}. 

Finally, we substitute \eqref{eq:t_same_given_tih} into \eqref{eq:t_diff_condition_tih} and obtain:
\begin{align*}
\bP(\tih \ne T ) = & \sum_{h=1}^{K+1} \bP(\tih \ne T \abs \tih=h ) \bP(\tih =h) \nln
\leq & \sum_{h=1}^{K+1} \lt[ 1 - \left(1- \frac{1}{\underline{\Delta}_G} \right)^{h-1} \rt] \bP(\tih =h) \nln
	= & \bE \left( 1- \left( 1- \frac{1}{\underline{\Delta}_G} \right)^{T_{\mbox{\tiny IH}}-1} \right) \\
	\overset{(a)}{\leq} &  1- \left( 1- \frac{1}{\underline{\Delta}_G} \right)^{\bE(T_{\mbox{\tiny IH}})-1}  \\
	\overset{(b)}{\leq} & 1- \left( 1- \frac{1}{\underline{\Delta}_G} \right)^{K} ,
\end{align*}
	where $(a)$ and $(b)$ are due to the Jensen's Inequality and $\bE(\tih) \leq K+1$, respectively. This completes the proof of Lemma \ref{lem:tih_t_diff1}.
\qed

\subsection{Proof of Proposition \ref{cliquedelay}}
\label{ap:cliquedelay}

Similar to the counter $L(t)$, define another counter $C^{o}_t$ such that for all $t \geq 1$, we have $L(t)+C^o_t=t$. Note that $C^o_t$ records the total time spent outside $\calV_D$ up to time $t$. 

\bpf
	Let us first find an upper bound on the expected total excursion time, that is, expected total time the agent spends outside $\calV_D$. For this purpose, we will use a coupling argument to analyze an alternative system in which transitions to $\calV \setminus \calV_D$ from both $\calV_D$ and $\calV \setminus \calV_D$ are more likely. Further, we will ignore the cases in which $D$ is hit while returning back to $\calV_D$ from $\calV \setminus \calV_D$. Then, the expected intentional goal-hitting time in this alternative system will provide an upper bound for the original $\tih\k$. Denote the number of vertices in each clique by $m=\frac{n}{k}$. 
	
	Under $\psi\k$, in each period $i \geq 1$, the probability that the state will make a transition from $\calV_D$ to $\calV \setminus \calV_D$ can be written as
	\begin{align*}
	\bP(X_{i+1} \notin \calV_D \abs X_i \in \calV_D ) = & \frac{(1-p_{L(i)}(1-\varepsilon))(k)}{m+k-1} := s_{i}. 
	\end{align*}
	
	Let $( \Omega, \mathcal{F}, \bP )$ be a probability space and let $U = \{U_i \}_{i \in \bN}$ be a sequence of i.i.d.~random variables with uniform distribution over $[0,1]$ defined in this space. Construct the sequence of independent Bernoulli random variables $\{C_i\}_{i \in \bN}$ as follows: let
	\begin{align*}
	{C_i} = & \begin{cases}
	1, & \mbox{ if } U_i< s_i, \\
	0, & \mbox{ otherwise,}
	\end{cases}
	\end{align*}
	where the event $\{C_i=1\}$ denotes that there is a transition leaving $\calV_D$ at time $i$. Then, we can represent the number of transitions leaving $\calV_D$ up to time $t$ by $Y_t = \sum_{i=1}^{t} C_i$.
	
	Similarly, construct another sequence of i.i.d.~Bernoulli random variables $\{ \tilde{C}_i \}_{i \in \bN}$ such that
	\begin{align*}
	\tilde{C_i} = & \begin{cases}
	1, & \mbox{ if } U_i < \frac{k}{m+k-1}, \\
	0, & \mbox{ otherwise,}
	\end{cases}
	\end{align*}
	and define $\tilde{Y_t} = \sum_{i=1}^{t} \tilde{C_i}$. Observe that whenever $Y_t$ increases by 1, so does $\tilde{Y}_t$ even though $Y_t$ might stay constant or increase whenever $\tilde{Y_t}$ increases by 1. This implies that $\bP(Y_t \leq \tilde{Y}_t, \forall t)=1$ and the expected number of transitions leaving $\calV_D$ up to time $t$, i.e., $\bE(Y_t)$, satisfies $ \bE(Y_t) \leq \bE(\tilde{Y_t})$.
	
	Next, define $Z_j$ as the time the agent spends outside $\calV_D$ when she leaves $\calV_D$ for the $j^{th}$ time. We know that under $\psi\k$ in each period $i \geq 1$, the probability that the state will stay in $\calV \setminus \calV_D$ is
	\begin{align*}
	\bP(X_{i+1} \notin \calV_D \abs X_i \notin \calV_D ) = & \frac{\varepsilon(m+k-2)}{m+k-1} .
	\end{align*}
	This implies $\{Z_j\}_{j \in \bN}$ is a sequence of i.i.d.~geometric random variables with success probability $1-\varepsilon+\frac{\varepsilon}{m+k-1}$. 
	
	Finally, we combine these results where we write $H=\tih\wf$ to simplify notation whenever $\tih\wf$ is a subscript. The expected total excursion time under $\psi\k$ satisfies:
	\begin{align*}
		\bE(C^o_{\mbox{\tiny H}}) = \bE \left( \sum_{j=1}^{Y_{\mbox{\tiny H}}} Z_j \right) \leq \bE \left(\sum_{j=1}^{\tilde{Y}_{\mbox{\tiny H}}} {Z_j} \right) \overset{(a)}{=} \bE(\tilde{Y}_{\mbox{\tiny H}}) \bE({Z_1}) 
		= \bE \left(\sum_{i=1}^{T^{\mbox{\tiny wf}}_{\mbox{\tiny IH}}} \tilde{C_i} \right) \bE({Z_1}) \overset{(b)}{=} \frac{\bE(\tih\wf) \bE(\tilde{C}_1)}{1-\varepsilon +\frac{\varepsilon}{m+k-1}} ,
		%= \frac{\bE(\tih\wf)(k-1)}{(1-\varepsilon)(m+k-2)},
	\end{align*}
	where $(a)$ follows from Wald's Identity upon observing that $\{{Z}_j\}_{j \in \bN}$ is an i.i.d.~sequence independent from the nonnegative integer valued random variable $\tilde{Y}_H$. Likewise, $(b)$ is due to Wald's Identity as $\{\tilde{C}_i\}_{i \in \bN}$ is an i.i.d.~sequence independent from the nonnegative integer valued $\tih\wf$.
	
	If we ignore the cases in which the goal is reached while going back to $\calV_D$ from another clique, the value we obtain can only be greater than the actual expected intentional goal-hitting time. Then, with time being indexed by $L(\cdot)$, the strategy $\psi\k$ evolves in an identical manner with $\psi\wf$ and the expected intentional goal-hitting time under $\psi\wf$ provides an upper bound for $\bE \left(L({T_{\mbox{\tiny IH}}\k}) \right)$. Thus, writing $\bE(\tih\k) = \bE \left(L({T_{\mbox{\tiny IH}}\k} )\right) + \bE \left(C^o_{T_{\mbox{\tiny IH}}\k} \right) $, we conclude
	\begin{align*}
		\bE(T\k) \leq \bE \left(L({T_{\mbox{\tiny IH}}\k}) \right) + \bE \left(C^o_{T_{\mbox{\tiny IH}}\k} \right)
		\leq  \bE(\tih\wf) + \frac{\bE(\tih\wf) \bE(\tilde{C}_1)}{1-\varepsilon+\frac{\varepsilon}{m+k-1}} = \bE(\tih\wf) + \frac{\bE(\tih\wf)(k)}{(1-\varepsilon)(m+k-1)+\varepsilon}. 
	\end{align*}
This completes the proof of Proposition \ref{cliquedelay}.
\qed

\subsection{Proof of Lemma \ref{lem3}}
\label{ap:lem3}

\bpf
	First, let us calculate the probability that the difference between the two goal-hitting times is equal to $d \in \{ 1, ... , K_n \}$. By conditioning on the value of $\tih^{\psi}$, we get
	\begin{align}
	\bP(\tih^{\psi} - T^{\psi} = d ) = & \sum_{h = 1+d}^{K_n+1} \bP(\tih^{\psi} - T^{\psi}=d \abs \tih^{\psi} =h ) \bP(\tih^{\psi} =h ) \nln
	\leq & \sum_{h = 1+d}^{K_n+1} \bP( T^{\psi}=h-d \abs \tih^{\psi} =h ) \nln
	= & \sum_{h = 1+d}^{K_n+1} \left(1- \frac{1}{n}\right)^{h-d-1} \frac{1}{n} \nln
	= & 1 - \left( 1- \frac{1}{n} \right)^{K_n-d} \nln
	 \leq & 1 - \left( 1- \frac{1}{n} \right)^{K_n}.\label{diffprob}
	\end{align}
	Next, we write the expected difference between the two goal-hitting times as below and use Eq.~\eqref{diffprob}:
	\begin{align*}
	\bE(\tih^{\psi} - T^{\psi}) = & \sum_{d=1}^{K_n} d \bP(\tih^{\psi} - T^{\psi} = d) \nln
	\leq & \sum_{d=1}^{K_n} d \left( 1 - \left( 1- \frac{1}{n} \right)^{K_n} \right)	\nln
	= & \left( 1 - \left( 1- \frac{1}{n} \right)^{K_n} \right) \frac{K_n(K_n+1)}{2} \nln
	= & \sigma_n.
	\end{align*}
	Thus, we have $\bE(\tih^{\psi} ) \leq \bE( T^{\psi}) + \sigma_n$ and $\lim\limits_{n \rightarrow \infty} \sigma_n = 0$, as $K_n \ll {n}$. This completes the proof. 
\qed

\subsection{Proof of Lemma \ref{lem1}}
\label{ap:lem1}
\bpf
	For any fixed agent strategy $\psi$, let 
	\begin{align*}
	t(\psi) \in \argmax_{t \in \bN} \bP(\tih^{\psi} =t).
	\end{align*} 
	Consider the adversary strategy $\tilde{\chi}$ where the adversary makes a prediction at $t(\psi)$ equal to the agent's state at that period, i.e., $\hat{D}_{t(\psi)} = X_{t(\psi)}$. Under $\tilde{\chi}$, the prediction risk of $\psi$ is given by
	\begin{align*}
	q(\psi, \tilde{\chi}) = \bP(T^{\psi} = t(\psi)) .
	\end{align*}
	The success event $\{ T = t(\psi) \}$ of the adversary can be written as the following union of events
	\begin{align*}
		\{T= t(\psi ) \} = \{ T = t(\psi), \tih> t(\psi) \} \cup \{ T= t(\psi), \tih = t(\psi) \},
	\end{align*}
	where under the first event the goal-hitting time is due to a random hit to the goal and under the latter it is due to an intentional hit. Ignoring the first event we have
	\begin{align*}
		\bP( T= t(\psi ) ) \geq \bP( T= t(\psi), \tih = t(\psi) ).
	\end{align*}
	Next, we condition on $\{ \tih = t(\psi ) \}$ and obtain:
	\begin{align*}
		\bP( T= t(\psi), \tih = t(\psi) ) = & \bP( T= t(\psi) \abs \tih = t(\psi) ) \bP( \tih = t(\psi) ) \nln
		\overset{(a)}{=} & \left( 1- \frac{1}{n} \right)^{t(\psi)-1} \tilde{q}_{\psi} \nln
		\overset{(b)}{\geq } & \left( 1- \frac{1}{n} \right)^{K_n} \tilde{q}_{\psi},
	\end{align*}
	where $(a)$ follows upon observing that conditional on $\{ \tih = t(\psi) \}$, for the goal-hitting time to coincide with the intentional goal-hitting time, it must be the case that there have been no random hits to the goal in the first $t(\psi)-1$ periods. Further, $(b)$ is due to $t(\psi) \leq K_n+1$. Finally, we let $\bar{\delta}_n = 1- \lt( 1- \frac{1}{n} \rt)^{K_n}$ and observe that $\lim\limits_{n \rightarrow \infty} \bar{\delta}_n =0$ holds, since $K_n \ll n$. Thus, we conclude $q^*(\psi) \geq q(\psi, \chi) \geq \lt( 1- \bar{\delta}_n \rt) \tilde{q}_{\psi}$.
\qed

\subsection{Proof of Lemma \ref{wf_optimal}} 
\label{app:optimal}
\bpf In order to prove that the Water-Filling Strategy solves the given optimization problem, we will first write the expected intentional goal-hitting time as follows,
\begin{align*}
\bE(\tih ) = \bE(\tih \,\big{|}\, \tih \geq t^*) \bP( \tih \geq t^*) + \bE(\tih \,\big{|}\, \tih < t^*) \bP( \tih < t^*).
\end{align*}

Conditional on the event $\{\tih \geq t^* \}$, where $t^* = \frac{1}{\bar q} - \frac{\varepsilon}{1-\varepsilon}$, the Water-Filling Strategy minimizes the conditional expectation $\bE(\tih \,\big{|}\, \tih \geq t^*)$. This is because the Water-Filling Strategy tries to reach the goal greedily at every period after $t^*$, by setting $p_t = 1$. Indeed, given that $\tih^{\psi}$ is strictly greater than $t^*-1$, the conditional expected value $\bE (\tih^{\psi} \,\big{|}\, \tih^{\psi} \geq t^*)$ needs to be at least $t^*-1 + \frac{1}{1-\varepsilon}$, where $\frac{1}{1-\varepsilon}$ is the stochastic shortest path diameter of the complete graph. However, the expression $t^*-1 + \frac{1}{1-\varepsilon}$ is exactly the conditional expectation of $\tih\wf$ under the same event,
\begin{align*}
\bE (\tih^{\psi} \,\big{|}\, \tih^{\psi} \geq t^*) \geq t^*-1 + \frac{1}{1-\varepsilon} = \bE (\tih\wf \,\big{|}\, \tih\wf \geq t^*).
\end{align*}
Hence, the term $\bE(\tih \,\big{|}\, \tih \geq t^*)$ is minimized under the Water-Filling Strategy.

Recall that under $\psi^{wf}_{\bar{q}}$, we have $\bP(\tih^{\psi}=t) = \bar{q} $ for all $t < t^*$. Thus, the probability that the intentional goal-hitting time is greater than $t^*$, i.e., $\bP(\tih^{\psi} \geq t^*) $, is minimized under the Water-Filling Strategy subject to the requirement that at any period we have $\bP(\tih^{\psi}=t) \leq \bar{q} $.

To see this, suppose that there exists some feasible strategy $\psi$ such that $\bP(\tih^{\psi} \geq t^*) < \bP(\tih\wf \geq t^*) = 1- \bar{q}t^* $. Equivalently, for $\psi$ we have
\begin{align*}
1- \bP(\tih^{\psi} \geq t^*) = \bP(\tih^{\psi} < t^*) > \bar{q} t^*.
\end{align*}
On the other hand, feasibility requires $\max\limits_{t \in \bN} \bP(\tih^{\psi}=t) \leq \bar{q}$. Thus, if $\psi$ is feasible, we must have 
\begin{align*}
\bP(\tih^{\psi} < t^*) = \bP(\tih^{\psi} \leq t^*-1) \leq \bar{q} (t^*-1),
\end{align*}
since otherwise the condition $ \max\limits_{t \in \bN} \bP(\tih^{\psi}=t) \leq \bar{q}$ is violated. However, we have reached a contradiction: $\bP(\tih^{\psi} < t^*) \leq \bar{q} (t^*-1)$ and $ \bP(\tih^{\psi} < t^*) > \bar{q} t^*$. Thus, we conclude that there does not exist any such strategy and that $\psi^{wf}_{\bar{q}}$ minimizes $\bP(\tih^{\psi} \geq t^*) $ among all feasible strategies. Also, note that $\psi^{wf}_{\bar{q}}$ maximizes $\bP(\tih^{\psi} < t^*) $ since the two expressions sum up to 1.

We have so far established that $\psi^{wf}_{\bar{q}}$ minimizes $ \bE(\tih^{\psi} \,\big{|}\, \tih^{\psi} \geq t^*) \bP( \tih^{\psi} \geq t^*) $. Consequently, if there exists a strategy $\psi$ such that $\bE(\tih^{\psi}) < \bE(\tih\wf)$, for this strategy we must have:
\begin{align}\label{ineq1}
\sum_{t = 1}^{t^*-1} t \bP(\tih^{\psi}=t) < \sum_{t = 1}^{t^*-1} t \bP(\tih\wf=t),
\end{align}
while satisfying $\sum_{t = 1}^{\infty} \bP(\tih^{\psi}=t) = 1$. We will now show that this is not possible. 

First, note that having $\bP(\tih^{\psi}=t) > \bP(\tih\wf=t) $ for some $t < t^*$ results in infeasibility since $ \bP(\tih\wf=t) = \bar{q}$. Then, for the strict inequality in Eq. (\ref{ineq1}) to hold, we must have $\bP(\tih^{\psi} = \tilde{t}) = \bP(\tih\wf= \tilde{t}) - \delta $ for some $\tilde{t} < t^*$ and $\delta>0$. Since $\bP(\tih^{\psi} < t^*) $ is maximal under $\psi^{wf}_{\bar{q}}$, there must exist a set of indices $\{ t_1, ..., t_k \}$ such that $k \geq 1$, $t_i \geq t^*$ for all $i=1,...,k$ and the following holds,
\begin{align*}
\bP(\tih^{\psi} = t_1) + ... + \bP(\tih^{\psi} = t_k )
= \bP(\tih\wf= t_1) + ... + \bP(\tih\wf= t_k) + \delta.
\end{align*}
Without loss of generality, let $k = 1$. For the expectations, because $\tilde{t} < t^* \leq t_1$, it follows that
\begin{align*}
\bE(\tih^{\psi}) = \bE(\tih\wf) - \delta \tilde{t} + \delta t_1 > \bE(\tih\wf).
\end{align*}

Accordingly, we can conclude that it is not possible to construct a strategy $\psi$ for which the expected intentional goal-hitting time is smaller than that under the Water-Filling Strategy without violating the constraint. For this reason, the Water-Filling Strategy solves the given optimization problem and we conclude the proof.
\qed

\subsection{Proof of Theorem \ref{thm:tx_lowerbound}} \label{ap:lower_bound}
\bpf
For completeness, we include the proof of the lower bound in Theorem 1 of \cite{tsitsiklis2018delay} and point out some minor changes from the prior work. 

Fix an agent strategy $\psi$ for which $\bE(T) \leq w$. Define 
\begin{align}
t(\psi) \in \arg\max_{t \in \N} \bP( T = t), 
\end{align}
and suppose the adversary's strategy, denoted by $\chi_t$, consists of predicting the goal vertex to be agent's state at time $t(\psi)$, i.e., $\hat{D} = X_{t(\psi)}$. We note that the offline adversary strategy $\chi_t$ only needs to choose a prediction, whereas the strategy defined in \cite{tsitsiklis2018delay} also needs to specify the timing of the prediction. This implies that the probability that the adversary succeeds is equal to the probability that the goal-hitting time $T$ is equal to $t(\psi)$. Hence, we have
\begin{align}
q(\psi, \chi_t) \geq \bP( T = t(\psi) ) = \max_{t \in \N} \bP(T= t) \overset{(a)}{\geq} \frac{1}{2 \bE(T) + 1} \overset{(b)}{\geq} \frac{1}{2w+1}.
\end{align}
where step $(b)$ follows from the assumption $\bE(T) \leq w $ and $(a)$ applies Lemma \ref{lem:lower_bound_markov} on $T$:
\begin{lemma}\label{lem:lower_bound_markov}
Let $Y$ be a random variable taking values in $\N$. Then, there exists $y \in \N$ such that
\begin{align}
\bP(Y=y) \geq \frac{1}{2 \bE(Y) + 1}.
\end{align}
\end{lemma}
Finally, we prove Lemma \ref{lem:lower_bound_markov} by contradiction. Let $\mu = \bE(Y)$. Suppose, for the sake of contradiction, that for all $y \in \N$, $\bP(Y=y) < \frac{1}{2w+1}$ holds. Then, we have
\begin{align*}
\bP(Y \geq i) = 1 - \bP( Y < i) = 1 - \sum_{j=1}^{i-1} \bP(Y=i) > 1- \frac{(i-1)}{2w+1}.
\end{align*}
Next, we derive $\bE(Y)$ as follows and obtain a contradiction:
\begin{align*}
\mu = \sum_{i=1}^{\infty} \bP(Y \geq i) \geq \sum_{i=1}^{ \lfloor 2 \mu + 1 \rfloor } \bP(Y \geq i) > \sum_{i=1}^{ \lfloor 2 \mu + 1 \rfloor } 1- \frac{(i-1)}{2w+1} = \lfloor 2 \mu + 1 \rfloor -\frac{\lfloor 2 \mu + 1 \rfloor - 1}{2} \geq \mu.
\end{align*}
This completes the proof of the lower bound in Theorem \ref{thm:tx_lowerbound}.
\qed

\ifx \useplain\undefined
\end{APPENDICES}
\fi

\end{document}